\documentclass[12pt,letterpaper,usenames,dvipsnames]{article}
\usepackage{jheppub}

\usepackage{dsfont}
\usepackage{amssymb}
\usepackage{latexsym}
\usepackage{braket}
\usepackage{graphicx}
\usepackage{subcaption}
\usepackage[utf8]{inputenc}
\usepackage{multirow}
\usepackage{graphicx,tikz}
\usepackage[framemethod=TikZ]{mdframed} 
\usepackage{tikz-cd} 
\usetikzlibrary{arrows,snakes,shapes.arrows,decorations.markings}
     \tikzset{>=triangle 90}
     \tikzstyle{bbc}=[draw,circle,fill=black,scale=.75]
     \tikzstyle{rc}=[circle,fill=red,scale=.6]
     \tikzstyle{wc}=[draw,circle,scale=.75]

\usetikzlibrary{calc}
\usepackage{hyperref}
\definecolor{darkred}{rgb}{0.8,0.1,0.1}
\hypersetup{colorlinks=true, linkcolor=darkred, citecolor=blue, linktoc=page}

\renewcommand{\thanks}[1]{\footnote{#1}}

\newcommand{\bea}{\begin{eqnarray}}
\newcommand{\eea}{\end{eqnarray}}

\newcommand{\be}{\begin{equation}}
\newcommand{\ee}{\end{equation}}

\usepackage{adjustbox}
\usepackage{multirow}
\usepackage{hhline}
\usepackage{arydshln}
\usepackage{lscape}

\def\a{\alpha}

\def\cC{{\cal C}}

\def\cF{{\cal F}}

\def\cH{{\cal H}}

\def\cN{{\cal N}}
\def\cO{{\cal O}}
\def\cP{{\cal P}}

\def\cR{{\cal R}}
\def\cT{{\cal T}}

\def\cW{{\cal W}}
\def\cX{{\cal X}}

\def\cZ{{\cal Z}}

\def\Z{{\mathbb Z}}

\def\half{ {1\over 2}}

\def\ZZ{{\mathbb Z}}
\def\RR{{\mathbb R}}
\def\NN{{\mathbb N}}
\def\CC{{\mathbb C}}

\def\mS{{\mathsf S}}
\def\mT{{\mathsf T}}
\def\mC{{\mathsf C}}
\def\mU{{\mathsf U}}

\usepackage{colortbl}
\definecolor{llightyellow}{rgb}{1.0, 0.95, 0.7}
\definecolor{llightblue}{rgb}{0.7, 0.9, 1.0}
\definecolor{llightpink}{rgb}{1.0, 0.85, 0.95}
\definecolor{llightgreen}{rgb}{0.7, 1.0, 0.4}
\colorlet{lightyellow}{llightyellow!50!white}
\colorlet{lightblue}{llightblue!50!white}
\colorlet{lightgreen}{llightgreen!50!white}
\colorlet{lightpink}{llightpink!50!white}

\def\no{\nonumber}
\def\text{\mathrm}
\usepackage{placeins}
\usepackage{afterpage}

\usetikzlibrary{decorations.pathreplacing,calligraphy}

\usepackage{pifont}
\usepackage{amssymb}
\def\tauYM{\tau_{\mathrm{YM}}}

\usepackage{mathtools}

\definecolor{dgreen}{rgb}{0, 0.55, 0}

\makeatletter
\def\@fpheader{\ }
\makeatother

\newcommand{\Tr}{\mathrm{Tr}}

  \usepackage{changepage}

\begin{document}
\begin{titlepage}

\begin{flushright}
\end{flushright}

\vskip 3cm

\begin{center}

{\Large \bfseries Non-Invertible Symmetries of $\cN=4$ SYM and\\\bigskip  Twisted Compactification}

\vskip 1cm
Justin Kaidi$^1$, Gabi Zafrir$^{1,2}$, and Yunqin Zheng$^{3,4}$
\vskip 1cm

\begin{centering}
\begin{tabular}{ll}
  $^1$&Simons Center for Geometry and Physics, \\& Stony Brook University, Stony Brook, NY 11794-3636, USA\\
  $^2$&C.~N.~Yang Institute for Theoretical Physics, \\& Stony Brook University, Stony Brook, NY 11794-3840, USA\\
  $^3$&Kavli Institute for the Physics and Mathematics of the Universe, \\
& University of Tokyo,  Kashiwa, Chiba 277-8583, Japan\\
$^4$&Institute for Solid State Physics, \\
&University of Tokyo,  Kashiwa, Chiba 277-8581, Japan\\
\end{tabular}
\end{centering}

\vskip 1cm

\end{center}

\noindent
Non-invertible symmetries have recently been understood to provide interesting contraints on RG flows of QFTs. In this work, we show how non-invertible symmetries can also be used to generate entirely \textit{new} RG flows, by means of so-called \textit{non-invertible twisted compactification}. We illustrate the idea in the example of twisted compactifications of 4d $\cN=4$ super-Yang-Mills (SYM) to three dimensions. After giving a catalogue of non-invertible symmetries descending from Montonen-Olive duality transformations of 4d $\cN=4$ SYM, we show that twisted compactification by non-invertible symmetries can be used to obtain 3d $\cN=6$ theories which appear otherwise unreachable if one restricts to twists by invertible symmetries.

\end{titlepage}

\setcounter{tocdepth}{3}
\tableofcontents

\section{Introduction}


Symmetries are a concept of fundamental importance to theoretical physics, and are well-known to give constraints on the dynamics of physical systems. This is true already in classical physics where by Noether's theorem (continuous) symmetries imply conservation laws that are highly non-trivial. Similar non-trivial restrictions exist in quantum physics, including Ward identities and selection rules. In quantum field theory, symmetries are also useful through their 't Hooft anomalies, which provide non-trivial constraints on renormalization group (RG) flow. 

Recently, there has been renewed interest in the study of symmetries in quantum theories. In particular, there is a growing realization that quantum theories allow for more general symmetries than just the familiar ones forming groups and acting on local operators. 
Central to this new understanding is the idea that each symmetry is associated with a topological operator/defect \cite{Gaiotto:2014kfa}. Specifically, given a continuous zero-form symmetry, we can generate a codimension-one operator by integrating the associated conserved current on a spatial slice. Current conservation then implies that the resulting operator does not change under continuous deformations, and hence is topological. Analogous operators should also exist for discrete symmetries, despite the lack of a conserved current in that case.

 Various properties of the symmetries can then be recast in the language of the associated topological defects, with e.g. the group law of the symmetry being manifested in the fusion properties of the topological defects. Specifically, given two defects associated with group elements $g_1$ and $g_2$, we can merge them to obtain a new topological defect associated with the group element $g_1 g_2$. The group property implies that for any such topological defect there exists another topological defect such that the fusion results in the trivial topological defect---in other words, the operator is invertible.

We can, however, also consider topological defects whose properties deviate from those associated with standard symmetries. This leads to the notion of ``generalized symmetries," and includes higher-form symmetries (for which the topological operator has codimension greater than one)
  and non-invertible symmetries (for which the associated topological operators do not possess an inverse, and as such do not form a group). The latter will be especially important for our purposes---they have been studied in two-dimensions in \cite{Fuchs:2002cm,Aasen:2016dop,Chang:2018iay,Bhardwaj:2017xup,Thorngren:2019iar,Komargodski:2020mxz,Thorngren:2021yso,Huang:2021zvu,Inamura:2021wuo,Huang:2021nvb, Feiguin:2006ydp, Inamura:2021szw, Vanhove:2021zop, Aasen:2020jwb} 
  and more recently in higher dimensions in \cite{Nguyen:2021yld,Koide:2021zxj,Kaidi:2021xfk,Choi:2021kmx,Roumpedakis:2022aik,Bhardwaj:2022yxj,Hayashi:2022fkw,Arias-Tamargo:2022nlf,Choi:2022zal}.
Despite going beyond our traditional notions of symmetry, the topological nature of these operators suggests that they give rise to constraints similar to those provided by ordinary symmetries, including constraints on RG flows \cite{Chang:2018iay,Thorngren:2019iar,Thorngren:2021yso,Komargodski:2020mxz,Choi:2021kmx,Choi:2022zal}.

Rather than using non-invertible symmetries to \textit{constrain} RG flows, in the current work we will aim to use non-invertible symmetries to generate entirely \textit{new} RG flows. This involves the notion of \textit{twisted compactification}. Concretely, given a QFT in $(d+1)$-dimensions, one can generate a new QFT in $d$-dimensions by taking one of the space directions to be a circle of radius $R$ and taking the limit $R\rightarrow 0$. If the original QFT possesses symmetries, then we can also introduce holonomies associated with these symmetries along the circle direction. This will change the resulting $(d-1)$-dimensional QFT, and when the symmetry is discrete is referred to as a twisted compactification. 

In this work we shall focus mainly on the special case of $4d$ $\mathcal{N}=4$ super Yang-Mills (SYM). This theory has a conformal manifold parameterized by a complex coupling $\tauYM$, and possesses a Montonen-Olive duality group $SL(2,\mathbb{Z})$ mapping the theory with a given value of $\tauYM$ to an equivalent theory at a generically different value. For special values of $\tauYM$ that remain unchanged under some subset of $SL(2,\mathbb{Z})$ transformations, a portion of this duality group may become a symmetry. 
Given such a symmetry, we can then consider compactifying $\mathcal{N}=4$ super Yang-Mills to $3d$ with a twist by it. If done appropriately, this leads to novel 3d theories preserving various amounts of SUSY, including $\mathcal{N}=6,4,$ and $2$ \cite{Ganor:2008hd,Ganor:2010md,Ganor:2012mu}.

There is however one complication. The $SL(2,\mathbb{Z})$ transformation in general changes the global form of the gauge group, i.e. it changes the spectrum of line operators in the theory \cite{Aharony:2013hda,Ang:2019txy}. This means that even for values of $\tauYM$ that are fixed by an $SL(2,\mathbb{Z})$ transformation, said transformation may not be a symmetry if the global form of the gauge group changes. While in some cases it is possible to find a global form that is invariant under a chosen transformation, as we shall soon see there are cases where no such global form exists.\footnote{See \cite{Evtikhiev:2020yix} for previous explorations of this, and \cite{Hayashi:2022fkw} for an analogous discussion in the context of the Cardy-Rabinovici model.} Thus, we would naively conclude that such transformations can never be used in a twisted compactification.

One of the main messages of this work is that these would-be symmetries can be salvaged by combining them with the gauging of a one-form symmetry, which maps the global form of the gauge group back to that of the original theory. The price we pay for this is that the resulting symmetry becomes non-invertible.  
We are thus led to consider twisted compactification by non-invertible symmetries.

In the current work, we shall classify the non-invertible symmetries of $\mathcal{N}=4$ SYM for all choices of gauge groups (see \cite{Choi:2022zal} for first steps in this direction), and having done so we will proceed to a discussion of twisted compactification by these non-invertible symmetries. This will allow us to construct exotic $\mathcal{N}=6$ SCFTs that would appear unreachable if one restricts to only twisted compactification by invertible symmetries.  Before getting into the details, let us outline some of the salient features of our program here.

\subsection*{Non-invertible defects} 

We begin by introducing the non-invertible symmetries that will be the focus of this work. These are based on the results of \cite{Kaidi:2021xfk,Choi:2021kmx,Choi:2022zal}, in which the authors identified a general construction for a class of non-invertible symmetries in four-dimensional gauge theories. The basic idea is to consider a theory with a $\ZZ_N^{(1)}$ one-form symmetry, and to introduce an operation $\sigma$ defined by gauging this symmetry (or a subgroup thereof). Associated to the operation $\sigma$ is a topological interface obtained by gauging $\ZZ_N^{(1)}$  in half of the space and imposing Dirichlet boundary conditions at the boundary. In certain situations the theories on both sides of the interface are equivalent, in which case the interface becomes a defect in a single theory. Checking the fusion rules of this defect shows immediately that it is non-invertible. 

An important point is that, while the theories on either side of the defect must be \textit{isomorphic} (i.e. equivalent), the isomorphism between the two need not be trivial. In other words, the two can differ by a non-trivial invertible transformation. In such cases it is best to think of the non-invertible defect not just as the defect defined by $\sigma$, but rather as the combination of $\sigma$ with the additional invertible defect. 

Let us illustrate this by means of a simple example, namely $\cN=4$ $SU(2)$ SYM theory. In this case $\sigma$ corresponds to gauging of the $\ZZ_2^{(1)}$ one-form symmetry, and acting with it gives a different global variant of the theory known as $SO(3)_+$, as will be discussed in  detail below.\footnote{Technically, this statement must be modified in the presence of non-trivial background gauge fields for the one-form symmetry. This will be discussed in the main text.} Importantly, the operation $\sigma$ does not change the value of the complex coupling $\tauYM$, so if we began with $SU(2)$ at $\tauYM$, then after applying $\sigma$ we will obtain $SO(3)_+$ at $\tauYM$. In general, $SO(3)_+$ is not equivalent to $SU(2)$, but at the special point $\tauYM=i$ it is a non-trivial fact that the two \textit{are} equivalent---this is the statement of S-duality \cite{Aharony:2013hda}. However, though the two theories are equivalent at $\tauYM=i$, there is still a non-trivial mapping between them, which in particular involves a reorganization of the spectrum of line operators. This non-trivial mapping is given by the modular $\mS$ transformation,
\bea
\mS\, SO(3)_+[\tauYM] = SU(2)[-1/\tauYM]~,
\eea
which is well-defined for any value of $\tauYM$.
We then conclude that there is a non-invertible symmetry at $\tauYM=i$, which is most cleanly understood as the composition of the $\sigma$ and $\mS$ actions. We will denote this composite non-invertible symmetry by $\cN := \sigma \mS$, and by abuse of notation will denote the corresponding defect by the same letter, occasionally also indicating the codimension-one manifold on which it lives, e.g. $\cN(M_3)$. See Figure \ref{fig:Ndef} for the schematic result. This non-invertible symmetry was first identified in \cite{Kaidi:2021xfk}.

\begin{figure}[!tbp]
\begin{center}
\[\hspace*{-0.8 cm}\begin{tikzpicture}[baseline=19,scale=0.8]
\draw[  thick] (0,-0.2)--(0,3.2);
\draw[  thick] (3,-0.2)--(3,3.2);
  
          \shade[line width=2pt, top color=red,opacity=0.4] 
    (0,0) to [out=90, in=-90]  (0,3)
    to [out=0,in=180] (3,3)
    to [out = -90, in =90] (3,0)
    to [out=190, in =0]  (0,0);
    
\node[left] at (-0.7,1.5) {$SU(2)[\tauYM]$};
    \node at (1.5,1.5) {$SO(3)_+[\tauYM]$};
    \node[right] at (3.7,1.5) {$SU(2)[-{1\over \tauYM}]$};
      \node[left] at (0,-0) {$\sigma$};
          \node[right] at (3,-0) {$\mS$};  
\end{tikzpicture}
\hspace{0.35 in}\Rightarrow\hspace{0.35 in}
\begin{tikzpicture}[baseline=19,scale=0.8]
\draw[red, thick] (0,-0.2)--(0,3.2);
  \node[left] at (-0.7,1.5) {$SU(2)[\tauYM]$};
    \node[right] at (+0.7,1.5) {$SU(2)[-{1\over \tauYM}]$};
      \node[left] at (0,0) {$\cN$};
\end{tikzpicture}
\]
\caption{At $\tauYM=i$, the $SU(2)$ theory has a non-invertible defect $\cN$, which can be understood as the composition of a defect $\sigma$ implementing gauging of the $\ZZ_2^{(1)}$ one-form symmetry, together with an invertible $\mS$ defect. }
\label{fig:Ndef}
\end{center}
\end{figure}

We now reproduce the fusion rules for the defect $\cN(M_3)$, which were originally obtained in \cite{Kaidi:2021xfk,Choi:2021kmx,Choi:2022zal}. For our purposes, we will only need the following portions of the fusion ring,
\bea
\label{eq:dualityfusionrules}
\cN(M_3) \times \overline{\cN}(M_3) &=& {1\over |H^0(M_3, \ZZ_N)|} \sum_{\Sigma \in H_2(M_3, \ZZ_N)} L(\Sigma)~,
\no\\
\cN(M_3) \times \cN(M_3) &=& {\cC \over |H^0(M_3, \ZZ_N)|} \sum_{\Sigma \in H_2(M_3, \ZZ_N)} L(\Sigma)~,
\no\\
\cN(M_3) \times L(\Sigma) &=& \cN(M_3)~,
\eea
where $\cC$ is the charge conjugation operator and $N$ is the order of the one-form symmetry group (namely $N=2$ for the case of $SU(2)$).

An exactly similar story holds in the case of triality. This can again be illustrated by the simple example of $\cN=4$ $SU(2)$ SYM. In particular, this theory has a triality defect at the point $\tauYM = e^{2 \pi i /3}$, which is the composition of $\sigma$ with the modular transformation $\mS \mT$ to give $\cN:=\sigma \tau \mS \mT$.\footnote{The operation $\tau$ represents stacking with an invertible phase, and will be introduced in the main text. It should not be confused with the complex coupling of the theory, which we denote by $\tauYM$.} The fusion rules for such defects were obtained in \cite{Choi:2022zal}, with the results differing depending on the (parity of) the order of the one-form symmetry.  When the order of the one-form symmetry is even, as for $SU(2)$, the relevant portions of the fusion rules are 
\bea\label{trialityfusion}
\overline{\cN}(M_3) \times \cN(M_3) &=&{1\over |H^0(M_3, \ZZ_N)|} \sum_{\Sigma \in H_2(M_3, \ZZ_N)} L(\Sigma)~,
\no\\
 \cN(M_3) \times \overline{\cN}(M_3) &=& {1\over |H^0(M_3, \ZZ_N)|} \sum_{\Sigma \in H_2(M_3, \ZZ_N)} (-1)^{Q(\Sigma)} L(\Sigma)~,
 \no\\
 \cN(M_3)\times \cN(M_3) &= & \cZ_{U(1)_N}(M_3)\, \overline{\cN}(M_3)~.
\eea
Above, the coefficient $\cZ_{\cT}(M_3)$ represents the value of the partition function of the TQFT $\cT$ on $M_3$.

\subsection*{Twisted compactification by non-invertible defects} 

We now give more details on the notion of twisted compactification by a non-invertible defect. 
We recall that standard twisted compactification involves choosing an element $g \in G$ and compactifying the theory with $g$ holonomy around the circle,
i.e. imposing
\bea
\cO(x, \theta+2 \pi) = g\, \cO(x, \theta)
\eea
for any operator $\cO(x, \theta)$ at a point $x \in M_{d}$ and $\theta \in S^1$.\footnote{The operator $\cO$ need not be local, and can be extended along some of the directions of $M_3$, in which case we should modify the argument $x$ appropriately. We will for simplicity not discuss operators which wrap $S^1$.} In more modern language, twisted  compactification amounts to compactification on a circle with the insertion of a codimension-one defect labelled by $g$, located at a point along the circle and extending in the remaining directions. 

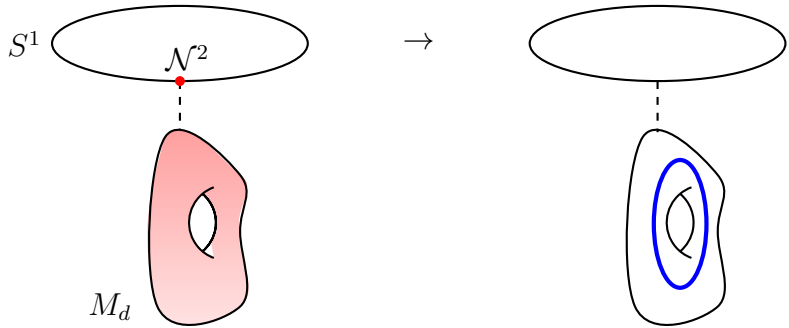
\begin{figure}[tbp]
\begin{center}

  \begin{tikzpicture}
\draw[thick] (0,0) circle [x radius=1.7cm, y radius=0.5cm];
\node[above] at (0.1,-0.5) {$\cN^2$};
\node[left] at (-1.7,0) {$S^1$};
\node[left] at (-0.5,-3.5) {$M_d$};
\draw[thick,dashed] (0,-0.5) to (0,-1.18);
\node[circle,draw=red,fill=red,scale=0.3] at (0,-0.5) {}; 
  \begin{scope}[rotate=-90,scale=0.5,xshift = 3 cm,yshift=-0.4cm]
    \shade[top color=red!40, bottom color=red!10]  
    (-0.5,0) to[out=140,in=-130] (0.5,2)
   to[out=60,in=200]  (2,2) 
   to[out=-10,in=130] (4,2)
   to[out=-50,in=30] (4,-0.2)
   to[out=195, in=-16] (-0.5,0);
    \draw[thick, smooth cycle,tension=.7] plot coordinates{(-0.5,0) (0.5,2) (2,2) (4,2) (4,-0.2)};
    \coordinate (A) at (1,1);
    \draw[thick, fill=white] (A) arc(140:40:1) (A) arc(-140:-20:1) (A) arc(-140:-160:1) (A) arc(140:40:1);
      \draw[ thick, white] (1.07,1) --(2.47,1);
      
    \end{scope}
   \begin{scope}[xshift = 1.25 in]
    \node[] at (0,0) {$\rightarrow$}; 
      \end{scope}
      
        \begin{scope}[xshift = 2.5 in]
    \draw[thick] (0,0) circle [x radius=1.7cm, y radius=0.5cm];
\draw[thick,dashed] (0,-0.5) to (0,-1.18);
  \begin{scope}[rotate=-90,scale=0.5,xshift = 3 cm,yshift=-0.4cm]
   \draw[ultra thick,blue] (1.8,1) circle [x radius=1.7cm, y radius=0.7cm];
    \draw[thick, smooth cycle,tension=.7] plot coordinates{(-0.5,0) (0.5,2) (2,2) (4,2) (4,-0.2)};
    \coordinate (A) at (1,1);
    \draw[thick, fill=white] (A) arc(140:40:1) (A) arc(-140:-20:1) (A) arc(-140:-160:1);
    \end{scope}
    \end{scope}

  \end{tikzpicture}
\caption{Left: an insertion of $\cN^2$ at a point on $S^1$ and wrapping $M_d$. Right: after having used the fusion rules, we may replace this with an insertion of the condensate (blue) on all two-cycles of $M_d$. This shows that twisted compactification by the square $\cN^2$ of a duality defect  is equivalent to normal compactification together with gauging in the remaining three-dimensions (up to charge conjugation). }
\label{fig:twistredundancy}
\end{center}
\end{figure}

In standard twisted compactifications, the defect inserted along the circle is invertible. 
If we consider a theory on $X_{d+1} = \Sigma_n \times M_{d-n+1}$ with symmetry group $G$, the set of all possible twisted compactifications on $\Sigma_n$ is given by $\mathrm{Hom}(\pi_1(\Sigma_n), \, G)$.  On the other hand, for non-invertible twisted compactifications the relevant structure is more subtle, and we will not give a detailed account of it here. Instead, in the current work we will restrict ourselves to the case of compactifaction on a single circle,  i.e. $X_{d+1} = S^1 \times M_d$. In this case the  set of non-invertible twisted compactifications is simply given by the set of distinct insertions of non-invertible defects along the circle.

Let us begin by considering the case of a theory with a single species of duality defect $\cN$, with fusion rules as given in (\ref{eq:dualityfusionrules}). One might naively envision twisting by $\cN$, $\cN^2$, $\cN^3$, etc. to obtain various new theories upon compactification, but most of these are in fact redundant. Indeed, by the fusion rules a compactification with $\cN^2$ twist is equivalent to the insertion of a fine mesh of defects on all two-cycles of $M_{d}$ (together with charge conjugation), with no insertions on the circle; c.f. Figure \ref{fig:twistredundancy}. This mesh of defects is referred to as a ``condensation defect," and is known to implement gauging \cite{Kaidi:2021xfk,Choi:2021kmx,Choi:2022zal,Roumpedakis:2022aik}. Thus the compactification with $\cN^2$ twist gives a $d$-dimensional theory which is simply a discrete gauging of the theory obtained by \textit{untwisted} circle compactification, and for that reason is rather uninteresting.\footnote{Note that upon circle compactification a one-form symmetry in $(d+1)$-dimensions gives rise to both one- and zero-form symmetries in $d$-dimensions. The insertion of the condensation defect in $M_{d}$ corresponds to gauging the zero-form symmetry in the compactified theory.} In a similar manner, the $\cN^3$ twisted compactification gives essentially identical results as the $\cN$ twisted compactification (up to discrete gauging and charge conjugation), and so on. For this reason, it is really only the $\cN$ compactification that will be of interest to us here.

\begin{figure}[tbp]
\begin{center}

  \begin{tikzpicture}[scale=0.8]
\draw[thick] (0,0) circle [x radius=1.7cm, y radius=0.5cm];
\node[below] at (0.1,-0.5) {$\cN$};
\node[] at (0,1) {$SU(2)$};
\node[circle,draw=red,fill=red,scale=0.3] at (0,-0.5) {}; 
   
    \begin{scope}[xshift = 1 in]
    \node[] at (0,0) {$\rightarrow$}; 
      \end{scope}
    
\begin{scope}[xshift=2 in]
\draw[thick] (0,0) circle [x radius=1.7cm, y radius=0.5cm];
\node[] at (0,1) {$SU(2)$};

\draw[thick,blue] ($(240:1.7 and 0.5)$) arc [start angle=240,   
                  end angle=300,
                  x radius=1.7cm, 
                  y radius=0.5cm];
                  
\node[above] at ($(240:1.7 and 0.5)$) {$\sigma$};
\node[circle,draw=red,fill=red,scale=0.3] at ($(240:1.7 and 0.5)$) {}; 
\node[above] at ($(300:1.7 and 0.5)$) {$\mS$};
\node[circle,draw=red,fill=red,scale=0.3] at ($(300:1.7 and 0.5)$) {}; 

\node[] at (0,-1) {$SO(3)_+$};
\end{scope}

    \begin{scope}[xshift = 3 in]
    \node[] at (0,0) {$\rightarrow$}; 
      \end{scope}
    
 \begin{scope}[xshift=4in]
    \draw[thick,blue] (0,0) circle [x radius=1.7cm, y radius=0.5cm];
\node[] at (0,1) {$SU(2)$};

\draw[thick] ($(120:1.7 and 0.5)$) arc [start angle=120,   
                  end angle=60,
                  x radius=1.7cm, 
                  y radius=0.5cm];
                  
\node[below] at ($(120:1.7 and 0.5)$) {$\sigma$};
\node[circle,draw=red,fill=red,scale=0.3] at ($(120:1.7 and 0.5)$) {}; 
\node[below] at ($(60:1.7 and 0.5)$) {$\mS$};
\node[circle,draw=red,fill=red,scale=0.3] at ($(60:1.7 and 0.5)$) {}; 

\node[] at (0,-1) {$SO(3)_+$};
\end{scope}

 \begin{scope}[xshift = 5 in]
    \node[] at (0,0) {$\rightarrow$}; 
      \end{scope}

   \begin{scope}[xshift=6in]  
    \draw[thick,blue] (0,0) circle [x radius=1.7cm, y radius=0.5cm];

\node[above] at ($(90:1.7 and 0.5)$) {$\cN$};
\node[circle,draw=red,fill=red,scale=0.3] at ($(90:1.7 and 0.5)$) {}; 
\node[] at (0,-1) {$SO(3)_+$};
\end{scope}

  \end{tikzpicture}
\caption{Twisted compactification of $SU(2)$ SYM and $SO(3)_+$ SYM by the non-invertible defect $\cN$ gives the same theory in three dimensions. This can be seen by splitting and moving the topological defects in the manner shown above.}
\label{fig:differentvars}
\end{center}
\end{figure}
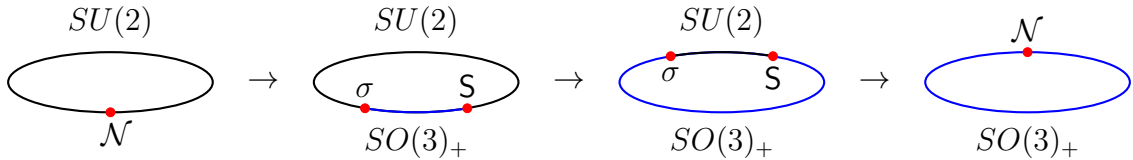

Analogous statements hold for triality defects. In that case we might suspect two distinct possibilities for twisted compactification, corresponding to insertion of $\cN(M_3)$ or $\overline{\cN}(M_3)$. These however will again give rise to the same 3d theories, up to stacking with a decoupled TQFT. To see this, consider first the insertion of $\cN(M_3)$ at a point on $S^1$. Moving an operator around the circle once produces an action of $\cN(M_3)$, whereas moving the operator around the circle \textit{twice} produces an action of $\cN(M_3)^2 = \cZ_\cT(M_3) \overline{\cN}(M_3)$. Hence this is the same as the insertion of $\overline{\cN}(M_3)$, up to stacking with $\cT$ in the copy of $M_3$ fibered over that point on $S^1$. Upon compactification, we obtain the same 3d theory as for an $\overline{\cN}(M_3)$ twisted compactification, up to stacking with the decoupled TQFT $\cT$.

Finally, note that given two $(d+1)$-dimensional theories in the same modular orbit, any $d$-dimensional theory which can be obtained via non-invertible twisted compactifications of the first theory can  also be obtained via (non-)invertible twisted compactification of the second theory. For example, any theory which can be obtained via twisted compactification of the $SU(2)$ theory can also be realized as a twisted compactification of the $SO(3)_+$ theory. The argument is most easily seen pictorially, as shown in Figure \ref{fig:differentvars}. In words, one decomposes the non-invertible defect $\cN$ into its elementary building blocks $\sigma$ and $\mS$ (as well as $\tau$ and $\mT$ in the case of triality), and then using the topological nature of each interface reorganizes them so as to expand the pockets of any particular element of the modular orbit to fill the circle.

\subsection*{Intrinsic vs. non-instrinsic non-invertibility}
 
We close this introduction by discussing an important and somewhat confusing point regarding ``invertibility" or ``non-invertiblity" of defects in a theory. The naive question that one would like to ask is whether, given a SYM theory $X$ at coupling $\tauYM$, the theory has non-invertible defects. Recall that the parameter $\tauYM$ is built from the theta-angle and gauge coupling via
\bea
\tauYM = {\theta \over 2\pi} + {8 \pi i \over g^2_{\mathrm{YM}}}~,
\eea
and hence to obtain a physically sensible theory one should restrict to $\mathrm{Im}\,\tauYM \geq 0$, i.e. $\tauYM$ takes values in the upper half-plane $\cH$. Typically, one further constrains $\tauYM$ to lie in the fundamental domain $\cF:=SL(2, \ZZ) \,\backslash\,\cH $, since each point outside of this domain can be mapped to a point inside the domain via an appropriate modular transformation.

However, in general the question of whether a given theory does or does not possess non-invertible symmetries depends on the value $\tauYM$ not just as an element of the fundamental domain, but rather as an element of an \textit{enlarged} fundamental domain, with the precise enlargement depending on the theory in question. Indeed, since a given global variant is (generically) not invariant under modular transformations, then we should not restrict to the quotient of $\cH$ by the full $SL(2,\ZZ)$, but rather to the quotient by the subgroup of $SL(2,\ZZ)$ mapping the given global variant to itself.

\begin{figure}[!tbp]
\begin{center}
\begin{tikzpicture}[baseline=0,scale = 0.7, baseline=-10]
 \node[below] (1) at (0,0) {$SU(2)$};
  \node[below] (2) at (6,0) {$SO(3)_+$};
   \node[below] (3) at (12,0) {$SO(3)_-$};
   
       \draw [thick, dgreen,{Latex[length=2.5mm]}-{Latex[length=2.5mm]}] (1) -- (2) node[midway,  above] {$\mS$};
        \draw [thick, dgreen,{Latex[length=2.5mm]}-{Latex[length=2.5mm]}] (2) -- (3) node[midway,  above] {$\mT$};
       
\draw[thick, dgreen,{Latex[length=2.5mm]}-{Latex[length=2.5mm]}] (1) to[out=135, in=225,loop] (1);
  \node[left]  at (-2.5,-0.5) {$\color{dgreen}{\mT}$};
  
\draw[thick, dgreen,{Latex[length=2.5mm]}-{Latex[length=2.5mm]}] (3) to[out=45, in=-45,loop] (3);
  \node[right]  at (14.5,-0.5) {$\color{dgreen}{\mS}$};

\end{tikzpicture} 
\caption{Modular transformations for theories with gauge algebra $\mathfrak{su}(2)$, with background gauge fields for the one-form symmetry turned off. }
\label{fig:su2usual}
\end{center}
\end{figure}
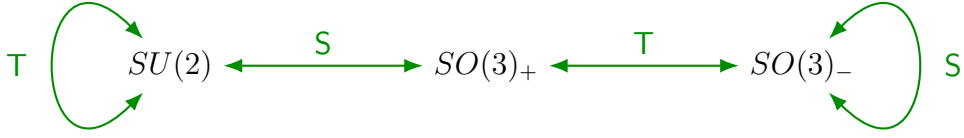

As a concrete example, consider $\cN=4$ SYM with gauge algebra $\mathfrak{su}(2)$. In this case there are three choices for the global structure of the gauge group, with the web of modular transformations between them given in Figure \ref{fig:su2usual} \cite{Aharony:2013hda}. If we fix ourselves to the $SU(2)$ variant, then we see that the modular transformations mapping it to itself are generated by $\mS$ and $\mS \mT^2 \mS$, which together generate the congruence subgroup $\Gamma_0(2) \subset SL(2,\ZZ)$.\footnote{Note that this result, and indeed all of Figure \ref{fig:su2usual}, is technically only valid in the absence of background gauge fields for the one-form symmetry. In the presence of background gauge fields, the correct figure is the one given in Figure \ref{fig:su2}. 
} Thus the distinct $SU(2)$ theories should be labelled by $\tauYM$ taking values in the enlarged domain $\widetilde{\cF}_{SU(2)}:=\Gamma_0(2)\, \backslash \, \cH$. 

 When considering $\tauYM$ as an element of $\widetilde{\cF}_{G}$, the question of whether or not $G$ SYM has non-invertible symmetries has a well-defined answer. 
In contrast, if we attempt to consider $\tauYM$ only as an element of the \textit{usual} fundamental domain $\cF$, then this question does \textit{not} (in general) have a well-defined answer. For example, as we have discussed above, the $SU(2)$ theory has a non-invertible symmetry at $\tauYM^* = i$. When treated as a point in the fundamental domain, this is equivalent to the point $ \tauYM^*|_{\mS\mT} = \half(i-1)$.  However, at $\tauYM^*|_{\mS\mT} $ the $SU(2)$ theory in fact does \textit{not} have a non-invertible symmetry. To see this, first note that under $\mS$ we have
\bea
\label{eq:SonSO3min}
\mS \, SO(3)_-[\tauYM] = SO(3)_-\left[- {1/ \tauYM}\right]
\eea
and thus at $ \tauYM^* = i$ the $SO(3)_-$ theory has an invertible $\mS$ symmetry. Noting that $SO(3)_-$ is mapped to $SU(2)$ under $\mS\mT$, 
we may apply $\mS\mT$ to both sides of (\ref{eq:SonSO3min}) to obtain 
\bea
SU(2) \left[\tauYM|_{\mS\mT}\right] = \mS\mT \, SO(3)_-[\tauYM]  = \mS\mT\mS\,SO(3)_-[\tauYM|_\mS] = \mS\mT\mS(\mS\mT)^{-1}\,SU(2)\left[\tauYM|_{\mS\mT\mS}\right] ~.
\no\\
\eea
Thus at the point $\tauYM^*|_{\mS\mT}= \tauYM^*|_{\mS\mT\mS} = \half (i-1)$, the $SU(2)$ theory has an invertible $(\mS\mT)\mS(\mS\mT)^{-1}$ symmetry. Furthermore, the original non-invertible symmetry $\cN = \sigma \mS$ no longer exists at this point, since $\mS$ no longer leaves $\tauYM$ invariant. Hence by switching between $\tauYM^* = i$ and $ \tauYM^*|_{\mS\mT} = \half(i-1)$, which are identical as points in $\cF$, the answer to the question ``Does $SU(2)$ have non-invertible symmetry?" has changed. In contrast, $\tauYM^*$ and $ \tauYM^*|_{\mS\mT}$ are {distinct} as points in the \textit{enlarged} fundamental domain $\widetilde{\cF}_{SU(2)}$, and so the question in that domain can still have a well-defined answer.

Since the enlarged fundamental domain $\widetilde{\cF}_G$ depends on the global variant under consideration, if we want the question ``Does $G$ SYM have non-invertible  symmetry?" to have a well-defined answer for \textit{all global variants $G$ simultaneously}, then we are forced to consider $\tauYM$ as an element of the union $\cup_G \widetilde{\cF}_G$. 

However, there is an alternative perspective. Instead of fixing ourselves to a particular global variant $G$, we can instead combine all global variants together to form a \textit{vector} of theories, e.g. $\overrightarrow{\mathfrak{su}(2)} := ( SU(2), SO(3)_+, SO(3)_-)$, which now behaves as a vector-valued modular function of $SL(2,\ZZ)$. From this perspective, we can now ask the question ``Does $\vec{\mathfrak{g}}$ have non-invertible symmetry?" and this question \textit{will} have a well-defined answer for $\tauYM$ as an element in $\cF$. This perspective is useful since it can save us from having to identify $\widetilde{\cF}_G$ for each global variant, which is not completely straightforward. 

Furthermore, in the second perspective more refined questions like ``Do $n$ of the elements of $\overrightarrow{\mathfrak{g}}$ have non-invertible symmetries?" also have well-defined answers in $\cF$, though the question of precisely which of the $n$ elements have the non-invertible symmetries will not. Turning again to the example of $\mathfrak{su}(2)$, at $\tauYM^* = i$ we have seen that $SU(2)$ and $SO(3)_+$ have non-invertible symmetries, whereas $SO(3)_-$ has an invertible symmetry. On the other hand, at $\tauYM^*|_{\mS \mT} = \half(1-i)$ we have seen that $SU(2)$ has an invertible symmetry, and one can show that  $SO(3)_+$ and $SO(3)_-$ have non-invertible symmetries. In both cases, two elements of $\overrightarrow{\mathfrak{su}(2)}$ have non-invertible symmetries, while one does not.  

To conclude, say that we have a global variant $G_1$ with non-invertible symmetry at some $\tauYM^*$, and a different global variant $G_2$ (in the same modular orbit) with only invertible symmetries at  $\tauYM^*$. Then there will be a point in $\widetilde{\cF}_{G_1}$ such that $G_1$ also has only invertible symmetries. In this case we will refer to the non-invertible symmetries of $G_1$ at $\tauYM^*$ as \textit{non-intrinsic}, reflecting the fact that they can be related to invertible symmetries of a related theory, and as such should not give rise to any constraints on the dynamics of the theory beyond those obtainable from invertible symmetries alone. 

In contrast, if there does not exist any such $G_2$ in the modular orbit of $G_1$, then we will refer to the non-invertible symmetry of $G_1$ as \textit{intrinsic}. The constraints which follow from intrinsic non-invertible defects may not follow from any arguments involving only invertible symmetries. 
 An example of an intrinsically non-invertible symmetry is the triality symmetry of the $SU(2)$ theory at $\tauYM = e^{2 \pi i /3}$, which we briefly mentioned above and will explore further in Section \ref{sec:su2}. Indeed, from Figure \ref{fig:su2usual} it is easy to see that no global variant is mapped to itself under the $\mS\mT$ transformation, and hence that none has an invertible symmetry. 

We should stress that by our definitions, even intrinsic non-invertible symmetries might be related to invertible symmetries if we consider more operations than just $\mS$ and $\mT$. In particular, we may use the $\sigma$ and $\tau$ transformations mentioned briefly above to try to relate non-invertible symmetries of one variant to invertible symmetries of a different global variant. If possible, then we again expect that any constraints on the dynamics of the theory obtained from the non-invertible symmetry can be obtained from the invertible symmetry of the other variant, and likewise that twisted compactifications done using the non-invertible symmetry can be related to standard twisted compactifications of the other variant. Another way of phrasing this is that in the case of non-intrinsic non-invertible symmetry of a given choice of global form and $\tauYM$, we can relate the non-invertible symmetry to an invertible symmetry of a theory with the same global form but different value of $\tauYM$ using the Montonen-Olive duality group. We can similarly consider the case where the non-invertible symmetry can be related to an invertible symmetry of a theory with the same value of $\tauYM$ but with different global form using the $\sigma$ and $\tau$ transformations. In any case, for the purposes of this paper we shall reserve the definition of ``intrinsic" versus ``non-intrinsic" non-invertible symmetries to denote cases where the potential relations to invertible symmetries involve only $\mS$ and $\mT$ transformations.  

\subsection*{Outline}
The rest of this paper is organized as follows. In Section \ref{sec:su} we begin our catalogue of non-invertible symmetries of $\cN=4$ SYM, starting with the case of $\mathfrak{su}(N)$ gauge algebra. We  move on to the case of $\mathfrak{so}(2N)$ gauge algebra in Section \ref{sec:so2}, and to the simply-laced exceptional cases in Section \ref{sec:excep}. The non-simply-laced case is complicated by the fact that the Montonen-Olive duality group is no longer $SL(2,\ZZ)$, but rather a certain Hecke group, and is treated in Section \ref{eq:nonsimplaced}. Then having catalogued all the non-invertible symmetries, we proceed to consider twisted compactifications in Section \ref{sec:twistcomp}. This will allow us to obtain novel 3d $\cN=6$ theories whose moduli spaces are labelled by exceptional complex reflection groups; see Table \ref{tab:newcasessumm} for a summary of the results. 
Finally, we include Appendix \ref{eq:Ndualfusion}, which gives an example of a class of non-intrinsic $N$-ality defects for $\mathfrak{su}(N)$ SYM theories at $\tauYM=1$. The point of this appendix is simply to give a further example of how non-intrinsic non-invertible defects can appear.

\section{$\mathfrak{su}(N)$ SYM}
\label{sec:su}

We begin our catalogue of non-invertible defects of $\cN=4$ SYM with the case of gauge algebra $\mathfrak{su}(N)$. After giving a thorough discussion of the $N=2$ and $N=3$ cases, we will provide the general results for prime $N$.

\subsection{$\mathfrak{su}(2)$}
\label{sec:su2}

We first discuss the case of gauge algebra $\mathfrak{su}(2)$. As is well-known, there exist three choices for the global structure of the gauge group, differing by their spectra of line operators \cite{Aharony:2013hda}. We denote these theories by $SU(2)$, $SO(3)_+$, and $SO(3)_-$. Important for our purposes will be the relation between these theories under the action of the Montonen-Olive duality group $SL(2,\ZZ)$. This data is usually captured in a diagram of the form shown in Figure \ref{fig:su2usual}, and can be found in e.g. \cite{Aharony:2013hda}.  We see that $SU(2)$ and $SO(3)_+$ are related by an $\mS$ transformation, while $SO(3)_\pm$ are related by $\mT$.\footnote{The $\mT$ transformation acts by shifting $\tauYM\to \tauYM+1$, while the $\mS$ transformation acts by $\tauYM\to -\frac{1}{\tauYM}$, accompanied by a map between electric fields and magnetic ones. In the absence of one-form and gravitational background fields, the Montonen-Olive duality of $\mathfrak{su}(2)$ theories implies the following identities between partition functions: $Z_{SU(2)}[\tauYM]= Z_{SU(2)}[\tauYM+1]$, $Z_{SO(3)_+}[\tauYM]= Z_{SU(2)}[-1/\tauYM]$, $Z_{SO(3)_+}[\tauYM]=Z_{SO(3)_-}[\tauYM+1]$, and $Z_{SO(3)_-}[\tauYM]= Z_{SO(3)_-}[-1/\tauYM]$. }

However, in the presence of background gauge fields for the $\ZZ^{(1)}_2$ one-form symmetry, the mappings in Figure \ref{fig:su2usual} are not quite precise. Instead, it is possible that the identifications hold only up to local counterterms (i.e. invertible phases) involving the background fields. Concretely, let us begin with the $SU(2)$ theory and couple to a two-form background gauge field $B$ for the $\ZZ_2^{(1)}$ one-form symmetry, where $B$ is normalized such that $B\sim B+2$. This can be done by replacing the $SU(2)$ field strength $F$ by $F'- \pi B \mathbf{1}$ where $F'$ is a $U(2)$ field strength and $\mathbf{1}$ is a two-dimensional identity matrix, and imposing the constraint $\Tr(F')= 2\pi B$ \cite{Gaiotto:2017yup}. Upon application of $\mT$, we have $\theta\rightarrow \theta + 2 \pi$, which produces a contribution to the action of the form
\bea\label{deltaS}
\delta S = {2 \pi \over 8 \pi^2} \int \Tr(F'- \pi B \mathbf{1}) (F'- \pi B \mathbf{1}) 
\eea
subject to the constraint $\Tr(F')= 2\pi B$. The right-hand side can be simplified to 
\begin{eqnarray}\label{deltaSpb}
\delta S = - \frac{\pi}{2}\int \cP(B)
\end{eqnarray}
with $\cP(B)$ the Pontryagin square of $B$, up to a second Chern number contribution that does not contribute to the path integral \cite{Gaiotto:2017yup}.  In this work, since we consider theories with fermions, the underlying spacetime manifold will be taken to be a spin manifold, and as a result $\cP(B)$ is even. Hence the sign in \eqref{deltaSpb} will not be important, and will be suppressed below. 
We thus conclude that under $\mT$  the $SU(2)$ theory is mapped to $SU(2)$ stacked with the counterterm \eqref{deltaSpb}. For simplicity, we adopt the notation $SU(2)_m$ where $m=0\mod 2$ is the $SU(2)$ theory without the counterterm, while $m=1\mod 2$ is the $SU(2)$ theory with the counterterm. We will likewise use the notation $SO(3)_{\pm,m}$.

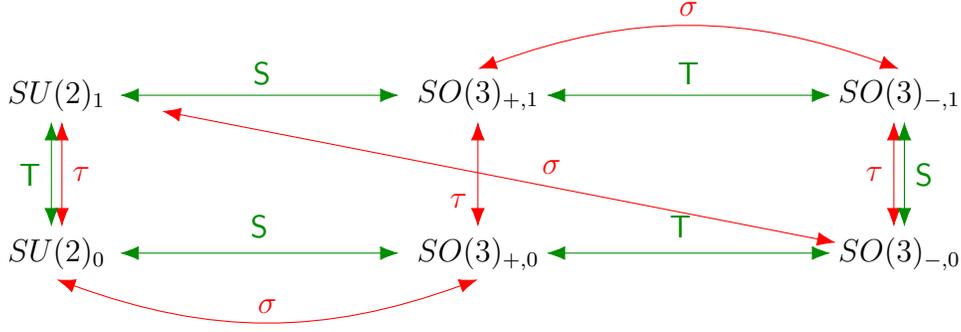
\begin{figure}[!tbp]
\begin{center}
\begin{tikzpicture}[baseline=0,scale = 0.7, baseline=-10]
 \node[below] (1) at (0,0) {$SU(2)_0$};
 \draw [red,{Latex[length=2.5mm]}-{Latex[length=2.5mm]}] (0.1,0) to (0.1,2);
  \node[right]  at (0.1,1) {$\color{red}{\tau}$};
  \draw [dgreen,{Latex[length=2.5mm]}-{Latex[length=2.5mm]}] (-0.1,0) to (-0.1,2);
  \node[left]  at (-0.1,1) {$\color{dgreen}{\mT}$};
  \node[above] (4) at (0,2) {$SU(2)_1$};
  
   \node[below] (2) at (8,0) {$SO(3)_{+,0}$};
 \draw [red,{Latex[length=2.5mm]}-{Latex[length=2.5mm]}] (8,0) to (8,2);
 \node[left] at (8,0.5) {$\color{red}{\tau}$};
  \node[above] (5)  at (8,2) {$SO(3)_{+,1}$};
  
 \node[below] (3) at (16,0) {$SO(3)_{-,0}$};
 \draw [red,{Latex[length=2.5mm]}-{Latex[length=2.5mm]}] (15.9,0) to (15.9,2);
  \draw [dgreen,{Latex[length=2.5mm]}-{Latex[length=2.5mm]}] (16.1,0) to (16.1,2);
 \node[left]  at (15.9,1) {$\color{red}{\tau}$};
  \node[right]  at (16.1,1) {$\color{dgreen}{{\mS}}$};
  \node[above] (6) at (16,2) {$SO(3)_{-,1}$};
  
 \draw[dgreen,{Latex[length=2.5mm]}-{Latex[length=2.5mm]}] (1.2,-0.5) to (6.5,-0.5);
   \node[] at (3.85,0) {$\color{dgreen}\mS$};
    \draw[dgreen,{Latex[length=2.5mm]}-{Latex[length=2.5mm]}] (9.3,-0.5) to (14.7,-0.5);
   \node[] at (11.85,0) {$\color{dgreen}\mT$};
   
     \draw[dgreen,{Latex[length=2.5mm]}-{Latex[length=2.5mm]}] (1.2,2.5) to (6.5,2.5);
      \node[above] at (3.9,2.5) {$\color{dgreen}\mS$};
       \draw[dgreen,{Latex[length=2.5mm]}-{Latex[length=2.5mm]}] (9.3,2.5) to (14.7,2.5);
   \node[above] at (12,2.5) {$\color{dgreen}\mT$};

 \draw[red,{Latex[length=2.5mm]}-{Latex[length=2.5mm]}] (0,-1) to[out=-20,in=200] (8,-1);
\node[] at (4,-1.5) {$\color{red} \sigma$};
              
   \draw[red,{Latex[length=2.5mm]}-{Latex[length=2.5mm]}] (2,2.2) to (14.8,-0.3);
\node[above] at (9.4,0.8) {$\color{red}{\sigma}$};
                
 \draw[red,{Latex[length=2.5mm]}-{Latex[length=2.5mm]}] (8,3) to[out=20,in=160] (16,3);
 \node[above] at (12,3.8) {$\color{red}{\sigma}$};   
\end{tikzpicture} 
\caption{Web of transformations for theories with gauge algebra $\mathfrak{su}(2)$. The transformations in green are the Montonen-Olive duality transformations, generating $SL(2, \ZZ)$. The transformations in red are topological manipulations generating $SL(2, \ZZ_2)$. }
\label{fig:su2}
\end{center}
\end{figure}

In order to understand the rest of the diagram in the presence of non-zero $B$, it is useful to introduce topological manipulations $\tau$ and $\sigma$ defined as follows \cite{Gaiotto:2014kfa}: 
\bea
\label{eq:su2sigtau}
&\tau:& \hspace{0.5 in} \text{stack\,\,with\,\,the\,\,counterterm}\,\,\frac{\pi}{2}\int \cP(B)~,
\no\\
&\sigma:& \hspace{0.5 in} \text{gauge\,\,}\ZZ_2^{(1)}\text{\,\,one\,\, form\,\,symmetry}~.
\eea
Clearly $\tau$ exchanges $SU(2)_m$ with $SU(2)_{m+1}$. On the other hand, $\sigma$ maps between $SU(2)_0 \leftrightarrow SO(3)_{+,0}$ and $SU(2)_1\leftrightarrow SO(3)_{-,0}$. More concretely, we can write 
\bea
\label{eq:defofsigma}
Z_{SO(3)_{+,0}}[\tauYM, B] &=& \sum_{b\in H^2(X_4, \ZZ_2)} Z_{SU(2)_0}[\tauYM, b] e^{i\pi \int b B}~,
\no\\
Z_{SO(3)_{-,0}}[\tauYM, B] &=& \sum_{b\in H^2(X_4, \ZZ_2)} Z_{SU(2)_0}[\tauYM, b] e^{i {\pi \over 2} \int \cP(b) }e^{i \pi \int b B}~.
\eea
For simplicity, we will suppress the overall normalization of the partition function.\footnote{The standard normalization of the partition function has pre-factor $|H^0(X_4,\ZZ_2)|/|H^1(X_4, \ZZ_2)|$ in \eqref{eq:defofsigma}, and is also subject to an Euler characteristic ambiguity $\lambda^{\chi(X_4)}$ for arbitrary $\lambda$. }
We may now fill in the rest of the diagram as follows. First, since $\sigma$ and $\tau$ generate $SL(2,\ZZ_2)$, they must satisfy $(\sigma\tau)^3 = \sigma^2 = 1$.\footnote{\label{ft:1footnote} What we mean by $(\sigma\tau)^3 = \sigma^2 = 1$ is that on the vector of theories $\overrightarrow{\mathfrak{su}(2)}$ defined in the introduction, the operations $(\sigma\tau)^3$ and $\sigma^2$ acts trivially. We do \textit{not} mean that the corresponding defects are trivial, and in general they will not be---rather, they will equal some appropriate condensation defect in accordance with (\ref{eq:dualityfusionrules}) and (\ref{trialityfusion}).} This allows us to conclude that $SO(3)_{+,1}$ and $SO(3)_{-,1}$ are related by $\sigma$---indeed, only then can we have a closed loop
\bea
SU(2)_0 \xrightarrow{\sigma} SO(3)_{+,0} \xrightarrow{\tau}SO(3)_{+,1}  \xrightarrow{\sigma}SO(3)_{-,1}\xrightarrow{\tau} SO(3)_{-,0} \xrightarrow{\sigma}SU(2)_1 \xrightarrow{\tau}SU(2)_0~
\no
\eea
corresponding to $(\sigma \tau)^3 = 1$. We next use commutativity of the $SL(2, \ZZ)$ and $SL(2,\ZZ_2)$ actions to fill in the remainder of the diagram. For example, we may consider the action of $\sigma \mT$ on $SU(2)_0$, which by the above considerations is fixed to $\sigma \mT SU(2)_0 = \sigma SU(2)_1 = SO(3)_{-,0}$. By commutativity, this should be equivalent to $\mT \sigma SU(2)_0 = \mT SO(3)_{+,0}$, and hence we conclude that $\mT$ maps $SO(3)_{+,0}$ to $SO(3)_{-,0}$, with no extra counterterm produced. By repeatedly imposing commutativity of $SL(2, \ZZ)$ and $SL(2,\ZZ_2)$, as well as the requirement that $(\mS\mT)^3 =1$, the web of modular transformations can be fixed unambiguously to the form given in Figure \ref{fig:su2}.

It is instructive to perform some consistency checks at the level of the partition function. To begin, let us check that the $\mS$ transformation of $SO(3)_{-,0}$ that we have obtained, namely $\mS\, SO(3)_{-,0} = SO(3)_{-,1}$ or in other words 
\bea
\label{eq:SO3m01anom}
Z_{SO(3)_{-,0}}[-1/\tauYM, B] = Z_{SO(3)_{-,1}}[\tauYM, B] = e^{i \pi \int {\cP(B) \over 2}}Z_{SO(3)_{-,0}}[\tauYM, B]~
\eea
is consistent. To see this, we begin by using (\ref{eq:defofsigma}) to write
\bea
Z_{SO(3)_{-,0}}[-1/\tauYM, B] &=& \sum_{b\in H^2(X_4, \ZZ_2)} Z_{SU(2)_0}[-1/\tauYM, b]\, e^{i \pi \int {\cP(b) \over 2}}e^{i \pi \int b B}~,
\no\\
&=&\sum_{b\in H^2(X_4, \ZZ_2)} Z_{SO(3)_{+,0}}[\tauYM, b] \,e^{i \pi \int {\cP(b) \over 2}}e^{i \pi \int b B}~, 
\eea
where we have made use of the $\mS$ transformation of $SU(2)_0$. Using (\ref{eq:defofsigma}) again allows us to write 
\bea
Z_{SO(3)_{-,0}}[-1/\tauYM, B] &=& \sum_{b, \hat{b}\in H^2(X_4, \ZZ_2)} Z_{SU(2)_{0}}[\tauYM, \hat b] \,e^{i \pi \int b (\hat b + B)}e^{i \pi \int {\cP(b) \over 2}}~, 
\no\\
&=& \sum_{b, \hat{b}\in H^2(X_4, \ZZ_2)} Z_{SU(2)_{0}}[\tauYM, \hat b] \,e^{i \pi \int {\cP(\hat b + B) \over 2}}e^{-i \pi \int {\cP(b+ \hat b + B) \over 2}}~, 
\eea
where we have used $\cP(a+b) - \cP(a)-\cP(b)=2 a b$ mod 4.
Since $b$ now appears in only one term in the integrand, we may integrate it out, giving \cite{Bhardwaj:2020ymp}
\bea
\sum_{b\in H^2(X_4, \ZZ_2)} e^{-i \pi \int {\cP(b+ \hat b + B) \over 2}} = \sum_{b\in H^2(X_4, \ZZ_2)} e^{-i \pi \int {\cP(b) \over 2}} = e^{2 \pi i {\mathsf{sig}(X_4) \over 16}} 
\eea
with $\mathsf{sig}(X_4)$ the signature of $X_4$. This is trivial on spin manifolds since $\mathsf{sig}(X_4) \in 16 \ZZ$ by Rokhlin's theorem. We thus conclude that 
\bea
Z_{SO(3)_{-,0}}[-1/\tauYM, B] &=& \sum_{\hat b\in H^2(X_4, \ZZ_2)} Z_{SU(2)_{0}}[\tauYM, \hat b] \,e^{i \pi \int {\cP(\hat b + B) \over 2}}~
\no\\
&=& e^{i \pi \int {\cP( B) \over 2}} \sum_{\hat b\in H^2(X_4, \ZZ_2)} Z_{SU(2)_{0}}[\tauYM, \hat b] \, e^{i \pi \int \hat b B} e^{i \pi \int {\cP(\hat b) \over 2}}
\no\\
&=& e^{i \pi \int {\cP(B) \over 2}}Z_{SO(3)_{-,0}}[\tauYM, B]~,
\eea
as desired. Similar manipulations can be used to confirm the consistency of the rest of the diagram. 

\paragraph{Non-invertible symmetries:} We now turn to the question of duality/triality defects in SYM theories of type $\mathfrak{su}(2)$. In fact, Figure \ref{fig:su2} allows us to straightforwardly read off the defects by constructing closed loops in the space of $\sigma, \tau, \mS,$ and $\mT$ transformations. The only closed loops involving exclusively $\{\sigma, \tau\}$ or $\{\mS, \mT\}$ are the trivial ones corresponding to $(\sigma \tau)^3=\sigma^2$ and $(\mS \mT)^{3} = \mS^2$. These act trivially on both the local and extended operators of the theory, as well as on the complex coupling $\tau_{\mathrm{YM}}$, and hence correspond to trivial (i.e. identity) defects. Any non-trivial defect must involve a mixture of $\{\sigma, \tau\}$ and $\{\mS, \mT\}$. The defects generated by $\{\mS,\mT\}$ are invertible, while those generated by $\{\sigma, \tau\}$ can be non-invertible \cite{Kaidi:2021xfk,Choi:2021kmx, Choi:2022zal,Koide:2021zxj}. 

Let us begin by considering the $SU(2)_0$ theory. Figure \ref{fig:su2} tells us that the combined transformation $\sigma \mS$ maps the theory back to itself, though it does so only up to an action on $\tauYM$,
\bea
Z_{SU(2)_0}[\tauYM, B^{(2)}]  \xrightarrow{ \,\,\,\,\sigma \mS\,\,\,\,\,} Z_{SU(2)_0}[-1/\tauYM, B^{(2)}] ~.
\eea
This means that when $\tauYM= i $, the operation $\sigma \mS$ is implemented by a defect in the theory. This is precisely the duality defect in $\cN=4$ $SU(2)$ YM at $\tauYM= i $ that was identified in \cite{Kaidi:2021xfk} and reviewed in the Introduction. 

\begin{table}
	\centering
	\begin{tabular}{| c | c |c |}
	\hline
		Theory & Defect  &$n$-ality \\
		\hline\hline
		$SU(2)_m, \, SO(3)_{+,m}$ & $\tau^m\sigma {\mS}\tau^{-m}$ & 2\\
		
		$SO(3)_{-,m}$ & $\tau^m  \tau {\mS} \tau^{-m}$ &1 \\ \hline

	\end{tabular}
	\caption{(Non-)invertible symmetries of $\mathfrak{su}(2)$ at $\tauYM= i $.}
	\label{tab:su2dutable}
\end{table}

It is now straightforward to extend this analysis to the other theories in Figure \ref{fig:su2}. Fixing $\tauYM= i $, we obtain the spectrum of duality defects illustrated in Table \ref{tab:su2dutable}. For $SU(2)_k$ and $SO(3)_{+,m}$ we obtain non-invertible defects $\tau^m \sigma \mS \tau^{-m}$, whereas for $SO(3)_{-,m}$ we obtain invertible defects of the form $\tau^m \tau \mS \tau^{-m}$. The latter is effectively just $\mS$, and hence corresponds to the invertible defect generating the $\ZZ_2^{(0)}$ $\mS$-transformation of $SO(3)_{-,m}$ at $\tauYM= i $. The presence of the extra factor of $\tau$ (remaining even for $m=0$) is indicative of a mixed anomaly between $\mS$ and the $\ZZ_2^{(1)}$ one-form symmetry, which was already exhibited in (\ref{eq:SO3m01anom}).  Note that in Table \ref{tab:su2dutable}, the number in the column labelled ``$n$-ality" is 2 in the case of a duality defect, 3 in the case of a triality defect, and 1 in the case of an invertible defect (of course, the invertible defects will also be ``modulo $n$" objects, in the sense that they will generate invertible $\ZZ_n$ symmetries. But for the purposes of this paper we will nevertheless assign such defects $n$-ality 1).

Since the  $SU(2)_m$ and $SO(3)_{+,m}$ theories can be related to the $SO(3)_{-,m}$ theory by modular transformations, we conclude that the non-invertible duality symmmetries discussed thus far can be related to (anomalous) invertible symmetries. In other words, these are examples of \textit{non-intrinsic} non-invertible symmetries, and as discussed in the Introduction they are not expected to give rise to any dynamical consequences beyond the ones implied by usual symmetries/anomalies. However, as we will now see, at different values of $\tauYM$ the theories do possess \textit{intrinsically} non-invertible defects, which can give rise to constraints going beyond those obtainable from invertible symmetries.\footnote{The general criteria for when a duality/triality defect is intrinsic (beyond just the case of $\cN=4$ SYM) will be explored in \cite{toappear}. }

\begin{table}
	\centering
	\begin{tabular}{| c | c |c |}
	\hline
		Theory & Defect  & $n$-ality \\
		\hline\hline
		$SU(2)_m, SO(3)_{-,m}$ & $\tau^m \sigma \tau{\mS\mT}\tau^{-m}$ & 3\\
		
		$SO(3)_{+,m}$ & $\tau^m \tau \sigma {\mS\mT}\tau^{-m}$ &3 \\ \hline

	\end{tabular}
	\caption{Non-invertible symmetries of $\mathfrak{su}(2)$ at $\tauYM= e^{2 \pi i/3}$.}
	\label{tab:su2tritable}
\end{table}

To obtain a non-trivial duality symmetry, it was crucial that we included an $\{\mS, \mT\}$ transformation, and furthermore that we were able to set $\tauYM$ to a fixed point of this actions. Restricting ourselves to the fundamental domain, the only other fixed points are then $\tauYM = i \infty$ and $\tauYM = e^{2 \pi i/3}$. The symmetries for the former will involve only $\mT$ and will be invertible, so we focus on the latter. The symmetries in this case involve the combination $\mS\mT$, and can be read off from Figure \ref{fig:su2} as before. The results are listed in Table \ref{tab:su2tritable}. There are two points worth noting: 
\begin{enumerate}
\item None of the symmetries obtained here are invertible---indeed, each involves a single action of $\sigma$, which is a non-invertible operation. Thus the non-invertible symmetries obtained in this way cannot be exchanged for invertible ones in a different variant, and the dynamical constraints obtained from such symmetries can go beyond those obtainable from anomalies and invertible symmetries. 

\item The non-invertible defects in this case no longer generate duality transformations, but rather \textit{triality} transformations. To see this, we note that acting three times by the first defect in Table \ref{tab:su2tritable} leaves the \emph{entire vector} of theories invariant by the $SL(2,\Z)$ and $SL(2,\Z_2)$ algebra $(\sigma \tau)^3=(\mS\mT)^3=1$, 
\begin{eqnarray}\label{211}
(\tau^m \sigma \tau{\mS\mT}\tau^{-m})^3= \tau^m (\sigma\tau)^3 \tau^{-m} (\mS\mT)^3 = 1
\end{eqnarray}
(see Footnote \ref{ft:1footnote} for the correct interpretation of this equation). 
Similar statements hold for the other defect in Table \ref{tab:su2tritable}. 
Such triality defects were discussed in \cite{Choi:2022zal}, where the fusion rules were also determined (see also \cite{Hayashi:2022fkw}). 
\end{enumerate}

\subsection{$\mathfrak{su}(3)$}
\label{sec.su3}

We now proceed to the case of  $\mathfrak{su}(3)$ SYM theory. The starting point is again to list the global variants for the theory. It is known that there are four such global variants: namely the $SU(3)$ theory as well as $PSU(3)_n$ where $n\in \Z_3$ labels the discrete theta angle.
The modular transformations between these theories can be found in Figure 4 of \cite{Aharony:2013hda}.

For our purposes, we will want more refined data about the modular transformations in the presence of non-trivial background gauge fields for the $\ZZ_3^{(1)}$ one-form symmetry. To obtain this data, we will as before introduce the following topological manipulations \cite{Gaiotto:2014kfa}
\bea
&\tau:& \hspace{0.5 in} \text{stack\,\,with\,\,a\,\,counterterm}\,\,{-\frac{2\pi}{3}\int BB}~,
\no\\
&\sigma:& \hspace{0.5 in} \text{gauge\,\,}\ZZ_3^{(1)}\text{\,\,one\,\,form\,\,symmetry}~,
\eea
which now generate $SL(2, \ZZ_3)$. They satisfy the relations $\tau^3=(\sigma\tau)^3=1$ and $\sigma^2=C$.  As before, we will denote 
\bea
SU(3)_m = \tau^m SU(3)~, \hspace{0.5 in} PSU(3)_m = \tau^m PSU(3)~,
\eea
and by  definition we have 
\bea
PSU(3)_{n,0} = \sigma \tau^n SU(3)_0~.
\eea
Importantly, unlike for $SL(2, \ZZ_2)\cong PSL(2,\ZZ_2)$, in the current case $\sigma$ does not square to the identity, but rather to the charge conjugation operation $C$ sending $B \rightarrow - B$. This can be seen at the level of the action by considering any  of the above theories $\cT$ and observing that
\bea
Z_{\sigma^2 \cT}[\tauYM, B] &= &\sum_{b, \hat b\in H^2(X_4, \ZZ_3)}Z_{\cT}[\tauYM, b] \,e ^{{2 \pi i \over 3}\int b \hat b} e ^{{2 \pi i \over 3}\int \hat b  \hat B}
\no\\
&=& \sum_{b\in H^2(X_4, \ZZ_3)}Z_{\cT}[\tauYM, b] \,\delta(b + B)
\no\\
&=& Z_{\cT}[\tauYM, -B]~.
\eea
From now on, the charge conjugate of a theory $\cT$ will be denoted as $\overline{\cT}$. Let us emphasize that charge conjugation, while being a zero-form symmetry, does not actually act on any local operators of the theory, and in particular will not affect the structure of the moduli space. This is because the charge conjugation defect is actually the condensation defect of the $\ZZ_3^{(3)}$ 1-form symmetry \cite{Roumpedakis:2022aik, Choi:2022zal}, which is transparent to any local operator.\footnote{Charge conjugation defined by gauging the one-form symmetry twice here should be distinguished from  charge conjugation defined by outer automorphisms of the gauge group, which \textit{can} act on local operators. For example, gauge groups with complex representations, like $SU(N)$ or $Spin(4N+2)$, possess such a $\Z_2$ charge conjugation symmetry, under which the adjoint scalars transform non-trivially. } This will be important when we consider twisted compactifications.

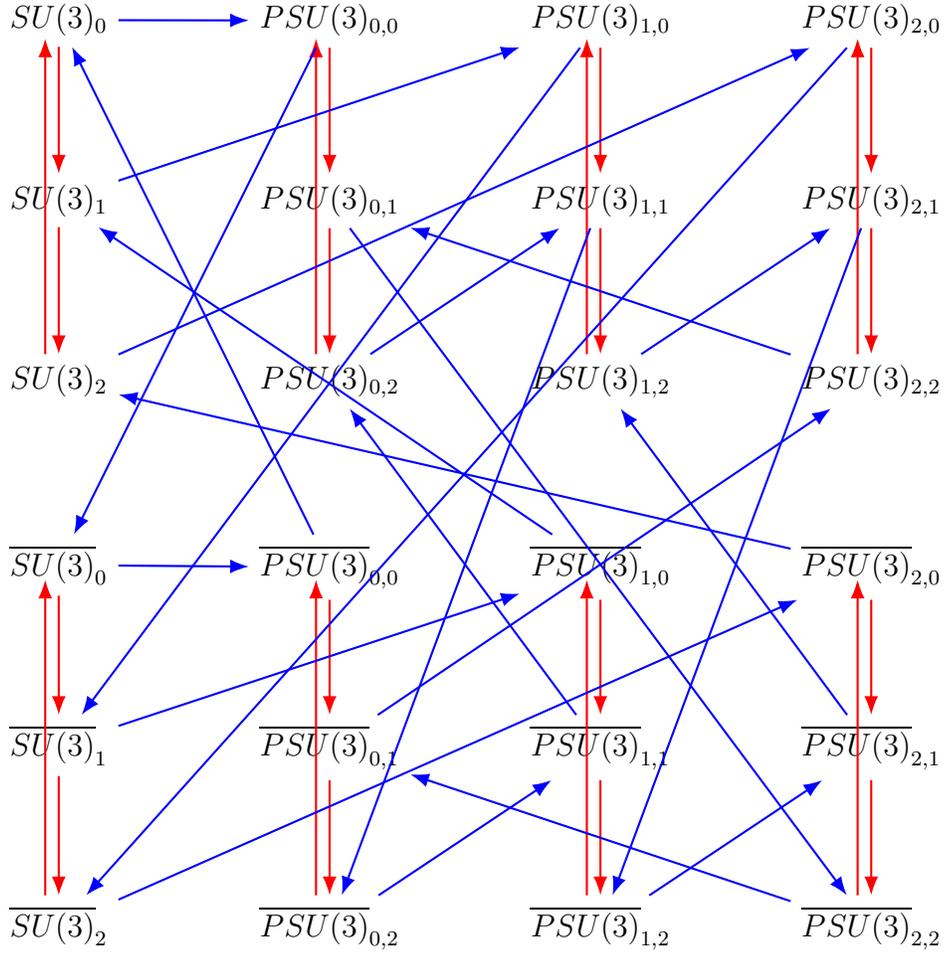
\begin{figure}[tbp]
\hspace*{4em}\begin{tikzpicture}[baseline=0,scale = 0.6, baseline=-10]

 \node[below] (1) at (0,0) {$SU(3)_{0}$};
 \node[below] (2) at (0,-4) {$SU(3)_{1}$};
 \node[below] (3) at (0,-8) {$SU(3)_{2}$};
  \node[below] (4) at (0,-12) {$\overline{SU(3)}_{0}$};
 \node[below] (5) at (0,-16) {$\overline{SU(3)}_{1}$};
 \node[below] (6) at (0,-20) {$\overline{SU(3)}_{2}$};
 
  \node[below] (7) at (6,0) {$PSU(3)_{0,0}$};
 \node[below] (8) at (6,-4) {$PSU(3)_{0,1}$};
 \node[below] (9) at (6,-8) {$PSU(3)_{0,2}$};
  \node[below] (10) at (6,-12) {$\overline{PSU(3)}_{0,0}$};
 \node[below] (11) at (6,-16) {$\overline{PSU(3)}_{0,1}$};
 \node[below] (12) at (6,-20) {$\overline{PSU(3)}_{0,2}$};
 
   \node[below] (13) at (12,0) {$PSU(3)_{1,0}$};
 \node[below] (14) at (12,-4) {$PSU(3)_{1,1}$};
 \node[below] (15) at (12,-8) {$PSU(3)_{1,2}$};
  \node[below] (16) at (12,-12) {$\overline{PSU(3)}_{1,0}$};
 \node[below] (17) at (12,-16) {$\overline{PSU(3)}_{1,1}$};
 \node[below] (18) at (12,-20) {$\overline{PSU(3)}_{1,2}$};
 
    \node[below] (19) at (18,0) {$PSU(3)_{2,0}$};
 \node[below] (20) at (18,-4) {$PSU(3)_{2,1}$};
 \node[below] (21) at (18,-8) {$PSU(3)_{2,2}$};
  \node[below] (22) at (18,-12) {$\overline{PSU(3)}_{2,0}$};
 \node[below] (23) at (18,-16) {$\overline{PSU(3)}_{2,1}$};
 \node[below] (24) at (18,-20) {$\overline{PSU(3)}_{2,2}$};
 
    
\draw [thick, red,-{Latex[length=2.5mm]}] (1) -- (2) ;
\draw [thick, red,-{Latex[length=2.5mm]}] (2) -- (3) ;
\draw [thick, red,-{Latex[length=2.5mm]}] (-0.3,-8) -- (-0.3,-1) ;
    
 \draw [thick, red,-{Latex[length=2.5mm]}] (4) -- (5) ;
\draw [thick, red,-{Latex[length=2.5mm]}] (5) -- (6) ;
\draw [thick, red,-{Latex[length=2.5mm]}] (-0.3,-20) -- (-0.3,-13) ;

  \draw [thick, red,-{Latex[length=2.5mm]}] (7) -- (8) ;
\draw [thick, red,-{Latex[length=2.5mm]}] (8) -- (9) ;
\draw [thick, red,-{Latex[length=2.5mm]}] (5.7,-8) -- (5.7,-1) ;

\draw [thick, red,-{Latex[length=2.5mm]}] (10) -- (11) ;
\draw [thick, red,-{Latex[length=2.5mm]}] (11) -- (12) ;
\draw [thick, red,-{Latex[length=2.5mm]}] (5.7,-20) -- (5.7,-13) ;

\draw [thick, red,-{Latex[length=2.5mm]}] (13) -- (14) ;
\draw [thick, red,-{Latex[length=2.5mm]}] (14) -- (15) ;
\draw [thick, red,-{Latex[length=2.5mm]}] (11.7,-8) -- (11.7,-1) ;

\draw [thick, red,-{Latex[length=2.5mm]}] (16) -- (17) ;
\draw [thick, red,-{Latex[length=2.5mm]}] (17) -- (18) ;
\draw [thick, red,-{Latex[length=2.5mm]}] (11.7,-20) -- (11.7,-13) ;

\draw [thick, red,-{Latex[length=2.5mm]}] (19) -- (20) ;
\draw [thick, red,-{Latex[length=2.5mm]}] (20) -- (21) ;
\draw [thick, red,-{Latex[length=2.5mm]}] (17.7,-8) -- (17.7,-1) ;

\draw [thick, red,-{Latex[length=2.5mm]}] (22) -- (23) ;
\draw [thick, red,-{Latex[length=2.5mm]}] (23) -- (24) ;
\draw [thick, red,-{Latex[length=2.5mm]}] (17.7,-20) -- (17.7,-13) ;

    
 \draw [thick, blue,-{Latex[length=2.5mm]}] (1) -- (7) ;
  \draw [thick, blue,-{Latex[length=2.5mm]}] (7) -- (4) ;
    \draw [thick, blue,-{Latex[length=2.5mm]}] (4) -- (10) ;
   \draw [thick, blue,-{Latex[length=2.5mm]}] (10) -- (1) ;
   
 \draw [thick, blue,-{Latex[length=2.5mm]}] (2) -- (13) ;
  \draw [thick, blue,-{Latex[length=2.5mm]}] (13) -- (5) ;
   \draw [thick, blue,-{Latex[length=2.5mm]}] (5) -- (16) ;
   \draw [thick, blue,-{Latex[length=2.5mm]}] (16) -- (2) ;
   
 \draw [thick, blue,-{Latex[length=2.5mm]}] (3) -- (19) ;
  \draw [thick, blue,-{Latex[length=2.5mm]}] (19) -- (6) ;
   \draw [thick, blue,-{Latex[length=2.5mm]}] (6) -- (22) ;
   \draw [thick, blue,-{Latex[length=2.5mm]}] (22) -- (3) ;
   
    \draw [thick, blue,-{Latex[length=2.5mm]}] (8) -- (24) ;
     \draw [thick, blue,-{Latex[length=2.5mm]}] (24) -- (11) ;    
   \draw [thick, blue,-{Latex[length=2.5mm]}] (11) -- (21) ; 
    \draw [thick, blue,-{Latex[length=2.5mm]}] (21) -- (8) ; 
    
      \draw [thick, blue,-{Latex[length=2.5mm]}] (17) -- (9) ;
     \draw [thick, blue,-{Latex[length=2.5mm]}] (12) -- (17) ;    
   \draw [thick, blue,-{Latex[length=2.5mm]}] (14) -- (12) ; 
    \draw [thick, blue,-{Latex[length=2.5mm]}] (9) -- (14) ; 
    
          \draw [thick, blue,-{Latex[length=2.5mm]}] (23) -- (15) ;
     \draw [thick, blue,-{Latex[length=2.5mm]}] (18) -- (23) ;    
   \draw [thick, blue,-{Latex[length=2.5mm]}] (20) -- (18) ; 
    \draw [thick, blue,-{Latex[length=2.5mm]}] (15) -- (20) ; 
\end{tikzpicture} 
\caption{Web of $SL(2,\ZZ_3)$ transformations for theories with gauge algebra $\mathfrak{su}(3)$. We have denoted the action of $\color{blue} \sigma$ in {\color{blue} blue} and the action of $\color{red} \tau$ in {\color{red} red}, where $\sigma$ and $\tau$ represent gauging the $\ZZ^{(1)}_3$ symmetry and coupling to an invertible phase, respectively.}
\label{fig:su31}
\end{figure}

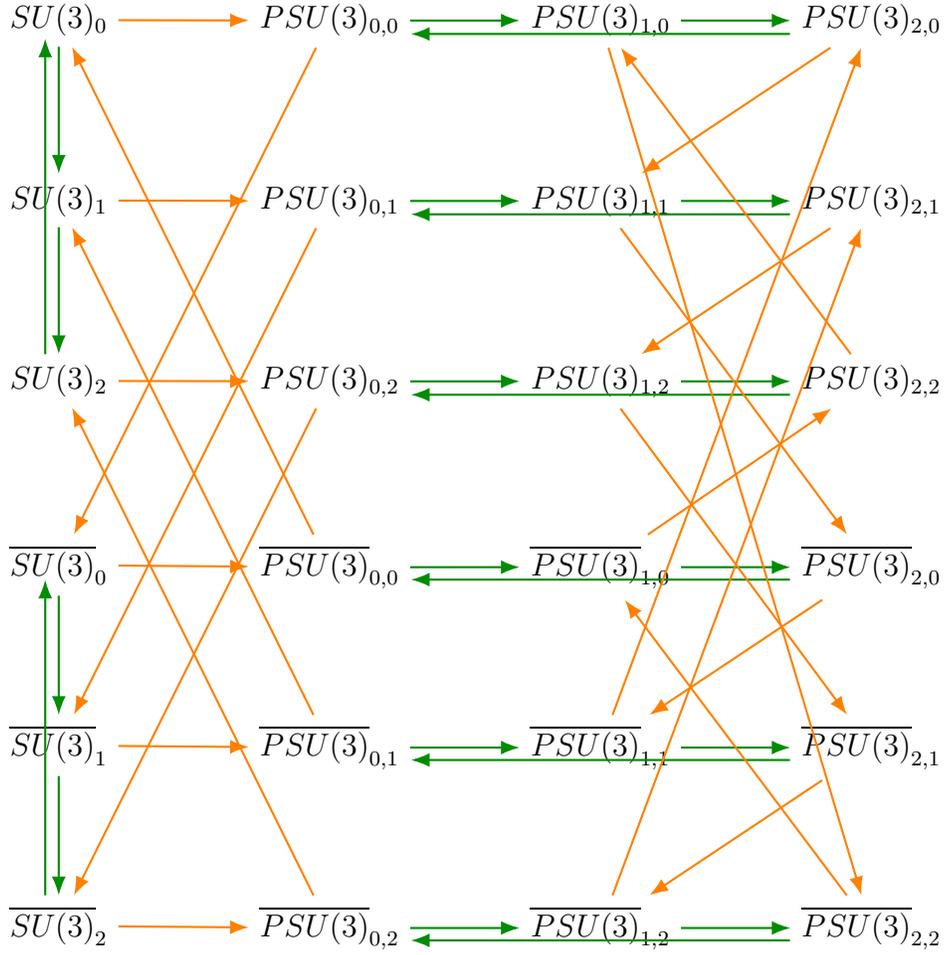
\begin{figure}[tbp]
\hspace*{4em}\begin{tikzpicture}[baseline=0,scale = 0.6, baseline=-10]

 \node[below] (1) at (0,0) {$SU(3)_{0}$};
 \node[below] (2) at (0,-4) {$SU(3)_{1}$};
 \node[below] (3) at (0,-8) {$SU(3)_{2}$};
  \node[below] (4) at (0,-12) {$\overline{SU(3)}_{0}$};
 \node[below] (5) at (0,-16) {$\overline{SU(3)}_{1}$};
 \node[below] (6) at (0,-20) {$\overline{SU(3)}_{2}$};
 
  \node[below] (7) at (6,0) {$PSU(3)_{0,0}$};
 \node[below] (8) at (6,-4) {$PSU(3)_{0,1}$};
 \node[below] (9) at (6,-8) {$PSU(3)_{0,2}$};
  \node[below] (10) at (6,-12) {$\overline{PSU(3)}_{0,0}$};
 \node[below] (11) at (6,-16) {$\overline{PSU(3)}_{0,1}$};
 \node[below] (12) at (6,-20) {$\overline{PSU(3)}_{0,2}$};
 
   \node[below] (13) at (12,0) {$PSU(3)_{1,0}$};
 \node[below] (14) at (12,-4) {$PSU(3)_{1,1}$};
 \node[below] (15) at (12,-8) {$PSU(3)_{1,2}$};
  \node[below] (16) at (12,-12) {$\overline{PSU(3)}_{1,0}$};
 \node[below] (17) at (12,-16) {$\overline{PSU(3)}_{1,1}$};
 \node[below] (18) at (12,-20) {$\overline{PSU(3)}_{1,2}$};
 
    \node[below] (19) at (18,0) {$PSU(3)_{2,0}$};
 \node[below] (20) at (18,-4) {$PSU(3)_{2,1}$};
 \node[below] (21) at (18,-8) {$PSU(3)_{2,2}$};
  \node[below] (22) at (18,-12) {$\overline{PSU(3)}_{2,0}$};
 \node[below] (23) at (18,-16) {$\overline{PSU(3)}_{2,1}$};
 \node[below] (24) at (18,-20) {$\overline{PSU(3)}_{2,2}$};
 
    
\draw [thick, dgreen,-{Latex[length=2.5mm]}] (1) -- (2) ;
\draw [thick, dgreen,-{Latex[length=2.5mm]}] (2) -- (3) ;
\draw [thick, dgreen,-{Latex[length=2.5mm]}] (-0.3,-8) -- (-0.3,-1) ;
    
 \draw [thick, dgreen,-{Latex[length=2.5mm]}] (4) -- (5) ;
\draw [thick, dgreen,-{Latex[length=2.5mm]}] (5) -- (6) ;
\draw [thick, dgreen,-{Latex[length=2.5mm]}] (-0.3,-20) -- (-0.3,-13) ;

\draw [thick, dgreen,-{Latex[length=2.5mm]}] (7) -- (13) ;
 \draw [thick, dgreen,-{Latex[length=2.5mm]}] (13) -- (19) ;
  \draw [thick, dgreen,-{Latex[length=2.5mm]}] (16.2,-0.9) -- (7.8,-0.9) ;
  
  \draw [thick, dgreen,-{Latex[length=2.5mm]}] (8) -- (14) ;
 \draw [thick, dgreen,-{Latex[length=2.5mm]}] (14) -- (20) ;
  \draw [thick, dgreen,-{Latex[length=2.5mm]}] (16.2,-4.9) -- (7.8,-4.9) ;
  
  \draw [thick, dgreen,-{Latex[length=2.5mm]}] (9) -- (15) ;
 \draw [thick, dgreen,-{Latex[length=2.5mm]}] (15) -- (21) ;
  \draw [thick, dgreen,-{Latex[length=2.5mm]}] (16.2,-8.9) -- (7.8,-8.9) ;
  
    \draw [thick, dgreen,-{Latex[length=2.5mm]}] (10) -- (16) ;
 \draw [thick, dgreen,-{Latex[length=2.5mm]}] (16) -- (22) ;
  \draw [thick, dgreen,-{Latex[length=2.5mm]}] (16.2,-13) -- (7.8,-13) ;
  
      \draw [thick, dgreen,-{Latex[length=2.5mm]}] (11) -- (17) ;
 \draw [thick, dgreen,-{Latex[length=2.5mm]}] (17) -- (23) ;
  \draw [thick, dgreen,-{Latex[length=2.5mm]}] (16.2,-17) -- (7.8,-17) ;
  
        \draw [thick, dgreen,-{Latex[length=2.5mm]}] (12) -- (18) ;
 \draw [thick, dgreen,-{Latex[length=2.5mm]}] (18) -- (24) ;
  \draw [thick, dgreen,-{Latex[length=2.5mm]}] (16.2,-21) -- (7.8,-21) ;
    
    \draw [thick, orange,-{Latex[length=2.5mm]}] (1) -- (7) ;
     \draw [thick, orange,-{Latex[length=2.5mm]}] (7) -- (4) ;
     \draw [thick, orange,-{Latex[length=2.5mm]}] (4) -- (10) ;
      \draw [thick, orange,-{Latex[length=2.5mm]}] (10) -- (1) ;
      
          \draw [thick, orange,-{Latex[length=2.5mm]}] (2) -- (8) ;
     \draw [thick, orange,-{Latex[length=2.5mm]}] (8) -- (5) ;
     \draw [thick, orange,-{Latex[length=2.5mm]}] (5) -- (11) ;
      \draw [thick, orange,-{Latex[length=2.5mm]}] (11) -- (2) ;
      
        \draw [thick, orange,-{Latex[length=2.5mm]}] (3) -- (9) ;
     \draw [thick, orange,-{Latex[length=2.5mm]}] (9) -- (6) ;
     \draw [thick, orange,-{Latex[length=2.5mm]}] (6) -- (12) ;
      \draw [thick, orange,-{Latex[length=2.5mm]}] (12) -- (3) ;
      
          \draw [thick, orange,-{Latex[length=2.5mm]}] (19) -- (14) ;
     \draw [thick, orange,-{Latex[length=2.5mm]}] (14) -- (22) ;
     \draw [thick, orange,-{Latex[length=2.5mm]}] (22) -- (17) ;
      \draw [thick, orange,-{Latex[length=2.5mm]}] (17) -- (19) ;
      
        \draw [thick, orange,-{Latex[length=2.5mm]}] (20) -- (15) ;
     \draw [thick, orange,-{Latex[length=2.5mm]}] (15) -- (23) ;
     \draw [thick, orange,-{Latex[length=2.5mm]}] (23) -- (18) ;
      \draw [thick, orange,-{Latex[length=2.5mm]}] (18) -- (20) ;
      
              \draw [thick, orange,-{Latex[length=2.5mm]}] (21) -- (13) ;
     \draw [thick, orange,-{Latex[length=2.5mm]}] (13) -- (24) ;
     \draw [thick, orange,-{Latex[length=2.5mm]}] (24) -- (16) ;
      \draw [thick, orange,-{Latex[length=2.5mm]}] (16) -- (21) ;
      
\end{tikzpicture} 
\caption{Web of $SL(2, \ZZ)$ transformations for theories with gauge algebra $\mathfrak{su}(3)$. We have denoted the action of $\color{orange} S$ in {\color{orange} orange} and the action of $\color{dgreen} T$ in {\color{dgreen} green}, where $S$ and $T$ represent the usual modular transformations.}
\label{fig:su32}
\end{figure}

Combining the defining relations of $SL(2, \ZZ_3)$ and $SL(2, \ZZ)$ with the commutativity of the two, as well as with the requirement that the diagram reduces to the one in Figure 4 of \cite{Aharony:2013hda} when the background $\ZZ_3^{(1)}$ gauge fields are turned off, it is possible to fill in the web of $\{\sigma, \tau\}$ and $\{\mS,\mT\}$ transformations. To minimize clutter, we have illustrated the two separately in Figures \ref{fig:su31} and \ref{fig:su32}.

\paragraph{Non-invertible symmetries:} These figures now allow us to obtain the list of duality and triality symmetries. As before, non-trivial defects must involve both $SL(2, \ZZ_3)$ and $SL(2, \ZZ)$ transformations. To obtain the desired symmetries, we must fix $\tauYM$ to the fixed points of $SL(2, \ZZ)$. Taking $\tauYM=i$, we obtain the spectrum of symmetries shown in Table \ref{tab:su3dutable}. Since $\sigma$ is now an order 4 operations, these defects would naively implement 4-ality, rather than duality. However, since $\sigma$ squares to charge conjugation, which is invertible, we prefer to think of these as duality symmetries extended by an invertible symmetry. We see in this case that all of the duality defects are non-invertible, and hence they are intrinsically non-invertible by our definitions. 

\begin{table}
	\centering
	\begin{tabular}{| c | c |c |}
	\hline
		Theory & Defect  & $n$-ality \\
		\hline\hline
		$SU(3)_m, \, PSU(3)_{0,m}$ & $\tau^m\sigma^3 {\mS}\tau^{-m}$ & 2\\
		
		$PSU(3)_{1,m} ,\, PSU(3)_{2,m-1}$ & $\tau^{m+1} \sigma {\mS} \tau^{2-m}$ &2 \\ \hline

	\end{tabular}
	\caption{Non-invertible symmetries of $\mathfrak{su}(3)$ at $\tauYM=i$.}
	\label{tab:su3dutable}
\end{table}

\begin{table}
	\centering
	\begin{tabular}{ |c | c |c |}
	\hline
		Theory & Defect  & $n$-ality \\
		\hline\hline
		$SU(3)_m, \, PSU(3)_{2,m}$ & $\tau^m \sigma^3 \tau^2 \mS \mT \tau^{-m}$ & 3\\
		
		$PSU(3)_{0,m}$ & $\tau^{m}\tau^2 \sigma^3 {\mS\mT} \tau^{-m}$ &3 \\
		
		$PSU(3)_{1,m}$ & $\tau^{m}\tau^2 {\mS\mT} \tau^{-m}$ &1 \\ \hline

	\end{tabular}
	\caption{(Non-)invertible symmetries of $\mathfrak{su}(3)$ at $\tauYM=e^{2 \pi i/3}$.}
	\label{tab:su3tritable}
\end{table}

On the other hand, fixing to $\tauYM=e^{2 \pi i /3}$ gives the spectrum of defects shown in Table \ref{tab:su3tritable}. For the $SU(3)_m$, $PSU(3)_{2,m}$, and $ PSU(3)_{0,m}$ theories we see that this gives rise to triality symmetries, whereas for $PSU(3)_{1,m}$ we obtain an invertible symmetry generated by $\mS\mT$. The fact that the defect is (for $m=0$) $\tau^2 {\mS\mT} $ and not simply $\mS\mT$ indicates that there is a mixed anomaly between $\mS\mT$ and $\ZZ_3^{(1)}$, precisely analogous to the one  between $\mS$ and $\ZZ_2^{(1)}$ in $SO(3)_-$, observed in (\ref{eq:SO3m01anom}). The same  path integral manipulations as before can be used to verify its existence.

\subsection{$\mathfrak{su}(N)$ for $N=\text{prime}$ and $N\geq 3$}
\label{sec:suN}

The discussions in Sections \ref{sec:su2} and \ref{sec.su3} can be straightforwardly generalized to $\mathfrak{su}(N)$ theories with $N$ a prime number. We will assume $N\geq 3$, so that we only need to consider odd $N$.  
The global variants of $\mathfrak{su}(N)$ theories are labelled by a gauge group, a discrete theta angle, and counterterms for background fields. Since $N$ is prime, there are only two distinct gauge groups, $SU(N)$ and $PSU(N)$. Introducing the label $n$ for the discrete theta term and $m$ for the counterterms, we have the following spectrum of global variants,
\begin{equation}\label{SUPSU}
\begin{split}
    Z_{SU(N)_m}[\tauYM, B] &:= Z_{SU(N)_0}[\tauYM, B]e^{\frac{2\pi i}{N} m \int \frac{N+1}{2}BB}~,\\
    Z_{PSU(N)_{n,m}}[\tauYM, B] &:= \sum_{b\in H^2(X_4, \ZZ_N)} Z_{SU(N)_0}[\tauYM, b]e^{\frac{2\pi i}{N} n  \int\frac{N+1}{2}bb+ \frac{2\pi i }{N}\int b B + \frac{2\pi i }{N} m  \int \frac{N+1}{2}BB}~.
\end{split}
\end{equation}
 As before, we introduce topological manipulations $\sigma$ and $\tau$ defined as \cite{Gaiotto:2014kfa} 
\begin{equation}\label{suNsigmatau}
\begin{array}{lrl}
    \sigma: ~~~& {SU(N)_m}[\tauYM,  B] &\to {PSU(N)_{m,0}}[\tauYM,  B]\\
    &{PSU(N)_{n,m}}[\tauYM,  B] &\to 
    \begin{cases}
    {SU(N)_n}[\tauYM, - B] & m=0\mod N\\
    {PSU(N)_{n-x,-m}}[\tauYM, - x  B] & m \neq 0\mod N
    \end{cases}\\\\
    \tau:~~~& {SU(N)_m}[\tauYM,  B] &\to {SU(N)_{m+1}}[\tauYM,  B]\\
    & {PSU(N)_{n,m}}[\tauYM,  B] &\to {PSU(N)_{n,m+1}}[\tauYM,  B] 
\end{array}
\end{equation}
where $x$ is the modular inverse of $m$, satisfying $m x = 1 \mod N$. 
On the other hand, under the modular $\mS$ and $\mT$ transformations the theories are mapped as follows 
\begin{equation}\label{modularSsuN}
\begin{array}{lrl}
     \mS: ~~~& {SU(N)_m}[\tauYM, B] &\to  {PSU(N)_{0,m}}[-1/\tauYM,  B]\\
    &{PSU(N)_{n,m}}[\tauYM,  B] &\to 
    \begin{cases}
    {SU(N)_m}[-1/\tauYM, - B], & n=0\mod N\\
    {PSU(N)_{-x,m n^2-n}}[-1/\tauYM, - x  B], & n \neq 0\mod N
    \end{cases}\\\\
    \mT:~~~& {SU(N)_m}[\tauYM,  B] &\to  {SU(N)_{n+1}}[\tauYM+1,  B]\\
    & {PSU(N)_{n,m}}[\tauYM,  B] &\to  {PSU(N)_{n+1,m}}[\tauYM+1,  B] 
\end{array}
\end{equation}
where $x$ is now taken to be the modular inverse of $n$, i.e. $x n=1\mod N$. By Montonen-Olive duality, the partition functions on the two sides of the transformations in \eqref{modularSsuN} are equal. 
When $N=3$, the rules \eqref{suNsigmatau} and \eqref{modularSsuN} are equivalent to Figures \ref{fig:su31} and \ref{fig:su32} respectively.

\subsubsection*{Duality symmetries at $\tauYM=i$}

We now begin by identifying the (non-)invertible symmetries at $\tauYM=i$. To do so, we start with a theory $\cX$ at generic coupling $\tauYM$ and perform a modular $\mS$ transformation \eqref{modularSsuN} to map to a (usually distinct) theory $\cX'$ at $-1/\tauYM$. We then look for a sequence of $\sigma$ and $\tau$ transformations that can map $\cX'$ at $-1/\tauYM$ back to $\cX$  at $-1/\tauYM$. Requiring $\cX$ to be invariant under this sequence of transformations constrains $\tauYM=i$.

For example, for $SU(N)_n$ we have
\begin{eqnarray}
\begin{split}
    {SU(N)_m}[\tauYM, B]  &\stackrel{\mS}{\longrightarrow} {PSU(N)_{0,m}}[-1/\tauYM,B]\\& \stackrel{\tau^{-m}}{\longrightarrow} {PSU(N)_{0,0}}[-1/\tauYM,B]\\ &\stackrel{\sigma^3}{\longrightarrow} {SU(N)_{0}}[-1/\tauYM, B]\\& \stackrel{\tau^{m}}{\longrightarrow} {SU(N)_m}[-1/\tauYM, B]~. 
\end{split}
\end{eqnarray}
Hence $SU(N)_{m}$ at $\tauYM=i$ is invariant under $\tau^m \sigma^3 \tau^{-m} S$. Since $(\tau^m \sigma^3 \tau^{-m} \mS)^2=1$ and $\tau^m \sigma^3 \tau^{-m} S$ involves non-trivial gauging, it is a non-invertible duality symmetry.

For $PSU(N)_{n,m}$ with $n=0\mod N$, we have
\begin{equation}
\begin{split}
    {PSU(N)_{0,m}}[\tauYM, B]&\stackrel{\mS}{\longrightarrow } {SU(N)_m}[-1/\tauYM, -B]\\& \stackrel{\tau^{-m}}{\longrightarrow} {SU(N)_0}[-1/\tauYM, -B]\\& \stackrel{\sigma^3}{\longrightarrow} {PSU(N)_{0,0}}[-1/\tauYM, B]\\& \stackrel{\tau^m}{\longrightarrow} {PSU(N)_{0,m}}[-1/\tauYM, B]
\end{split}
\end{equation}
Hence $PSU(N)_{0,m}$ at $\tauYM=i$ is invariant under $\tau^m \sigma^3 \tau^{-m} \mS$, which is also a non-invertible duality symmetry.

For $PSU(N)_{n,m}$ with $n\neq 0\mod N$, we have
\begin{equation}
\begin{split}
    {PSU(N)_{n,m}}[\tauYM, B]&\stackrel{\mS}{\longrightarrow} {PSU(N)_{-x,mn^2-n}}[-1/\tauYM, -xB]\\& \stackrel{\tau^{-m}}{\longrightarrow} {PSU(N)_{-x,-n}}[-1/\tauYM, -xB]\\& \stackrel{\sigma}{\longrightarrow} {PSU(N)_{0,n}}[-1/\tauYM, -B]\\& \stackrel{\tau^{-n}}{\longrightarrow} {PSU(N)_{0,0}}[-1/\tauYM, -B]\\& \stackrel{\sigma}{\longrightarrow} {SU(N)_0}[-1/\tauYM, B]\\& \stackrel{\tau^{n}}{\longrightarrow} {SU(N)_n}[-1/\tauYM, B]\\& \stackrel{\sigma}{\longrightarrow} {PSU(N)_{n,0}}[-1/\tauYM, B]\\& \stackrel{\tau^{m}}{\longrightarrow} {PSU(N)_{n,m}}[-1/\tauYM, B]~,
\end{split}
\end{equation}
where $x$ is the modular inverse of $n$. Hence $PSU(N)_{n,m}$ for $n\neq 0\mod N$ at $\tauYM=i$ is invariant under $\tau^m \sigma \tau^n \sigma \tau^{-n} \sigma  \tau^{-m} \mS$. Since $(\tau^m \sigma \tau^n \sigma \tau^{-n} \sigma  \tau^{-m} \mS)^2=1$, this is an operation of order two.  The symmetry involves non-trivial gauging  and hence is non-invertible (except when $N=2$ and $n=1$, where it reduces to $\tau \mS$ and is invertible).  
We summarize the symmetries in Table \ref{tab:suNdutable}. One can confirm that in the special case of $N=3$, the non-invertible duality symmetries we find here reduce to those listed in Table \ref{tab:su3dutable}.

\begin{table}
	\centering
	\begin{tabular}{| c | c |c| }
	\hline
		Theory & Defect  & $n$-ality \\
		\hline\hline
		$SU(N)_m, \, PSU(N)_{0,m}$ & $\tau^m\sigma^3 \tau^{-m} \mS$ & 2\\
		
		$PSU(N)_{n,m},~~ (n\neq 0\mod N)$ & $\tau^m \sigma \tau^n \sigma \tau^{-n} \sigma  \tau^{-m} \mS$ &2 \\ \hline
		

	\end{tabular}
	\caption{Non-invertible symmetries of $\mathfrak{su}(N)$ with $N$ prime and $N\geq 3$ at $\tauYM=i$.
	}
	\label{tab:suNdutable}
\end{table}

\subsubsection*{Triality symmetries at $\tauYM=e^{2\pi i/3}$}

We next identify the modular defects at $\tauYM=e^{2\pi i/3}$. The algorithm is similar to the previous case. For $SU(N)_m$, we have
\begin{eqnarray}
\begin{split}
    {SU(N)_m}[\tauYM, B]& \stackrel{\mT}{\longrightarrow} {SU(N)_{m+1}}[\tauYM+1, B]\\& \stackrel{\mS}{\longrightarrow} {PSU(N)_{0,m+1}}[-1/(\tauYM+1), B]\\& \stackrel{\tau^{-m-1}}{\longrightarrow} {PSU(N)_{0,0}}[-1/(\tauYM+1), B]\\& \stackrel{\sigma^3}{\longrightarrow} {SU(N)_{0}}[-1/(\tauYM+1), B]\\& \stackrel{\tau^{m}}{\longrightarrow} {SU(N)_{m}}[-1/(\tauYM+1), B]~.
\end{split}
\end{eqnarray}
Hence the $SU(N)_m$ theory at $\tauYM=e^{2\pi i/3}$ is invariant under $\tau^m \sigma^3 \tau^{-m-1} \mS\mT$. This is a non-invertible triality symmetry since $(\tau^m \sigma^3 \tau^{-m-1})^3=1$ by the $SL(2,\ZZ_{N})$ algebra.

For $PSU(N)_{-1, m}$, we have
\begin{equation}
\begin{split}
    {PSU(N)_{-1,m}}[\tauYM, B] &\stackrel{\mT}{\longrightarrow} {PSU(N)_{0,m}}[\tauYM+1, B]\\& \stackrel{\mS}{\longrightarrow} {SU(N)_{m}}[-1/(\tauYM+1), -B]\\& \stackrel{\tau^{-m-1}}{\longrightarrow} {SU(N)_{-1}}[-1/(\tauYM+1), -B]\\& \stackrel{\sigma^3}{\longrightarrow} {PSU(N)_{-1,0}}[-1/(\tauYM+1), B]\\& \stackrel{\tau^{m}}{\longrightarrow} {PSU(N)_{-1,m}}[-1/(\tauYM+1), B]~.
\end{split}
\end{equation}
Hence $PSU(N)_{-1,m}$ at $\tauYM=e^{2\pi i/3}$ is invariant under $\tau^m \sigma^3 \tau^{-m-1} \mS\mT$, which is again a non-invertible triality symmetry. 

\begin{table}
	\centering
	\begin{tabular}{| c | c |c |}
	\hline
		Theory & Defects  & $n$-ality \\
		\hline\hline
		$SU(N)_m, \, PSU(N)_{-1,m}$ & $\tau^m\sigma^3 \tau^{-m-1} \mS\mT$ & 3\\
		
		$PSU(N)_{n,m},~~ (n\neq -1\mod N, ~(n,N)\neq (1,3))$ & $\tau^m \sigma \tau^n \sigma \tau^{-n-1} \sigma  \tau^{-m} \mS\mT$ &3 \\

        $PSU(3)_{1,m}$ & $\tau^{-1} \mS\mT$ & 1\\ \hline
        
	\end{tabular}
	\caption{(Non-)invertible symmetries of $\mathfrak{su}(N)$ with $N$ prime at $\tauYM=e^{2\pi i/3}$.}
	\label{tab:suNtritable}
\end{table}

Finally for $PSU(N)_{n,m}$ with $n\neq -1\mod N$, we have
\begin{equation}
\begin{split}
    {PSU(N)_{n,m}}[\tauYM, B]&\stackrel{\mT}{\longrightarrow} {PSU(N)_{n,m}}[\tauYM+1, B]\\&\stackrel{\mS}{\longrightarrow} {PSU(N)_{-x,m(n+1)^2-(n+1)}}[-1/(\tauYM+1), -xB] \\&\stackrel{\tau^{-m}}{\longrightarrow} {PSU(N)_{-x,-n-1}}[-1/(\tauYM+1), -xB]\\& \stackrel{\sigma}{\longrightarrow} {PSU(N)_{0,n+1}}[-1/(\tauYM+1), -B]\\& \stackrel{\tau^{-n-1}}{\longrightarrow} {PSU(N)_{0,0}}[-1/(\tauYM+1), -B]\\& \stackrel{\sigma}{\longrightarrow} {SU(N)_0}[-1/(\tauYM+1), B]\\& \stackrel{\tau^{n}}{\longrightarrow} {SU(N)_n}[-1/(\tauYM+1), B]\\& \stackrel{\sigma}{\longrightarrow} {PSU(N)_{n,0}}[-1/(\tauYM+1), B]\\& \stackrel{\tau^{m}}{\longrightarrow} {PSU(N)_{n,m}}[-1/(\tauYM+1), B]~.
\end{split}
\end{equation}
Hence $PSU(N)_{n,m}$ for $n\neq -1\mod N$ at $\tauYM=e^{2\pi i/3}$ is invariant under $\tau^m \sigma \tau^n \sigma \tau^{-n-1} \sigma  \tau^{-m} \mS\mT$. Since $(\tau^m \sigma \tau^n \sigma \tau^{-n-1} \sigma  \tau^{-m} \mS\mT)^3=1$, it realizes a symmetry of order three, and since it involves non-trivial gauging it is non-invertible (except when $N=3$ and $n=1$, in which case the symmetry reduces to $\tau^{-1}$ which is invertible). Thus we conclude that $\tau^m \sigma \tau^n \sigma \tau^{-n-1} \sigma  \tau^{-m} \mS\mT$ is a non-invertible triality symmetry. We enumerate the symmetries in Table \ref{tab:suNtritable}. One can confirm that when $N=3$, the symmetries here reduce to those listed in Table \ref{tab:su3tritable}.

\section{$\mathfrak{so}(2n)$ SYM}
\label{sec:so2}

We next consider the case of $\mathfrak{so}(N)$ SYM for $N$ even. The case of $N$ odd is non-simply-laced, and will be discussed in detail in Section \ref{eq:nonsimplaced}. Among the algebras with $N$ even, it is useful to split the discussion into the cases of $\mathfrak{so}(4k+2)$ and $\mathfrak{so}(4k)$ for $k \in \NN$.

\subsection{$\mathfrak{so}(4k+2)$}
\label{so4k2}

When $N=4k+2$, the $\mathfrak{so}(N)$ theory has a $\ZZ_4^{(1)}$ center symmetry. In the absence of background gauge fields for $\ZZ_4^{(1)}$, there are known to be six variants of the theory, and the corresponding modular transformations can be found in Figure 6 of \cite{Aharony:2013hda}. The six variants are denoted by $Spin(N)$, $SO(N)_\pm$, and $PSO(N)_n$ for $n=0,1,2,3$. 
As usual, we will want to understand the behavior of the theory under modular transformations after having been coupled to background gauge fields for $\ZZ_4^{(1)}$. To do so, it is once again useful to introduce operations 
\bea
&\tau:& \hspace{0.5 in} \text{stack\,\,with\,\,invertible\,\,phase}\,\,{\frac{2\pi}{8}\int \cP(B)}~,
\no\\
&\sigma:& \hspace{0.5 in} \text{gauge\,\,}\ZZ_4^{(1)}\text{\,\,one\,\,form\,\,symmetry}~.
\eea
which now generate $SL(2, \ZZ_4)$.\footnote{Note that there is also an operation $\sigma_2$ corresponding to gauging the $\ZZ_2^{(1)}$ subgroup of $\ZZ_4^{(1)}$. When acting on $Spin(N)_0$ one gets the $SO(N)_{+,0}$ theory, which has a $\Z_2^{(1)}\times \Z_2^{(1)}$ one-form symmetry and a mixed anomaly between them. One can also introduce various generalized $\tau$ operations corresponding to stacking with counterterms associated with various $\Z_2^{(1)}$ background fields. We will briefly touch upon this possibility later, but will not consider it in detail.} As in the case of $\mathfrak{su}(3)$, the operation $\sigma$ is an order four operation squaring to charge conjugation.

We may now proceed in the standard way by using the defining relations of $SL(2, \ZZ_4)$ and $SL(2, \ZZ)$, together with commutativity and reduction to the known diagram in the absence of background gauge fields for $\ZZ_4^{(1)}$, to construct the relevant diagrams. Since the full diagrams are rather complicated, we exhibit only the results after quotienting by charge conjugation, which are in Figures \ref{fig:spin4n+2} and \ref{fig:spin4n+22}. We should point out that as drawn, Figure \ref{fig:spin4n+22} holds for the case of $N=4k+2$ with $k$ odd. The diagram for $k$ even is obtained by simply reversing all of the $\mT$ arrows. 

The consistency of these diagrams mostly follows from the same discussion as for the $\mathfrak{su}(N)$ theories with prime $N$ in Section \ref{sec:su}. However, there is a new feature due to the fact that the order of the one-form symmetry is not prime: there are now theories participating in the web, namely $SO(N)_{-,n}$, which can be obtained by gauging a $\Z_2^{(1)}$ subgroup of the $\Z_4^{(1)}$ one-form symmetry of $Spin(N)$.  As an example, consider the relation between ${SO}(N)_{-,0}$ and $PSO(N)_{0,2}$. By Figure \ref{fig:spin4n+2}, the two are seen to be related by $ PSO(N)_{0,2} = \sigma {SO}(N)_{-,0} $.\footnote{Recall that Figure \ref{fig:spin4n+2} is the quotient of the full diagram by $C$, and thus from this figure alone it is actually unclear whether we should have $ PSO(N)_{0,2} = \sigma {SO}(N)_{-,0} $ or $ PSO(N)_{0,2} = \sigma^3 {SO}(N)_{-,0} $. It turns out that the former is correct.} We can derive this result as follows.

To begin, note that ${SO}(N)_{-,0}$ can be obtained from $Spin(N)_{0}$ via 
\bea
\label{eq:SONexp}
Z_{SO(N)_{-,0}}[\tau, B] = \sum_{\substack{\tilde b\in C^2(X_4,\ZZ_2)\\ \delta \tilde b = \beta (B_e)}} Z_{Spin(N)_0}[\tau, 2 \tilde b + B_e] e^{i {2 \pi \over 4} \int \cP(\tilde b) } e^{i \pi \int \tilde b  B_m}  e^{i {2 \pi \over 4} \int \cP(B_m) }~.
\eea 
Let us explain this expression. First, $2\tilde b + B_e$ is a $\ZZ_4$ gauge field coupled to the $Spin(N)$ theory, which can be constructed from $\ZZ_2$ gauge fields $\tilde b$ and $B_e$ by requiring the dynamical gauge field $\tilde b$ to satisfy the nontrivial bundle constraint $\delta \tilde b= \beta (B_e)$, where $\beta: H^2(X, \ZZ_2) \rightarrow H^3(X, \ZZ_2)$ is the Bockstein homomorphism. $B_m$ is the background

\begin{landscape}
\begin{figure}
\begin{center}
\begin{tikzpicture}[baseline=0,scale = 0.6, yshift=1cm]
 \node[below] (1) at (0,0) {$Spin(N)_0$};
  \node[below] (2) at (0,-4) {$Spin(N)_1$};
   \node[below] (3) at (0,-8) {$Spin(N)_2$};
    \node[below] (4) at (0,-12) {$Spin(N)_3$};
    
     \node[below] (5) at (6,0) {$PSO(N)_{0,0}$};
  \node[below] (6) at (6,-4) {$PSO(N)_{1,0}$};
   \node[below] (7) at (6,-8) {$PSO(N)_{2,0}$};
    \node[below] (8) at (6,-12) {$PSO(N)_{3,0}$};
    
     \node[below] (9) at (12,0) {$PSO(N)_{0,1}$};
  \node[below] (10) at (12,-4) {$PSO(N)_{1,1}$};
   \node[below] (11) at (12,-8) {$PSO(N)_{2,1}$};
    \node[below] (12) at (12,-12) {$PSO(N)_{3,1}$};
    
     \node[below] (13) at (18,0){$PSO(N)_{0,2}$};
  \node[below] (14) at (18,-4) {$PSO(N)_{1,2}$};
   \node[below] (15) at (18,-8) {$PSO(N)_{2,2}$};
    \node[below] (16) at (18,-12) {$PSO(N)_{3,2}$};
    
     \node[below] (17) at (24,0) {$PSO(N)_{0,3}$};
  \node[below] (18) at (24,-4) {$PSO(N)_{1,3}$};
   \node[below] (19) at (24,-8) {$PSO(N)_{2,3}$};
    \node[below] (20) at (24,-12) {$PSO(N)_{3,3}$};
 
      \node[below] (21) at (30,0) {$SO(N)_{-,0}$};
  \node[below] (22) at (30,-4) {$SO(N)_{-,3}$};
   \node[below] (23) at (30,-8) {$SO(N)_{-,2}$};
    \node[below] (24) at (30,-12) {$SO(N)_{-,1}$};
 
  \draw [thick, red,-{Latex[length=2.5mm]}] (1) -- (2) node[midway,  right] {$\tau$};
 \draw [thick, red,-{Latex[length=2.5mm]}] (2) -- (3) node[midway,  right] {$\tau$};
  \draw [thick, red,-{Latex[length=2.5mm]}] (3) -- (4) node[midway,  right] {$\tau$};

    \draw [thick, red,{Latex[length=2.5mm]}-] (21) -- (22) node[midway,  right] {$\tau$};
 \draw [thick, red,{Latex[length=2.5mm]}-] (22) -- (23) node[midway,  right] {$\tau$};
  \draw [thick, red,{Latex[length=2.5mm]}-] (23) -- (24) node[midway,  right] {$\tau$};
  
\draw [thick, red,-{Latex[length=2.5mm]}] (5) -- (9) node[midway,  above] {$\tau$};
\draw [thick, red,-{Latex[length=2.5mm]}] (6) -- (10) node[midway,  above] {$\tau$};
\draw [thick, red,-{Latex[length=2.5mm]}] (7) -- (11) node[midway,  above] {$\tau$};
\draw [thick, red,-{Latex[length=2.5mm]}] (8) -- (12) node[midway,  above] {$\tau$};

\draw [thick, red,-{Latex[length=2.5mm]}] (9) -- (13) node[midway,  above] {$\tau$};
\draw [thick, red,-{Latex[length=2.5mm]}] (10) -- (14) node[midway,  above] {$\tau$};
\draw [thick, red,-{Latex[length=2.5mm]}] (11) -- (15) node[midway,  above] {$\tau$};
\draw [thick, red,-{Latex[length=2.5mm]}] (12) -- (16) node[midway,  above] {$\tau$};
  
  \draw [thick, red,-{Latex[length=2.5mm]}] (13) -- (17) node[midway,  above] {$\tau$};
\draw [thick, red,-{Latex[length=2.5mm]}] (14) -- (18) node[midway,  above] {$\tau$};
\draw [thick, red,-{Latex[length=2.5mm]}] (15) -- (19) node[midway,  above] {$\tau$};
\draw [thick, red,-{Latex[length=2.5mm]}] (16) -- (20) node[midway,  above] {$\tau$};
  
   \draw[thick, red,-{Latex[length=2.5mm]}] (4) to[out=140,in=220] (1);
   \node[left] at (-2,-6) {$\color{red} \tau$};
   
\draw[thick,red,-{Latex[length=2.5mm]}] (21) to[out=320,in=40] (24);
 \node[right] at (32,-6) {$\color{red} \tau$};
   
\draw[thick,red,-{Latex[length=2.5mm]}] (17) to[out=190,in=350] (5);
\node[] at (15,-2) {$\color{red} \mathsf{\tau}$};

\draw[thick,red,-{Latex[length=2.5mm]}] (18) to[out=190,in=350] (6);
\node[] at (15,-6) {$\color{red} \mathsf{\tau}$};

\draw[thick,red,-{Latex[length=2.5mm]}] (19) to[out=190,in=350] (7);
\node[] at (15,-10) {$\color{red} \mathsf{\tau}$};
   
\draw[thick,red,-{Latex[length=2.5mm]}] (20) to[out=190,in=350] (8);
\node[] at (15,-14) {$\color{red} \mathsf{\tau}$};
              
 
\draw [thick, red,{Latex[length=2.5mm]}-{Latex[length=2.5mm]}] (1) -- (5) node[midway,  above] {$\sigma$};
\draw [thick, red,{Latex[length=2.5mm]}-{Latex[length=2.5mm]}] (2) -- (6) node[midway,  above] {$\sigma$};
\draw [thick, red,{Latex[length=2.5mm]}-{Latex[length=2.5mm]}] (3) -- (7) node[midway,  above] {$\sigma$};
\draw [thick, red,{Latex[length=2.5mm]}-{Latex[length=2.5mm]}] (4) -- (8) node[midway,  above] {$\sigma$};
  
\draw [thick, red,{Latex[length=2.5mm]}-{Latex[length=2.5mm]}] (13) to[out=30,in=150] (21);
\node[] at (24,1) {$\color{red} \sigma$};

\draw [thick, red,{Latex[length=2.5mm]}-{Latex[length=2.5mm]}] (14) to[out=30,in=150] (22);
\node[] at (24,-3) {$\color{red} \sigma$};

\draw [thick, red,{Latex[length=2.5mm]}-{Latex[length=2.5mm]}] (15) to[out=30,in=150] (23);
\node[] at (24,-7) {$\color{red} \sigma$};

\draw [thick, red,{Latex[length=2.5mm]}-{Latex[length=2.5mm]}] (16) to[out=30,in=150] (24);
\node[] at (24,-11) {$\color{red} \sigma$};

\draw [thick, red,{Latex[length=2.5mm]}-{Latex[length=2.5mm]}] (10) -- (17) node[midway,  below] {$\sigma$};
\draw [thick, red,{Latex[length=2.5mm]}-{Latex[length=2.5mm]}] (11) -- (18) node[midway,  below] {$\sigma$};
\draw [thick, red,{Latex[length=2.5mm]}-{Latex[length=2.5mm]}] (12) -- (19) node[midway,  below] {$\sigma$};
  
 \draw [thick, red,*-{Latex[length=2.5mm]}] (16,2) -- (9) node[midway,  above] {$\sigma$};
  \draw [thick, red,{Latex[length=2.5mm]}-*]  (20)--(20,-15) node[midway,  below] {$\sigma$};
 
\end{tikzpicture} 
\caption{Web of transformations (quotiented by charge conjugation) for theories with gauge algebra $\mathfrak{so}(4k+2)$. Here $\tau$ represents coupling to an invertible phase, while $\sigma$ corresponds to gauging the $\ZZ_4$ center symmetry. Note that $Spin(N)_m$ represents $Spin(N)$ coupled to $m$ copies of the primitive invertible phase, and likewise for the other theories.}
\label{fig:spin4n+2}
\end{center}
\end{figure}
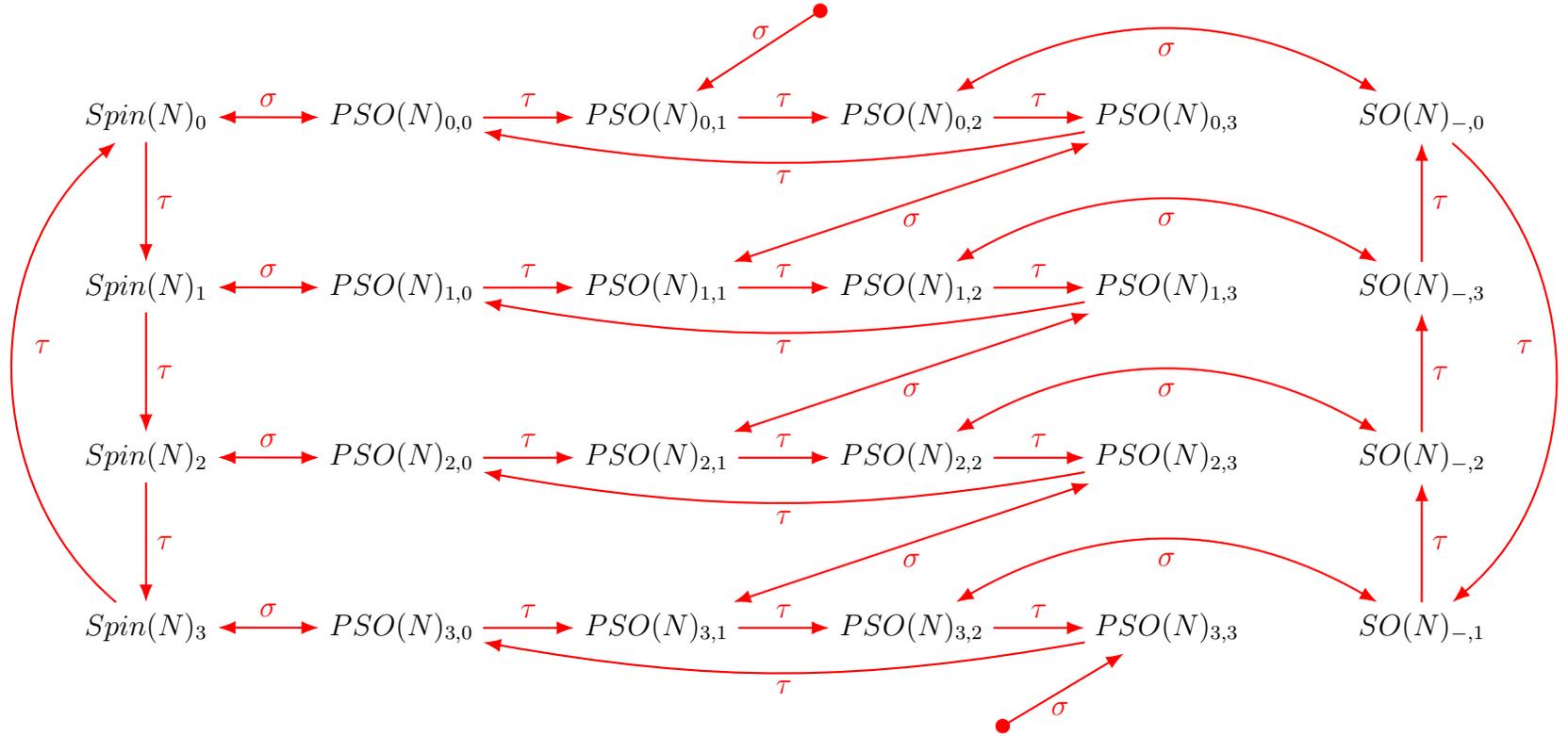
\end{landscape}

\begin{figure}[t]
\hspace*{-3.6em}\begin{tikzpicture}[baseline=0,scale = 0.6, baseline=-10]

 \node[below] (1) at (0,0) {$PSO(N)_{0,0}$};
\node[below] (2) at (0,-4) {$Spin(N)_{0}$};
\node[below] (3) at (0,-8) {$PSO(N)_{0,1}$};
 \node[below] (4) at (0,-12) {$Spin(N)_{1}$};
 \node[below] (5) at (0,-16) {$PSO(N)_{0,2}$};
\node[below] (6) at (0,-20) {$Spin(N)_{2}$};
\node[below] (7) at (0,-24) {$PSO(N)_{0,3}$};
\node[below] (8) at (0,-28) {$Spin(N)_{3}$};

 \node[below] (9) at (8,0) {$PSO(N)_{1,0}$};
\node[below] (10) at (8,-4) {$PSO(N)_{3,3}$};
\node[below] (11) at (8,-8) {$PSO(N)_{1,1}$};
 \node[below] (12) at (8,-12) {$PSO(N)_{3,0}$};
 \node[below] (13) at (8,-16) {$PSO(N)_{1,2}$};
\node[below] (14) at (8,-20) {$PSO(N)_{3,1}$};
\node[below] (15) at (8,-24) {$PSO(N)_{1,3}$};
\node[below] (16) at (8,-28) {$PSO(N)_{3,2}$};
        
 \node[below] (17) at (16,0) {$PSO(N)_{2,0}$};
\node[below] (18) at (16,-4) {$\mathrm{SO}(N)_{-,0}$};
\node[below] (19) at (16,-8) {$PSO(N)_{2,1}$};
 \node[below] (20) at (16,-12) {$\mathrm{SO}(N)_{-,1}$};
 \node[below] (21) at (16,-16) {$PSO(N)_{2,2}$};
\node[below] (22) at (16,-20) {$\mathrm{SO}(N)_{-,2}$};
\node[below] (23) at (16,-24) {$PSO(N)_{2,3}$};
\node[below] (24) at (16,-28) {$\mathrm{SO}(N)_{-,3}$};

    \draw [thick, dgreen,-{Latex[length=2.5mm]}] (1) -- (9) node[midway,  below] {$\mT$};
  \draw [thick, dgreen,-{Latex[length=2.5mm]}] (9) -- (17) node[midway,  below] {$\mT$};
   \draw [thick, dgreen,-{Latex[length=2.5mm]}] (17) -- (12) node[midway,  right] {$\mT$};
     \draw [thick, dgreen,-{Latex[length=2.5mm]}] (12) -- (1) node[midway,  left] {$\mT$};

    \draw [thick, dgreen,-{Latex[length=2.5mm]}] (3) -- (11) node[midway,  below] {$\mT$};
  \draw [thick, dgreen,-{Latex[length=2.5mm]}] (11) -- (19) node[midway,  below] {$\mT$};
    \draw [thick, dgreen,-{Latex[length=2.5mm]}] (19) -- (14) node[midway,  right] {$\mT$};
     \draw [thick, dgreen,-{Latex[length=2.5mm]}] (14) -- (3) node[midway,  left] {$\mT$};

\draw [thick, dgreen,-{Latex[length=2.5mm]}] (5) -- (13) node[midway,  below] {$\mT$};
  \draw [thick, dgreen,-{Latex[length=2.5mm]}] (13) -- (21) node[midway,  below] {$\mT$};
   \draw [thick, dgreen,-{Latex[length=2.5mm]}] (21) -- (16) node[midway,  right] {$\mT$};
     \draw [thick, dgreen,-{Latex[length=2.5mm]}] (16) -- (5) node[midway,  left] {$\mT$};

\draw [thick, dgreen,-{Latex[length=2.5mm]}] (7) -- (15) node[midway,  below] {$\mT$};
  \draw [thick, dgreen,-{Latex[length=2.5mm]}] (15) -- (23) node[midway,  below] {$\mT$};
    \draw [thick, dgreen,-o] (23) -- (13,-29) node[midway,  right] {$\mT$};
      \draw [thick, dgreen,{Latex[length=2.5mm]}-*] (7) -- (3,-29) node[midway,  left] {$\mT$};

  \draw [thick, dgreen,o-{Latex[length=2.5mm]}] (12,0.5) -- (10) node[midway,  right] {$\mT$};
    \draw [thick, dgreen,*-] (4,0.5) -- (10) node[midway,  left] {$\mT$};

       \draw[dgreen,-{Latex[length=2.5mm]}] (2) to[out=230,in=130] (4);
   \node[left] at (-2,-8.5) {$\color{dgreen} \mT$};
         \draw[dgreen,-{Latex[length=2.5mm]}] (4) to[out=230,in=130] (6);
   \node[left] at (-2,-16.5) {$\color{dgreen} \mT$};
    \draw[dgreen,-{Latex[length=2.5mm]}] (6) to[out=230,in=130] (8);
   \node[left] at (-2,-24.5) {$\color{dgreen} \mT$};
    \draw[dgreen,-{Latex[length=2.5mm]}] (8) to[out=140,in=220] (2);
   \node[left] at (-4.7,-16.5) {$\color{dgreen} \mT$};

    \draw[dgreen,-{Latex[length=2.5mm]}] (18) to[out=-50,in=50] (20);
   \node[right] at (18,-8.5) {$\color{dgreen} \mT$};
        \draw[dgreen,-{Latex[length=2.5mm]}] (20) to[out=-50,in=50] (22);
   \node[right] at (18,-16.5) {$\color{dgreen} \mT$};
    \draw[dgreen,-{Latex[length=2.5mm]}] (22) to[out=-50,in=50] (24);
   \node[right] at (18,-24.5) {$\color{dgreen} \mT$};
    \draw[dgreen,-{Latex[length=2.5mm]}] (24) to[out=40,in=-40] (18);
   \node[left] at (21.7,-16.5) {$\color{dgreen} \mT$};
       
 
 \draw[dgreen,{Latex[length=2.5mm]}-{Latex[length=2.5mm]}] (1) -- (2) node[midway,  right] {$\mS$};
  \draw[dgreen,{Latex[length=2.5mm]}-{Latex[length=2.5mm]}] (3) -- (4) node[midway,  right] {$\mS$};
   \draw[dgreen,{Latex[length=2.5mm]}-{Latex[length=2.5mm]}] (5) -- (6) node[midway,  right] {$\mS$};
 \draw[dgreen,{Latex[length=2.5mm]}-{Latex[length=2.5mm]}] (7) -- (8) node[midway,  right] {$\mS$};
 
  \draw[dgreen,{Latex[length=2.5mm]}-{Latex[length=2.5mm]}] (9) -- (10) node[midway,  right] {$\mS$};
  \draw[dgreen,{Latex[length=2.5mm]}-{Latex[length=2.5mm]}] (11) -- (12) node[midway,  right] {$\mS$};
   \draw[dgreen,{Latex[length=2.5mm]}-{Latex[length=2.5mm]}] (13) -- (14) node[midway,  right] {$\mS$};
 \draw[dgreen,{Latex[length=2.5mm]}-{Latex[length=2.5mm]}] (15) -- (16) node[midway,  right] {$\mS$};
 
   \draw[dgreen,{Latex[length=2.5mm]}-{Latex[length=2.5mm]}] (17) -- (18) node[midway,  right] {$\mS$};
  \draw[dgreen,{Latex[length=2.5mm]}-{Latex[length=2.5mm]}] (19) -- (20) node[midway,  right] {$\mS$};
   \draw[dgreen,{Latex[length=2.5mm]}-{Latex[length=2.5mm]}] (21) -- (22) node[midway,  right] {$\mS$};
 \draw[dgreen,{Latex[length=2.5mm]}-{Latex[length=2.5mm]}] (23) -- (24) node[midway,  right] {$\mS$};
 
\end{tikzpicture} 
\caption{Web of transformations (quotiented by charge conjugation) for theories with gauge algebra $\mathfrak{so}(4k+2)$ with $k$ odd. To obtain the web for $k$ even, one simply reverses all of the $\mT$ arrows. Note that $\mS$ and $\mT$ denote the standard modular transformations, while $Spin(N)_m$ denotes $Spin(N)$ coupled to $m$ copies of the primitive invertible phase, and likewise for the other theories.}
\label{fig:spin4n+22}
\end{figure}
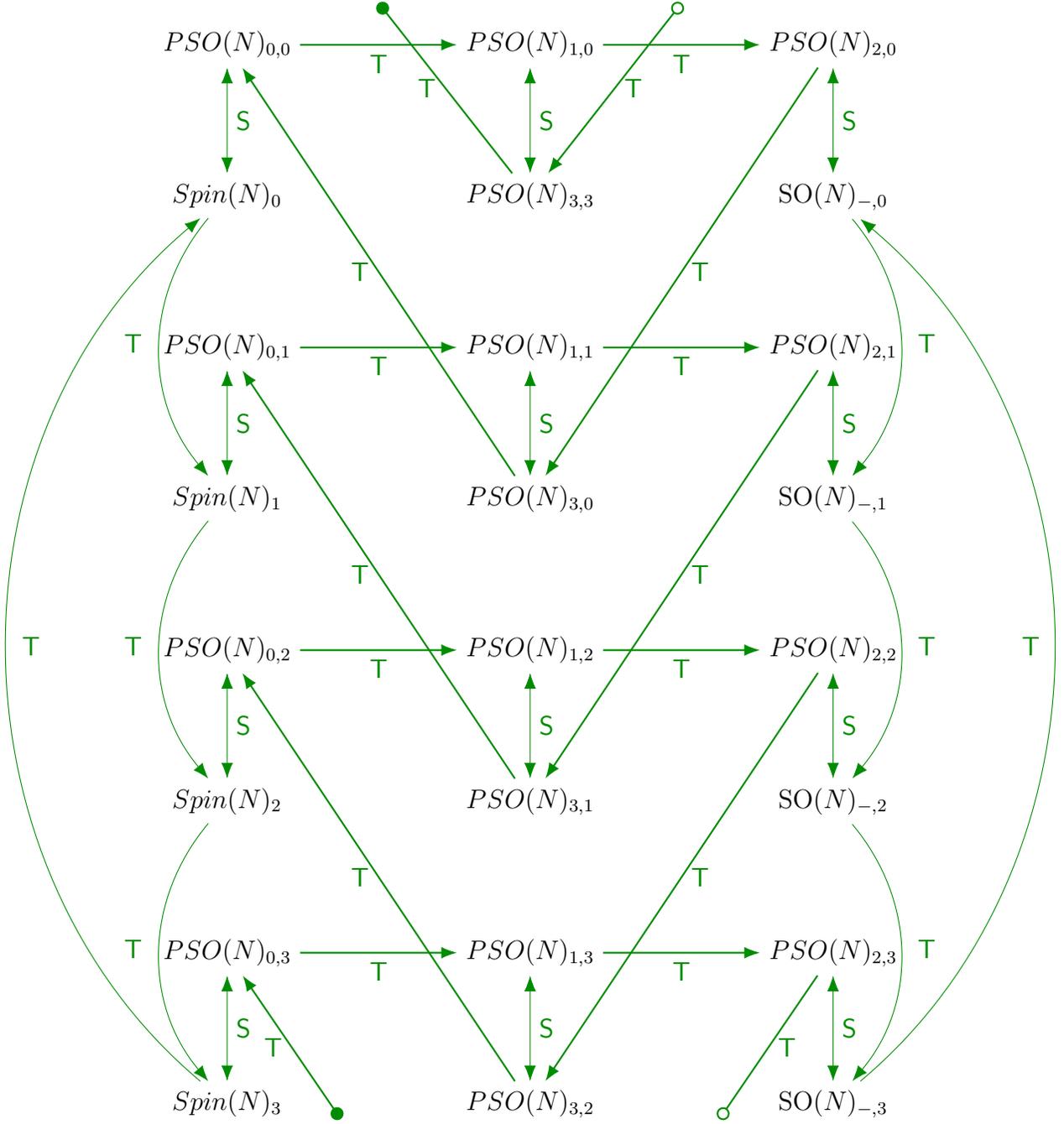

\afterpage{\FloatBarrier}

\noindent  field for the emergent $\ZZ_2$ one-form symmetry after gauging the $\Z_2$  normal subgroup of $\Z_4$.

Although the global symmetry of $SO(N)_{-,0}$ naively appears to be $\Z_2^{(1)}\times \Z_2^{(1)}$ with background fields $B_e$ and $B_m$, it is actually $\Z_4^{(1)}$ \cite{Hsin:2020nts}.  To see this, we extend all the topological terms in \eqref{eq:SONexp} to a bounding 5d manifold $X_5$, and demand that the 5d bulk does not depend on the dynamical gauge field. In particular, the 5d dependence of \eqref{eq:SONexp} is 
\bea
\mathrm{exp} \left[ {i \pi\int_{X_5} \left(  \tilde b  \delta \tilde b + (\delta \tilde b ) B_m + \tilde b  \delta B_m + B_m  \delta B_m\right) }\right]~. 
\eea
Making use of the bundle constraint   $\delta \tilde b= \beta (B_e)$, only the first and the third terms depend linearly on the dynamical field $\tilde b$, and we require them to vanish. This enforces the bundle constraint between background fields to be
\bea
\label{eq:B2mB2erel}
\delta B_m = \beta (B_e)~.
\eea
This implies that $B_m$ and $B_e$ combine into a flat $\Z_4$ two-form gauge field $2B_m + B_e$.
The second and the last terms only depend on the background fields $B_e$ and $B_m$, and they annihilate each other thanks to the bundle constraint \eqref{eq:B2mB2erel}.  So the $SO(N)_{-, 0}$ theory \eqref{eq:SONexp} has a non-anomalous $\Z_4^{(1)}$ one-form global symmetry.

We now note from Figure \ref{fig:spin4n+2} that $Spin(N)_{0} = \tau^2\sigma^{-1} PSO(N)_{0,2}$, which at the level of partition functions means that 
\bea
Z_{Spin(N)_0}[\tau, B] = \sum_{b \in H^2(X, \ZZ_4)} Z_{PSO(N)_{0,2}}[\tau, b] \, e^{- {2 \pi i \over 4} \int \cP(b) - {2 \pi i \over 4} \int b  B}~.
\eea
Plugging this into (\ref{eq:SONexp}) gives 
\bea
Z_{{SO}(N)_{-,0}}[\tau, B] &=& \sum_{\substack{ \tilde b \in C^2(X,\ZZ_2) \\\delta \tilde b = \beta ( B_e) } } \sum_{b \in H^2(X, \ZZ_4)}Z_{PSO(N)_{0,2}}[\tau, b] \, e^{- {2 \pi i \over 4} \int \cP(b) - {2 \pi i \over 4} \int b  (2 \tilde b + B_e)}
\no\\
&\vphantom{.}& \hspace{2in} \times \,e^{i {2 \pi \over 4} \int \cP(\tilde b) + i \pi \int \tilde b  B_m + i {2 \pi \over 4} \int \cP(B_m) } ~.
\eea
Regrouping the Pontryagin squares and integrating out $\tilde b$ gives 
\bea
Z_{{SO}(N)_{-,0}}[\tau, B]&=& \sum_{b \in H^2(X, \ZZ_4)} Z_{PSO(N)_{0,2}}[\tau, b]  \, e^{i {2\pi \over 4} \int \cP(B_m - b)} e^{i {2\pi \over 4} \int \cP(B_m)} e^{- i {2\pi \over 4} \int b  B_e} e^{- i {2\pi \over 4} \cP(b)}
\no\\
&=&\sum_{b \in H^2(X, \ZZ_4)} Z_{PSO(N)_{0,2}}[\tau, b]  \,  e^{- i {2\pi \over 4} \int b  B} ~.
\eea
The right-hand side is none other than $\sigma^3 PSO(N)_{0,2}$, which finally reproduces the result in the diagram.

Having produced a consistent diagram, we may now read off the various modular defects straightforwardly, with the results given in Table \ref{tab:so4N+2dutable} for $\tauYM=i$ and in Table \ref{tab:so4N+2tritable} for $\tauYM=e^{2\pi i/3}$. We see that all of the symmetries in this case are intrinsically non-invertible.

Finally, we wish to comment on the $SO(N)_{+}$ theory, which is an additional variant of the $\mathfrak{so}(N)$ theories that we have not bothered to include in Figures \ref{fig:spin4n+2} and \ref{fig:spin4n+22}. Unlike the variants considered thus far, the $SO(N)_{+}$ theory has a $\ZZ^{(1)}_2\times \ZZ^{(1)}_2$ 1-form symmetries with a mixed anomaly. As such, it sits in its own $SL(2, \ZZ)$ orbit and is in fact completely invariant under $\mS$ and $\mT$, up to stacking with appropriate invertible phases. The $SO(N)_{+}$ can be related to the variants with $\ZZ^{(1)}_4$ one-form symmetry by the gauging of various $\ZZ_2^{(1)}$ subgroups of its one-form symmetry. Specifically, the group $\ZZ^{(1)}_2\times \ZZ^{(1)}_2$ has three $\ZZ^{(1)}_2$ subgroups, corresponding to the two individual factors and the diagonal $\ZZ^{(1)}_2$, and we expect to be able to stack an invertible phase for each of these. This will give eight different variants of the $SO(N)_{+}$ theory, differing solely by invertible phases. 

\begin{table}[t]
	\centering
	\begin{tabular}{| c | c |c |}
	\hline
		Theory & Defect  & $n$-ality  \\
		\hline\hline
		$Spin(N)_m,\, PSO(N)_{0,m}$ & $\tau^m\sigma^{3} {\mS}\tau^{-m}$ & 2\\
		
		$ PSO(N)_{2,m} ,\, SO(N)_{-,m}$ & $\tau^{2+m} \sigma^3 {\mS}\tau^{2-m}$ &2 \\
	
		$ PSO(N)_{n,m}, \, n=1,3$ & $\tau^{m} \sigma \tau^n \sigma \tau^{-n} \sigma \mS \tau^{-m}$ & 2 \\
\hline
	\end{tabular}
	\caption{Non-invertible symmetries of $\mathfrak{so}(N)$ with $N\in 4\ZZ+2$ at $\tauYM=i$.}
	\label{tab:so4N+2dutable}
\end{table}

\begin{table}[t]
	\centering
	\begin{tabular}{| c | c |c |}
	\hline
		Theory & Defect  & $n$-ality \\
		\hline\hline
		$Spin(N)_m,\, PSO(N)_{3,m} $ & $\tau^m \sigma \tau^{-1}\mS\mT\tau^{-m}$ & 3\\
		$PSO(N)_{0,m} $ & $\tau^m \tau^{-1} \sigma \mS\mT\tau^{-m}$ & 3\\
		$PSO(N)_{1,m}, \, SO(N)_{-,m} $ & $\tau^{m+2}  \sigma \tau^{-1}\mS\mT\tau^{-m}$ & 3\\
		$PSO(N)_{2,m} $ & $\tau^{m-1}  \sigma \tau^{2}\mS\mT\tau^{-m}$ & 3\\ \hline
	\end{tabular}
	\caption{Non-invertible symmetries of $\mathfrak{so}(N)$ with $N\in 4\ZZ+2$ at $\tauYM=e^{2 \pi i/3}$.}
	\label{tab:so4N+2tritable}
\end{table}

We can transition between $SO(N)_{+}$ and the other theories by gauging $\ZZ^{(1)}_2$ subgroups of its one-form symmetry group. Since there are eight variants of $SO(N)_{+}$ and three distinct $\ZZ_2$ subgroups, this leads to 24 possible variations. This precisely matches with the 24 different theories participating in the $SL(2, \ZZ_4)$ and $SL(2, \ZZ)$ orbits in Figures \ref{fig:spin4n+2} and \ref{fig:spin4n+22}. This process should be reversible by gauging the $\ZZ^{(1)}_2$ subgroup of the $\ZZ^{(1)}_4$ one-form symmetry of the latter family of theories. 
As $SO(N)_{+}$ is invariant under $\mS$ and $\mT$ up to stacking with invertible phases, we expect all its self-duality defects to be invertible, and thus we shall not work out the precise mapping here. This suggests that the non-invertible defects of the other variants can all be related to the invertible defects of $SO(N)_{+}$ by gauging $\ZZ^{(1)}_2$ subgroups of their $\ZZ^{(1)}_4$ one-form symmetry group. 

We should also mention that it is possible to stack the $\mathfrak{so}(N)$ variants with $\ZZ^{(1)}_4$ one-form symmetry with an invertible phase for the $\ZZ^{(1)}_2$ subgroup of the one-form symmetry. This would lead to an additional 24 variants differing by this invertible phase. We note, though, that due to the results of \cite{Ang:2019txy} gauging the $\ZZ^{(1)}_2$ subgroup of the one-form symmetry in these variants should actually map us back to a variant with $\ZZ^{(1)}_4$ one-form symmetry, which we have already made use of in \eqref{eq:SONexp}. As such, this class of theories should be disconnected from the $SO(N)_{+}$ theory, but  are related to themselves and the variants without the invertible phase under gauging of subgroups of the one-form symmetry. While one might be able to use such relations to generate non-invertible symmetries, we expect the resulting symmetries to be equivalent to the ones we already found, and thus shall not consider them here. 

\subsection{$\mathfrak{so}(4k)$}
\label{sec:so4nsec}

We next proceed to the case of $\mathfrak{so}(N)$ SYM for $N=4k$, which has a $\ZZ_2^{(1)} \times \ZZ_2^{(1)}$ one-form symmetry. This case is significantly more complicated than the previous ones, due to the large number of global variants. Even without considering couplings to background gauge fields, there are already 15 variants of the theory, which can be split into $Spin(N)$, $SO(N)_\pm$, $Ss(N)_\pm$, $Sc(N)_\pm$, and eight variants of $PSO(N)$ differing by discrete theta angles. When counterterms are accounted for, this number grows considerably. 

Let us begin the discussion by considering $Spin(N)$. The most general local counterterm is of the form 
\bea\label{countertermZ2Z2}
S^{\mathrm{ct}}_{m_1, m_2, m_3}[B_1, \, B_2] =  {\pi \over 2} \int\left[m_1 \cP(B_1) + m_2  \cP(B_2) + m_3  \cP(B_1+B_2)   \right]
\eea
with $m_i \in \{0,1\}$, and hence there are eight variants of the $Spin(N)$ theory which we denote by $Spin(N)_{m_1, m_2, m_3}$. 

We next consider the $Ss(N)$ theories. These are obtained by gauging the first $\ZZ_2^{(1)}$ of the symmetry group $\Z_2^{(1)}\times \Z_2^{(1)}$ of $Spin(N)$, with or without discrete torsion. In full generality we might expect $2^{3+3}=64$ variants of the theory, given by 
\begin{eqnarray}
\begin{split}
    &Z_{Ss(N)_{n_1,n_2,n_3;m_1,m_2,m_3}}[\tau, B_1, B_2] \\& \hspace{0.8cm}=e^{i S^{\mathrm{ct}}_{m_1, m_2, m_3}[B_1, \, B_2] } 
\sum_{b_1\in H^2(X_4, \Z_2)} Z_{Spin(N)_{0,0,0}}[\tau, b_1, B_2] e^{i \pi \int b_1 \cup B_1} e^{i S^{\mathrm{ct}}_{n_1, n_2, n_3}[b_1, \, B_2] }~.\,\,
\end{split}
\end{eqnarray}
However, noting that $n_2$ can be absorbed by shifting $m_2$ and $n_3$ can be absorbed by shifting $B_1 \rightarrow B_1 + n_3 B_2$, we conclude that only $n_1$ is physically relevant, corresponding to the difference between $Ss(N)_+$ and $Ss(N)_-$. We will in general denote these theories by $Ss(N)_{n_1;m_1,m_2,m_3}$, and we see there are 16 distinct variants. 

Exactly analogous statements hold for $Sc(N)$ and $SO(N)$, both of which are obtained from $Spin(N)$ by gauging a single $\ZZ_2^{(1)}$. 
On the other hand, for $PSO(N)$ which is obtained by gauging the full $\ZZ_2^{(1)} \times \ZZ_2^{(1)}$ of $Spin(N)$, there are indeed 64 distinct variants which we denote by $PSO(N)_{n_1, n_2, n_3; m_1, m_2, m_3}$. Note that in the notation of \cite{Aharony:2013hda}, the eight choices of discrete theta angles correspond to the choice of triplet $(n_1, n_2, n_3)$. The labels $(m_1, m_2, m_3)$ carry the additional data about local counterterms. In total then, we have $8 + 3 \times 16 + 64 = 120$ variants of the theory which must be kept track of.

In order to better understand these theories, we should also specify the topological manipulations acting on them, including gauging either the entire $\Z_2^{(1)}\times \Z_2^{(1)}$ one-form symmetry or a subgroup thereof, and stacking with a counterterm of the form \eqref{countertermZ2Z2}. Given any theory $\cT$ with symmetry $\Z_2^{(1)}\times \Z_2^{(1)}$, one can define the following six operations analogous to the $\sigma$ and $\tau$ in \eqref{suNsigmatau}: 
\begin{equation}
\renewcommand{\arraystretch}{1.2}
\begin{array}{lll}
    &\sigma_1: {\cT}[B_1, B_2] \to \sum_{b\in H^2(X_4, \Z_2)}{\cT}[b, B_2]e^{i \pi \int b B_1}, 
    & \tau_1: {\cT}[B_1, B_2] \to {\cT}[B_1, B_2] e^{\frac{\pi i}{2}\int \cP(B_1)}\\
    &\sigma_2: {\cT}[B_1, B_2] \to \sum_{b\in H^2(X_4, \Z_2)}{\cT}[B_1, b]e^{i \pi \int b B_2},
    & \tau_2: {\cT}[B_1, B_2] \to {\cT}[B_1, B_2] e^{\frac{\pi i}{2}\int \cP(B_2)}\\
    &\sigma_3: {\cT}[B_1, B_2] \to \sum_{b\in H^2(X_4, \Z_2)}{\cT}[b+B_1, b]e^{i \pi \int b B_2}, 
    & \tau_3: {\cT}[B_1, B_2] \to {\cT}[B_1, B_2] e^{i \pi \int B_1 B_2}\\
\end{array}
\end{equation}
The algebra between $\{\sigma_{1,2,3}, \tau_{1,2,3}\}$ is a generalization of the $SL(2,\Z_2)$ algebra, though we will not determine the entire set of relations between them here.  Instead, let us just highlight a special combination $\lambda := \sigma_3 \sigma_2 \tau_3 \sigma_3 \sigma_1$ which turns out to be an invertible defect exchanging the two $\Z_2^{(1)}$ factors. Concretely, 
\begin{eqnarray}\label{exchangeB1B2}
\lambda:\,\, \cT[B_1, B_2] \to \cT[B_2, B_1]~.
\end{eqnarray}
This will be useful below.

Since there are 120 varieties of  $\mathfrak{so}(4k)$ theories, we will not attempt to find the entire set of duality/triality defects for each variant of the theory. Instead, we will focus on $Spin(8)_{0,0,0}$ theory only, as an illustrative example. Defects for other variants can be constructed analogously.

\subsection*{Defects in $Spin(8)_{0,0,0}$ theory}

The modular $\mS$ and $\mT$ transformations acting on the $Spin(8)_{0,0,0}$ theory are \cite{Aharony:2013hda}
\begin{eqnarray}
\begin{split}
    \mS: \hspace{1cm}& Spin(8)_{0,0,0}[\tauYM, B_1, B_2]\to PSO(8)_{0,0,0;0,0,0}[-1/\tauYM, B_1, B_2]~,\\
    \mT:\hspace{1cm} & Spin(8)_{0,0,0}[\tauYM, B_1, B_2]\to Spin(8)_{0,0,1}[\tauYM+1, B_1, B_2]~.\\
\end{split}
\end{eqnarray}
Recall that $\mathfrak{so}(8)$ SYM is especially interesting since in this case there is an additional triality operation $\mU$ at any value of $\tauYM$ \cite{Aharony:2013hda, Choi:2021kmx, Choi:2022zal}. This operation cyclically permutes the three $\Z_2^{(1)}$ subgroups of $\Z_2^{(1)}\times \Z_2^{(1)}$, which we will hereafter denote by $\Z_{2,S}^{(1)}, \Z_{2,C}^{(1)}$, and $\Z_{2,V}^{(1)}$. In terms of the background fields, $\mU$ is defined as
\begin{eqnarray}
\mU: \hspace{1cm} Spin(8)_{0,0,0}[\tauYM, B_1, B_2]\to Spin(8)_{0,0,0}[\tauYM, B_2, B_1+B_2]~.
\end{eqnarray}
Below we will identify a duality defect at $\tauYM=i$ (involving $\mS$) and two triality defects at $\tauYM=e^{2\pi i/3}$ (involving $\mS\mT$ and $\mU\mS\mT$ respectively), following similar steps as in Section \ref{sec:suN}.

We begin with the duality defect. Starting with $Spin(8)_{0,0,0}$, we first perform $\mS$, and then search for a sequence of topological manipulations that map the resulting theory back to $Spin(8)_{0,0,0}$ at $-1/\tauYM$. Concretely, we have
\begin{eqnarray}
\begin{split}
    Spin(8)_{0,0,0}[\tauYM, B_1, B_2] &\xrightarrow{\mS} PSO(8)_{0,0,0;0,0,0}[-1/\tauYM, B_1, B_2] \\
    &\xrightarrow{\sigma_1} Sc(8)_{0;0,0,0}[-1/\tauYM, B_1, B_2]\\
    &\xrightarrow{\sigma_2} Spin(8)_{0,0,0}[-1/\tauYM, B_1, B_2]~.
\end{split}
\end{eqnarray}
Requiring that the initial and the final theories coincide, we find that the $Spin(8)_{0,0,0}$ theory at $\tauYM=i$ has a symmetry $\sigma_2\sigma_1 \mS$. This symmetry is non-invertible since it involves gauging \cite{Kaidi:2021xfk,Choi:2021kmx,Choi:2022zal}. The symmetry is also of order two, since gauging the two $\Z_2^{(1)}$ factors twice amounts to doing nothing. Hence we conclude that $Spin(8)_{0,0,0}$ SYM at $\tauYM=i$ has a non-invertible duality symmetry.

We proceed to the triality defect involving $\mS \mT$. Following the same steps as before, we find 
\begin{eqnarray}
\begin{split}
    Spin(8)_{0,0,0}[\tauYM, B_1, B_2] & \xrightarrow{\mT} Spin(8)_{0,0,1}[\tauYM+1, B_1, B_2]\\
    &\xrightarrow{\mS} PSO(8)_{0,0,0;0,0,1}[-1/(\tauYM+1), B_1, B_2]\\
    &\xrightarrow{\tau_3} PSO(8)_{0,0,0;0,0,0}[-1/(\tauYM+1), B_1, B_2]\\
    &\xrightarrow{\sigma_1} Sc(8)_{0;0,0,0}[-1/(\tauYM+1), B_1, B_2]\\
    &\xrightarrow{\sigma_2} Spin(8)_{0,0,0}[-1/(\tauYM+1), B_1, B_2]~.\\
\end{split}
\end{eqnarray}
Thus the $Spin(8)_{0,0,0}$ theory at $\tauYM=e^{2\pi i/3}$ is invariant under $\sigma_2\sigma_1 \tau_3 \mS\mT$. To see that this is a triality, we imagine applying $\sigma_2\sigma_1\tau_3$ to an arbitrary theory three times. This can be seen to give
\begin{eqnarray}
(\sigma_2\sigma_1\tau_3)^3: \hspace{1cm} \cT[B_1, B_2]\to \cT[B_2, B_1]~,
\end{eqnarray}
i.e. it simply exchanges $B_1$ and $B_2$. Hence fusing $\sigma_2\sigma_1 \tau_3 \mS\mT$ three times is an invertible defect, and hence $\sigma_2\sigma_1 \tau_3 \mS\mT$ is a triality defect.

Finally, we consider the triality defect which involves $\mU\mS\mT$. Applying the same steps as above, we find 
\begin{eqnarray}
\begin{split}
    Spin(8)_{0,0,0}[\tauYM, B_1, B_2] & \xrightarrow{\mS\mT}  PSO(8)_{0,0,0;0,0,1}[-1/(\tauYM+1), B_1, B_2]\\
    &\xrightarrow{\mU} PSO(8)_{0,0,0;0,0,1}[-1/(\tauYM+1), B_1+B_2, B_1]\\
    &\xrightarrow{\tau_3} PSO(8)_{0,0,0;0,0,0}[-1/(\tauYM+1), B_1+B_2, B_1]\\
    &\xrightarrow{\sigma_2} Sc(8)_{0;0,0,1}[-1/(\tauYM+1), B_2, B_1]\\
    &\xrightarrow{\tau_3} Sc(8)_{0;0,0,0}[-1/(\tauYM+1), B_2, B_1]\\
    &\xrightarrow{\lambda} Sc(8)_{0;0,0,0}[-1/(\tauYM+1), B_1, B_2]\\
    &\xrightarrow{\sigma_2} Spin(8)_{0;0,0,0}[-1/(\tauYM+1), B_1, B_2]~.\\
\end{split}
\end{eqnarray}
Thus the $Spin(8)_{0,0,0}$ theory at $\tauYM=e^{2\pi i/3}$ is invariant under $\sigma_2 \lambda\tau_3 \sigma_2 \tau_3 \mU\mS\mT$, where $\lambda$ is defined as in \eqref{exchangeB1B2}. Again, it is a triality defect since applying $\sigma_2 \lambda\tau_3 \sigma_2 \tau_3 $ three times yields an invertible defect
\begin{eqnarray}
(\sigma_2 \lambda\tau_3 \sigma_2 \tau_3 )^3: \hspace{1cm} \cT[B_1, B_2]\to \cT[B_1, B_1+B_2]e^{i \pi \int B_1 B_2}
\end{eqnarray}
On the other hand, it is also straightforward to see that $\sigma_2 \lambda\tau_3 \sigma_2 \tau_3 \mU\mS\mT$ itself is non-invertible since it involves non-trivial gauging. We conclude that the $Spin(8)_{0,0,0}$ theory at $\tauYM=e^{2\pi i/3}$  has a second non-invertible triality defect given by $\sigma_2 \lambda\tau_3 \sigma_2 \tau_3 \mU\mS\mT$. We summarize the results in Table \ref{tab:spin8defects}.  

Let us close by asking whether these non-invertible defects can be related to invertible ones. From the orbit under $SL(2,\ZZ)$ given in Figure 7 of \cite{Aharony:2013hda}, we see that there is a global variant of the theory that is $\mS$ invariant and that sits in the same modular orbit as $Spin(8)$. This suggests that the duality defect is not intrinsically non-invertible. The one based on $\mS\mT$, however, appears to be intrinsically non-invertible, at least by the definitions given in the Introduction. The defect based on $\mU\mS\mT$ is also intrinsically non-invertible, and in fact cannot be related to invertible defects via $\sigma$ or $\tau$ either.

\begin{table}[]
    \centering
    \begin{tabular}{|c|c |c|c|}
    \hline
        Theory & $\tauYM$ & Defect & $n$-ality   \\
         \hline\hline
         $Spin(8)_{0,0,0}$ & $i$ & $\sigma_2\sigma_1 \mS$ & 2\\
         $Spin(8)_{0,0,0}$ & $e^{2\pi i /3}$ & $\sigma_2\sigma_1  \tau_3 \mS\mT$ & 3\\
         $Spin(8)_{0,0,0}$ & $e^{2\pi i /3}$ & $\sigma_2 \lambda\tau_3 \sigma_2 \tau_3 \mU\mS\mT$ & 3\\
         \hline
    \end{tabular}
    \caption{Non-invertible symmetries of $Spin(8)_{0,0,0}$ SYM. }
    \label{tab:spin8defects}
\end{table}

\section{Exceptional cases}
\label{sec:excep}

We now briefly discuss the exceptional cases. Fortunately, the results for these cases are already implicit in the discussions in Section \ref{sec:su}. Indeed, throughout our discussions above, the only ingredients which we have used are the modular group $SL(2, \ZZ)$ and the relevant one-form symmetry. The one-form symmetries of $\mathfrak{e}_6$, $\mathfrak{e}_7$, and $\mathfrak{e}_8$ are respectively $\ZZ_3^{(1)}$, $\ZZ_2^{(1)}$, and trivial, and hence we conclude that the spectrum of modular defects for $\mathfrak{e}_6$ and $\mathfrak{e}_7$ are the same as for $\mathfrak{su}(3)$ and $\mathfrak{su}(2)$ respectively, whereas for $\mathfrak{e}_8$ all modular defects should be invertible.

To be slightly more concrete, the set of global variants for $\mathfrak{e}_7$ can be obtained as follows. One global variant is just $E_7$ itself, where we allow for Wilson lines in all representations, and hence (by Dirac quantization) for monopoles with charges in the weight lattice of $E_7/\mathbb{Z}_2$. The two other variants are associated with $E_7/\mathbb{Z}_2$, but differ by the electric charges carried by the monopole associated with the weights of the fundamental representation $\mathbf{56}$ of $E_7$. The first option is to have the monopole without electric charge, while the second is to allow a dynonic monopole with both electric and magnetic charges associated with the weights of the $\mathbf{56}$. Dirac quantization then fixes the spectrum. We shall refer to these two choices as $(E_7/\mathbb{Z}_2)_+$ and $(E_7/\mathbb{Z}_2)_-$, respectively. As expected, this gives the same set of global variants as in the $\mathfrak{su}(2)$ case, and the analysis continues to be the same in the presence of a background gauge field for $\ZZ_2^{(1)}$. We thus conclude that the spectrum of modular defects is as in Table \ref{tab:e7dutritable}. The results for $\mathfrak{e}_6$ are likewise obtained by making the appropriate relabelings in the $\mathfrak{su}(3)$ tables, i.e. Tables \ref{tab:su3dutable} and \ref{tab:su3tritable}. For completeness, we reproduce this in Table \ref{tab:e6dutritable}.

\begin{table}[!t]
	\hspace{-0.1 in}\begin{tabular}{| c | c |c |}
	\hline
		Theory & Defect  &$n$-ality \\
		\hline\hline
		$(E_7)_m, \, (E_7/\ZZ_2)_{+,m}$ & $\tau^m\sigma {\mS}\tau^{-m}$ & 2\\
		
		$(E_7/\ZZ_2)_{-,m}$ & $\tau^m  \tau {\mS} \tau^{-m}$ &1 \\ \hline
	\end{tabular}
	\quad
	\begin{tabular}{ |c | c |c |}
	\hline
		Theory & Defect  & $n$-ality \\
		\hline\hline
		$(E_7)_m, (E_7/\ZZ_2)_{-,m}$ & $\tau^m \sigma \tau{\mS\mT}\tau^{-m}$ & 3\\
		
		$(E_7/\ZZ_2)_{+,m}$ & $\tau^m \tau \sigma {\mS\mT}\tau^{-m}$ &3 \\ \hline

	\end{tabular}

	\caption{(Non-)invertible symmetries of $\mathfrak{e}_7$ at $\tauYM= i $ (left) and $\tauYM= e^{2\pi i/3}$ (right).}
	\label{tab:e7dutritable}
\end{table}

\begin{table}[!t]
	\hspace{-0.4 in}\begin{tabular}{| c | c |c |}
	\hline
		Theory & Defect  &$n$-ality \\
		\hline\hline
		$(E_6)_m, \, (E_6/\ZZ_3)_{0,m}$ & $\tau^m\sigma^3 {\mS}\tau^{-m}$ & 2\\
		
		$(E_6/\ZZ_3)_{1,m} ,\, (E_6/\ZZ_3)_{2,m-1}$ & $\tau^{m+1} \sigma {\mS} \tau^{2-m}$ &2 \\ \hline

	\end{tabular}
	\quad \begin{tabular}{ |c | c |c |}
	\hline
		Theory & Defect  & $n$-ality \\
		\hline\hline
		$(E_6)_m, \, (E_6/\ZZ_3)_{2,m}$ & $\tau^m \sigma^3 \tau^2 \mS \mT \tau^{-m}$ & 3\\
		
		$(E_6/\ZZ_3)_{0,m}$ & $\tau^{m}\tau^2 \sigma^3 {\mS\mT} \tau^{-m}$ &3 \\
		
		$(E_6/\ZZ_3)_{1,m}$ & $\tau^{m}\tau^2 {\mS\mT} \tau^{-m}$ &1 \\ \hline

	\end{tabular}

	\caption{(Non-)invertible symmetries of $\mathfrak{e}_6$ at $\tauYM= i $ (left) and $\tauYM= e^{2\pi i/3}$ (right).}
	\label{tab:e6dutritable}
\end{table}

\section{Non-simply-laced cases}
\label{eq:nonsimplaced}

Finally, we consider the case of non-simply-laced gauge groups. The only non-simply-laced algebras with a non-trivial center are $\mathfrak{usp}(2N)$ and $\mathfrak{so}(2N+1)$, and hence these are the only non-simply-laced groups where non-invertible defects of the type discussed in this paper can arise. Since S-duality relates the two algebras, we need to discuss both at the same time.

For non-simply-laced groups, the discussion regarding S-duality that we have given above needs to be slightly modified. Indeed, it is known that for theories with non-simply-laced gauge algebras, S-duality identifies the theory with algebra $\mathfrak{g}$ at coupling $\tauYM$ to a theory with algebra $\mathfrak{g}^\vee$ at coupling $\widetilde{\mS}\tauYM$, where $\widetilde{\mS}$ is defined as \cite{Vafa:1997mh,Argyres:2006qr}
\bea
\label{eq:Stildedef}
\widetilde \mS:\,\, \tauYM \rightarrow - {1\over n_\mathfrak{g} \tauYM}~, \hspace{0.5 in} n_\mathfrak{g} = \left\{ \begin{matrix} \,\,2 &\hspace{0.5in} \mathfrak{g}= \mathfrak{b}_N,\, \mathfrak{c}_N, \mathfrak{f}_4 \\ \,\,3 &  \mathfrak{g}=\mathfrak{g}_2  \end{matrix} \right. ~. 
\eea
As an element of $SL(2, \RR)$, this can be written as
\bea
\widetilde{\mS} = \left( \begin{matrix} 0 & 1/\sqrt{n_\mathfrak{g}} \\ - \sqrt{n_\mathfrak{g}} & 0 \end{matrix} \right)~.
\eea
Noting that $\widetilde{\mS}\mT\widetilde{\mS}$ is equivalent to $\mS\mT^{n_\mathfrak{g}}\mS$, and that $\mS\mT^{n_\mathfrak{g}}\mS$ together with $\mT$ generates the congruence subgroup $\Gamma_0(n_\mathfrak{g})\subset SL(2,\ZZ)$, we conclude that the full Montonen-Olive duality group in the non-simply-laced case is an extension of $\Gamma_0(n_\mathfrak{g})$ by $\widetilde{\mS}$. This gives a so-called ``Hecke group" $\cH(n_\mathfrak{g})$ defined by the relations 
\bea
\label{HeckeRel}
(\widetilde{\mS})^2 = (\widetilde{\mS} \mT)^{2n_\mathfrak{g}} = \mC~, \hspace{0.5in} \mC^2 =1 ~.
\eea
At special points on the boundary of the fundamental domain of the Hecke group, part of the group becomes an enhanced symmetry of the theory. For the cases of $\mathfrak{usp}(2N)$ and $\mathfrak{so}(2N+1)$ of interest to us here, these points (and the resulting symmetries) are
\bea
\label{eq:f4enhancedsymm}
 \tauYM &=& {i \over \sqrt{2}}: \hspace{0.45in} \widetilde{\ZZ}_4 := \langle \widetilde{\mS} \rangle~,
 \no\\
  \tauYM &=& {i \pm 1\over 2}: \qquad \widetilde{\ZZ}_8 := \langle \widetilde{\mS} \mT\rangle \quad \mathrm{or} \quad  \ZZ_{4} := \langle (\widetilde{\mS} \mT)^2\rangle~.
\eea
Note that from \eqref{HeckeRel}, and because the groups we consider here are real, the faithful symmetries are actually only a $\widetilde{\mathbb{Z}}_2$ (coming from $\widetilde{\ZZ}_4$) or $\mathbb{Z}_4$.

\begin{figure}[tbp]
\hspace*{2em}\begin{tikzpicture}[baseline=0,scale = 0.6, baseline=-10]

 \node[below] (1) at (0,0) {$Spin(2N+1)_{1}$};
 \node[below] (2) at (0,-4) {$Spin(2N+1)_{0}$};
 \node[below] (3) at (0,-8) {$SO(2N+1)_{+,0}$};
  \node[below] (4) at (0,-12) {$SO(2N+1)_{+,1}$};
  
 \node[below] (5) at (6,0) {$PUSp(2N)_{+,1}$};
 \node[below] (6) at (6,-4) {$PUSp(2N)_{+,0}$};
 \node[below] (7) at (6,-8) {$USp(2N)_{0}$};
  \node[below] (8) at (6,-12) {$USp(2N)_{1}$};
 
  \node[below] (9) at (12,0) {$SO(2N+1)_{-,1}$};
 \node[below] (10) at (12,-4) {$SO(2N+1)_{-,0}$};
 \node[below] (11) at (12,-8) {$PUSp(2N)_{-,1}$};
  \node[below] (12) at (12,-12) {$PUSp(2N)_{-,0}$};

    \draw[thick,dgreen,{Latex[length=2.5mm]}-{Latex[length=2.5mm]}] (1) to[out=150, in=210,loop] (1);
 \node[left] at (-3.4,-.5) {$\color{dgreen} \mT$};
 
     \draw[thick,dgreen,{Latex[length=2.5mm]}-{Latex[length=2.5mm]}] (2) to[out=150, in=210,loop] (2);
 \node[left] at (-3.4,-4.5) {$\color{dgreen} \mT$};
 
      \draw[thick,dgreen,{Latex[length=2.5mm]}-{Latex[length=2.5mm]}] (3) to[out=150, in=210,loop] (3);
 \node[left] at (-3.4,-8.5) {$\color{dgreen} \mT$};
 
  \draw[thick,dgreen,{Latex[length=2.5mm]}-{Latex[length=2.5mm]}] (4) to[out=150, in=210,loop] (4);
 \node[left] at (-3.4,-12.5) {$\color{dgreen} \mT$};

   \draw[thick,dgreen,{Latex[length=2.5mm]}-{Latex[length=2.5mm]}] (5) to[out=60, in=120,loop] (5);
 \node[right] at (8.5,-4) {$\color{dgreen} \mT$};
 
   \draw[thick,dgreen,{Latex[length=2.5mm]}-{Latex[length=2.5mm]}] (6) to[out=30, in=-30,loop] (6);
 \node[below] at (6.1,2.4) {$\color{dgreen} \mT$};
 
  \draw[thick,dgreen,{Latex[length=2.5mm]}-{Latex[length=2.5mm]}] (7) to[out=30, in=-30,loop] (7);
 \node[right] at (8.5,-8.5) {$\color{dgreen} \mT$};
 
     \draw[thick,dgreen,{Latex[length=2.5mm]}-{Latex[length=2.5mm]}] (8) to[out=240, in=300,loop] (8);
 \node[below] at (6,-14.6) {$\color{dgreen} \mT$};
 
    \draw[thick,dgreen,{Latex[length=2.5mm]}-{Latex[length=2.5mm]}] (9) to[out=60, in=120,loop] (9);
 \node[below] at (12.1,2.4) {$\color{dgreen} \mT$};

   \draw[thick,dgreen,{Latex[length=2.5mm]}-{Latex[length=2.5mm]}] (10) to[out=30, in=-30,loop] (10);
 \node[right] at (14.5,-4.5) {$\color{dgreen} \mT$};
 
  \draw[thick,dgreen,{Latex[length=2.5mm]}-{Latex[length=2.5mm]}] (11) to[out=30, in=-30,loop] (11);
 \node[right] at (14.5,-8.5) {$\color{dgreen} \mT$};
 
      \draw[thick,dgreen,{Latex[length=2.5mm]}-{Latex[length=2.5mm]}] (12) to[out=240, in=300,loop] (12);
 \node[below] at (12,-14.6) {$\color{dgreen} \mT$};
 
    
    \draw [thick, dgreen,{Latex[length=2.5mm]}-{Latex[length=2.5mm]}] (1) -- (5) node[midway,  below] {$\widetilde \mS$};
     \draw [thick, dgreen,{Latex[length=2.5mm]}-{Latex[length=2.5mm]}] (2) -- (6) node[midway,  below] {$\widetilde \mS$};
     \draw [thick, dgreen,{Latex[length=2.5mm]}-{Latex[length=2.5mm]}] (3) -- (7) node[midway,  below] {$\widetilde \mS$};
      \draw [thick, dgreen,{Latex[length=2.5mm]}-{Latex[length=2.5mm]}] (4) -- (8) node[midway,  below] {$\widetilde \mS$};
    
     \draw [thick, dgreen,{Latex[length=2.5mm]}-{Latex[length=2.5mm]}] (10) -- (11) node[midway,  right] {$\widetilde \mS$};
     \draw[thick,dgreen,{Latex[length=2.5mm]}-{Latex[length=2.5mm]}] (9) to[out=-20, in=20] (12);
 \node[right] at (15.5,-6.6) {$\color{dgreen} \widetilde \mS$};

    
    \draw [thick, red,{Latex[length=2.5mm]}-{Latex[length=2.5mm]}] (1) -- (2) node[midway,  right] {$\tau$};
     \draw [thick, red,{Latex[length=2.5mm]}-{Latex[length=2.5mm]}] (3) -- (4) node[midway,  right] {$\tau$};
     \draw [thick, red,{Latex[length=2.5mm]}-{Latex[length=2.5mm]}] (5) -- (6) node[midway,  right] {$\tau$};
      \draw [thick, red,{Latex[length=2.5mm]}-{Latex[length=2.5mm]}] (7) -- (8) node[midway,  right] {$\tau$};
        \draw [thick, red,{Latex[length=2.5mm]}-{Latex[length=2.5mm]}] (9) -- (10) node[midway,  right] {$\tau$};
      \draw [thick, red,{Latex[length=2.5mm]}-{Latex[length=2.5mm]}] (11) -- (12) node[midway,  right] {$\tau$};
      
 \draw [thick, red,{Latex[length=2.5mm]}-{Latex[length=2.5mm]}] (2) -- (3) node[midway,  right] {$\sigma$};
  \draw [thick, red,{Latex[length=2.5mm]}-{Latex[length=2.5mm]}] (6) -- (7) node[midway,  right] {$\sigma$};
  \draw [thick, red,{Latex[length=2.5mm]}-{Latex[length=2.5mm]}] (5) -- (11) node[midway,  below] {$\sigma$};
  \draw [thick, red,{Latex[length=2.5mm]}-{Latex[length=2.5mm]}] (8) -- (12) node[midway,  above] {$\sigma$};
  
    \draw [thick, red,{Latex[length=2.5mm]}-{Latex[length=2.5mm]}] (1) to[out=70, in=120]  (10);
    \node[above] at (3,2) {$\color{red} \sigma$};
    
     \draw [thick, red,{Latex[length=2.5mm]}-*] (4)--(-1.5,-14.5) node[midway,  right] {$\sigma$};

       \draw [thick, red,*-{Latex[length=2.5mm]}] (13.5,1.4)--(9) node[midway,  right] {$\sigma$};
\end{tikzpicture} 
\caption{Web of transformations for theories with gauge algebra $\mathfrak{usp}(2N)$ for $N$ even.}
\label{fig:usp2Neven}
\end{figure}
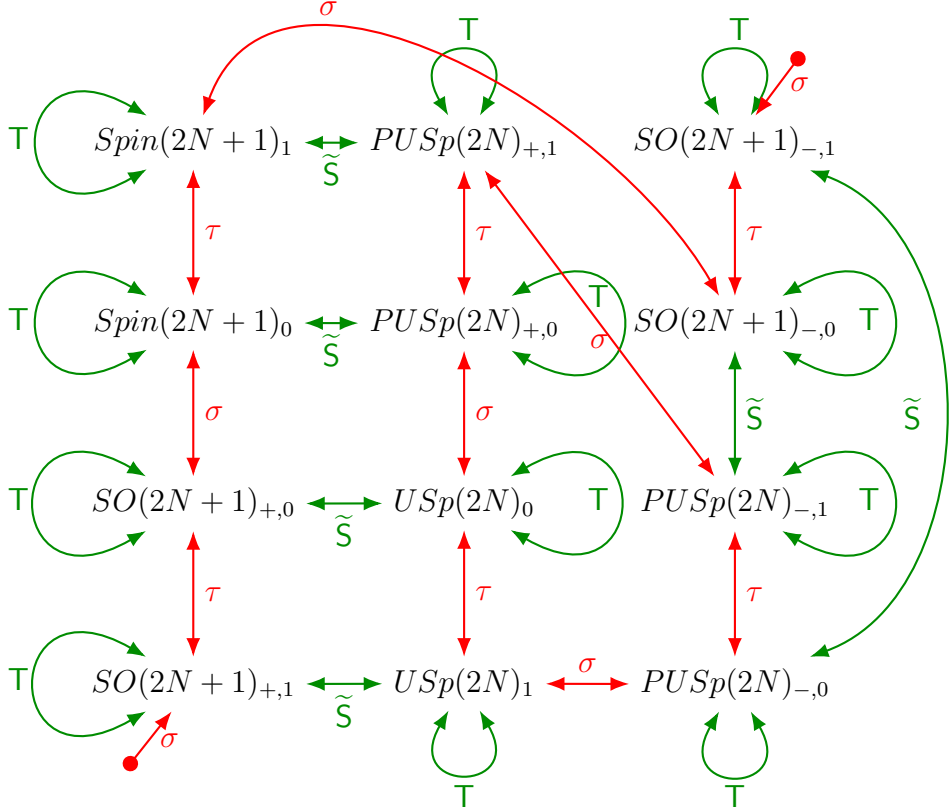

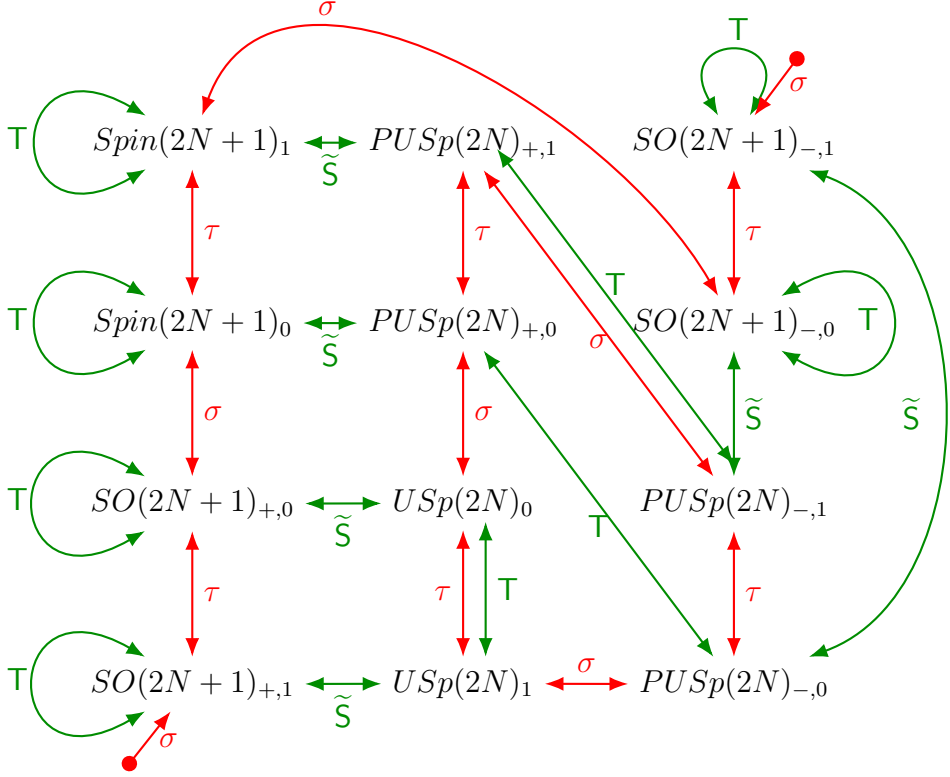
\begin{figure}[tbp]
\hspace*{2em}\begin{tikzpicture}[baseline=0,scale = 0.6, baseline=-10]

 \node[below] (1) at (0,0) {$Spin(2N+1)_{1}$};
 \node[below] (2) at (0,-4) {$Spin(2N+1)_{0}$};
 \node[below] (3) at (0,-8) {$SO(2N+1)_{+,0}$};
  \node[below] (4) at (0,-12) {$SO(2N+1)_{+,1}$};
  
 \node[below] (5) at (6,0) {$PUSp(2N)_{+,1}$};
 \node[below] (6) at (6,-4) {$PUSp(2N)_{+,0}$};
 \node[below] (7) at (6,-8) {$USp(2N)_{0}$};
  \node[below] (8) at (6,-12) {$USp(2N)_{1}$};
 
  \node[below] (9) at (12,0) {$SO(2N+1)_{-,1}$};
 \node[below] (10) at (12,-4) {$SO(2N+1)_{-,0}$};
 \node[below] (11) at (12,-8) {$PUSp(2N)_{-,1}$};
  \node[below] (12) at (12,-12) {$PUSp(2N)_{-,0}$};

   \draw[thick,dgreen,{Latex[length=2.5mm]}-{Latex[length=2.5mm]}] (1) to[out=150, in=210,loop] (1);
 \node[left] at (-3.4,-.5) {$\color{dgreen} \mT$};
 
     \draw[thick,dgreen,{Latex[length=2.5mm]}-{Latex[length=2.5mm]}] (2) to[out=150, in=210,loop] (2);
 \node[left] at (-3.4,-4.5) {$\color{dgreen} \mT$};
 
      \draw[thick,dgreen,{Latex[length=2.5mm]}-{Latex[length=2.5mm]}] (3) to[out=150, in=210,loop] (3);
 \node[left] at (-3.4,-8.5) {$\color{dgreen} \mT$};
 
  \draw[thick,dgreen,{Latex[length=2.5mm]}-{Latex[length=2.5mm]}] (4) to[out=150, in=210,loop] (4);
 \node[left] at (-3.4,-12.5) {$\color{dgreen} \mT$};

    \draw[thick,dgreen,{Latex[length=2.5mm]}-{Latex[length=2.5mm]}] (9) to[out=60, in=120,loop] (9);
 \node[below] at (12.1,2.4) {$\color{dgreen} \mT$};

   \draw[thick,dgreen,{Latex[length=2.5mm]}-{Latex[length=2.5mm]}] (10) to[out=30, in=-30,loop] (10);
 \node[right] at (14.5,-4.5) {$\color{dgreen} \mT$};

 \draw [thick, dgreen,{Latex[length=2.5mm]}-{Latex[length=2.5mm]}] (6.5,-9) -- (6.5,-12) node[midway,  right] {$\mT$};
  \draw [thick, dgreen,{Latex[length=2.5mm]}-{Latex[length=2.5mm]}] (6) -- (12) node[midway,  below] {$\mT$};
    \draw [thick, dgreen,{Latex[length=2.5mm]}-{Latex[length=2.5mm]}] (6.75,-0.733) -- (12,-7.733) node[midway,  above] {$\mT$};
 
    
    \draw [thick, dgreen,{Latex[length=2.5mm]}-{Latex[length=2.5mm]}] (1) -- (5) node[midway,  below] {$\widetilde \mS$};
     \draw [thick, dgreen,{Latex[length=2.5mm]}-{Latex[length=2.5mm]}] (2) -- (6) node[midway,  below] {$\widetilde \mS$};
     \draw [thick, dgreen,{Latex[length=2.5mm]}-{Latex[length=2.5mm]}] (3) -- (7) node[midway,  below] {$\widetilde \mS$};
      \draw [thick, dgreen,{Latex[length=2.5mm]}-{Latex[length=2.5mm]}] (4) -- (8) node[midway,  below] {$\widetilde \mS$};
    
     \draw [thick, dgreen,{Latex[length=2.5mm]}-{Latex[length=2.5mm]}] (10) -- (11) node[midway,  right] {$\widetilde \mS$};
     \draw[thick,dgreen,{Latex[length=2.5mm]}-{Latex[length=2.5mm]}] (9) to[out=-20, in=20] (12);
 \node[right] at (15.5,-6.6) {$\color{dgreen} \widetilde \mS$};

    
    \draw [thick, red,{Latex[length=2.5mm]}-{Latex[length=2.5mm]}] (1) -- (2) node[midway,  right] {$\tau$};
     \draw [thick, red,{Latex[length=2.5mm]}-{Latex[length=2.5mm]}] (3) -- (4) node[midway,  right] {$\tau$};
     \draw [thick, red,{Latex[length=2.5mm]}-{Latex[length=2.5mm]}] (5) -- (6) node[midway,  right] {$\tau$};
      \draw [thick, red,{Latex[length=2.5mm]}-{Latex[length=2.5mm]}] (7) -- (8) node[midway,  left] {$\tau$};
        \draw [thick, red,{Latex[length=2.5mm]}-{Latex[length=2.5mm]}] (9) -- (10) node[midway,  right] {$\tau$};
      \draw [thick, red,{Latex[length=2.5mm]}-{Latex[length=2.5mm]}] (11) -- (12) node[midway,  right] {$\tau$};
      
 \draw [thick, red,{Latex[length=2.5mm]}-{Latex[length=2.5mm]}] (2) -- (3) node[midway,  right] {$\sigma$};
  \draw [thick, red,{Latex[length=2.5mm]}-{Latex[length=2.5mm]}] (6) -- (7) node[midway,  right] {$\sigma$};
  \draw [thick, red,{Latex[length=2.5mm]}-{Latex[length=2.5mm]}] (5) -- (11) node[midway,  below] {$\sigma$};
  \draw [thick, red,{Latex[length=2.5mm]}-{Latex[length=2.5mm]}] (8) -- (12) node[midway,  above] {$\sigma$};
  
    \draw [thick, red,{Latex[length=2.5mm]}-{Latex[length=2.5mm]}] (1) to[out=70, in=120]  (10);
    \node[above] at (3,2) {$\color{red} \sigma$};
    
     \draw [thick, red,{Latex[length=2.5mm]}-*] (4)--(-1.5,-14.5) node[midway,  right] {$\sigma$};

       \draw [thick, red,*-{Latex[length=2.5mm]}] (13.5,1.4)--(9) node[midway,  right] {$\sigma$};
\end{tikzpicture} 
\caption{Web of transformations for theories with gauge algebra $\mathfrak{usp}(2N)$ for $N$ odd.}
\label{fig:usp2Nodd}
\end{figure}

Having understood the relevant modular group, we may now proceed with our usual strategy of constructing the appropriate diagram and reading off the modular defects. Fortunately, since the center of both $\mathfrak{usp}(2N)$ and $\mathfrak{so}(2N+1)$ are $\ZZ_2$, we have the same $SL(2,\mathbb{Z}_2)$ transformations (\ref{eq:su2sigtau}) as in the case of $\mathfrak{su}(2)$. Again in analogy with the $SU(2)$ case, there are three variants per group, leading to a total of six groups in the relevant diagrams: $USp(2N)$, $PUSp(2N)_+$, $PUSp(2N)_-$, $Spin(2N+1)$, $SO(2N+1)_+$ and $SO(2N+1)_-$.\footnote{We use the shorthand notation $PUSp(2N)$ for $USp(2N)/\mathbb{Z}_2$} As in the previous cases, the difference between the $+$ and $-$ choice in the non-simply-connected theory is whether the basic 't Hooft line carries electric charges or not (for $PUSp(2N)_-$ the electric charges are in the fundamental representation while for $SO(N)_-$ they are in the spinor representation). With background gauge fields for the $\ZZ_2^{(1)}$ symmetry turned on, we obtain the diagram in Figure \ref{fig:usp2Neven} for $N$ even, and in Figure \ref{fig:usp2Nodd} for $N$ odd.

We next identify potential topological defects generated by discrete subgroups of the Hecke group. For this we need to find subgroups mapping $\mathfrak{usp}$ to $\mathfrak{usp}$ or $\mathfrak{so}$ to $\mathfrak{so}$. For generic $N$ and generic global variant, this leaves us only with the $\mathbb{Z}_2 = \langle(\widetilde{S} T)^2 \rangle$ at $\tauYM = {i \pm 1 \over {2}}$. This indeed leads to a symmetry, which is invertible for $N$ even since no factors of $\tau$ or $\sigma$ are needed to map back to the original theory, c.f. Figure \ref{fig:usp2Neven}. On the other hand, in the case of $N$ odd this symmetry can be non-invertible. We list the results in Table \ref{tab:usp}. 

\begin{table}[t]
	\centering
	\begin{tabular}{ |c | c |c |}
	\hline
		Theory & Defect  & $n$-ality \\
		\hline\hline
		$Spin(2N+1)_m,\, SO(2N+1)_{-,m} $ & $\tau^m \sigma (\widetilde{\mS}\mT)^2\tau^{-m}$ & 2\\
		$SO(2N+1)_{+,m}, \, USp(2N)_m $ & $\tau^m \tau (\widetilde{\mS}\mT)^2\tau^{-m}$ & 1\\
		$PUSp(2N)_{\pm,m} $ & $\tau^{m}\tau  \sigma \tau (\widetilde{\mS}\mT)^2\tau^{-m}$ & 2\\ \hline
	\end{tabular}
	\caption{(Non-)invertible symmetries of $\mathfrak{usp}(2N)$ and $\mathfrak{so}(2N+1)$ at $\tauYM={1\pm i \over 2}$ for $N$ odd. For $N$ even, all defects are invertible and just given by $(\widetilde{\mS}\mT)^2$ itself.}
	\label{tab:usp}
\end{table}

\subsection*{Case of $USp(4)$}

\begin{figure}[!tbp]
\begin{center}
\begin{tikzpicture}[baseline=0,scale = 0.6, baseline=-10]
 \node[below] (1) at (0,0) {$USp(4)_1$};
  \node[below] (2) at (0,-4) {$USp(4)_0$};

    \node[below] (3) at (5,0) {$PUSp(4)_{+,1}$};
       \node[below] (4) at (5,-4) {$PUSp(4)_{+,0}$};

     \node[below] (5) at (10,0) {$PUSp(4)_{-,0}$};
     \node[below] (6) at (10,-4) {$PUSp(4)_{-,1}$};
     
\draw[thick,dgreen,{Latex[length=2.5mm]}-{Latex[length=2.5mm]}] (1) to[out=150, in=210,loop] (1);
 \node[left] at (-3.4,-.5) {$\color{dgreen} \mT$};
 
 \draw[thick,dgreen,{Latex[length=2.5mm]}-{Latex[length=2.5mm]}] (2) to[out=150, in=210,loop] (2);
 \node[left] at (-3.4,-4.5) {$\color{dgreen} \mT$};
 
 \draw[thick,dgreen,{Latex[length=2.5mm]}-{Latex[length=2.5mm]}] (5) to[out=30, in=-30,loop] (5);
 \node[right] at (13.5,-.5) {$\color{dgreen} \mT$};
 
  \draw[thick,dgreen,{Latex[length=2.5mm]}-{Latex[length=2.5mm]}] (6) to[out=30, in=-30,loop] (6);
 \node[right] at (13.5,-4.5) {$\color{dgreen} \mT$};
 
   \draw[thick,dgreen,{Latex[length=2.5mm]}-{Latex[length=2.5mm]}] (3) to[out=120, in=60,loop] (3);
 \node[above] at (5,1.4) {$\color{dgreen} \mT$};
 
   \draw[thick,dgreen,{Latex[length=2.5mm]}-{Latex[length=2.5mm]}] (4) to[out=-60, in=-120,loop] (4);
 \node[below] at (5,-6.6) {$\color{dgreen} \mT$};
 
 \draw [thick, dgreen,{Latex[length=2.5mm]}-{Latex[length=2.5mm]}] (1) -- (3) node[midway,  below] {$\widetilde \mS$};
\draw [thick, dgreen,{Latex[length=2.5mm]}-{Latex[length=2.5mm]}] (1.3,-4.75) -- (3.3,-4.75) node[midway,  below] {$\widetilde \mS$};
\draw [thick, dgreen,{Latex[length=2.5mm]}-{Latex[length=2.5mm]}] (5) -- (6) node[midway,  right] {$\widetilde \mS$};

  \draw [thick, red,{Latex[length=2.5mm]}-{Latex[length=2.5mm]}] (1) -- (2) node[midway,  left] {$\tau$};
    \draw [thick, red,{Latex[length=2.5mm]}-{Latex[length=2.5mm]}] (3) -- (4) node[midway,  left] {$\tau$};
     \draw [thick, red,{Latex[length=2.5mm]}-{Latex[length=2.5mm]}] (9.5,-1) -- (9.5,-4) node[midway,  left] {$\tau$};

    \draw [thick, red,{Latex[length=2.5mm]}-{Latex[length=2.5mm]}] (3) -- (6) node[midway,  below] {$\sigma$};
     \draw [thick, red,{Latex[length=2.5mm]}-{Latex[length=2.5mm]}] (1.3,-4.25) -- (3.3,-4.25) node[midway,  above] {$\sigma$};
  
  \draw[thick,red,{Latex[length=2.5mm]}-{Latex[length=2.5mm]}] (1) to[out=60, in=120] (5);
 \node[above] at (5,2.5) {$\color{red} \sigma$};
  
\end{tikzpicture} 
\caption{Web of transformations for theories with gauge algebra $\mathfrak{usp}(4)$.}
    \label{USp4Orbit}
\end{center}
\end{figure}
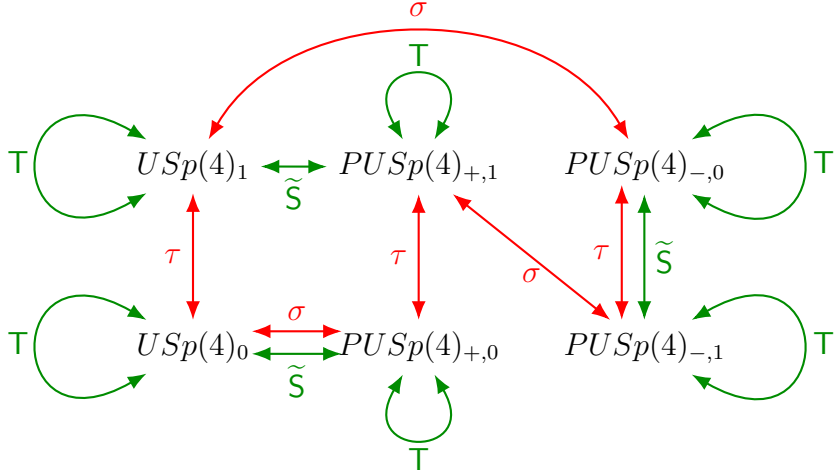

The case of $N=2$ requires special attention since in this case $\mathfrak{usp}(4)\cong \mathfrak{so}(5)$ and hence in addition to the  $(\widetilde{\mS} \mT)^2$ transformation studied above, the $\widetilde{\mS}$ transformation at $\tauYM = {i \over \sqrt{2}} $ can also give rise to symmetries. The diagram for this case is as usual obtained using the defining relations of $SL(2, \ZZ_2)$ and $SL(2,\ZZ)$, together with commutativity of the two and the requirement that the appropriate diagram of \cite{Aharony:2013hda} is reproduced when background fields for $\ZZ_2^{(1)}$ are turned off. The result is given in Figure \ref{USp4Orbit}. From this, we may obtain the set of modular defects at $\tauYM = {i \over \sqrt{2}} $, which are as shown in Table \ref{USp4table}. In particular, we see that we get an invertible $\mathbb{Z}_2$ symmetry for $PUSp(4)_{-,0}$ and $PUSp(4)_{-,1}$, while for the other cases we have a non-invertible defect. 

\begin{table}
	\centering
	\begin{tabular}{|c | c|c |}
		\hline
		Theory & Symmetry & $n$-ality \\
	\hline
		\hline
		$USp(4)_m$, $PUSp(4)_{+,m}$ & $\tau^m \sigma \widetilde{\mS} \tau^{-m}$ &2\\
		
		$PUSp(4)_{-,m}$ & $\tau^m \tau \widetilde{\mS} \tau^{-m}$&1\\ \hline
		
	\end{tabular}
	\caption{(Non-)invertible symmetries for the case of $\mathfrak{usp}(4)$ at $\tauYM = i / \sqrt{2}$.}
	\label{USp4table}
\end{table}

\section{Non-invertible Twisted Compactification}
\label{sec:twistcomp}

 In this section we will consider twisted compactifications of 4d $\cN=4$ using the non-invertible defects obtained above. We will then argue that these give rise to novel 3d $\cN=6$ theories, summarized in Table \ref{tab:newcasessumm}. As mentioned in the Introduction, we will focus on the case of four-manifolds of the form $X_4 = S^1 \times M_3$, with the twist around $S^1$ chosen so as to obtain a 3d theory with $\cN=6$ supersymmetry. Of course, the general idea of twisted compactification by non-invertible defects is not limited to 4d $\cN=4$, and it would be interesting to extend our analysis to find new  theories (and reproduce known ones) in various dimensions and with varying amounts of supersymmetry.

 \begin{table}[tp]
\begin{center}
\begin{tabular}{|c|c|c|}
\hline
$\mathfrak{g}$ & $k$ & $\Gamma$ 
\\
\hline\hline
$\mathfrak{so}(8)$ & 3 & $G_4$ 
\\
$\mathfrak{f}_4$ & 4 & $G_8$ 
\\
& 4$'$ & $G_{12}$ 
\\
$\mathfrak{e}_6$ & 3 & $G_{25}$ 
\\
& 6 & $G_{5}$ 
\\
$\mathfrak{e}_7$ & 4 & $G_{8}$ 
\\
 & 3 & $G_{26}$ 
 \\
$\mathfrak{e}_8$ & 4 & $G_{31}$ 
\\
 & 3 & $G_{32}$ 
 \\\hline
\end{tabular}
\end{center}
\caption{List of exotic 3d $\cN=6$ theories that can be obtained via twisted compactification by $\cN_k \cR_k$ of 4d $\cN=4$ SYM with gauge algebra $\mathfrak{g}$. The exotic theories are characterized by their moduli spaces, which are of the form $\CC^{4r}/ \Gamma$, with $\Gamma$ a complex reflection group. Note that the $\mathfrak{g} = \mathfrak{f}_4$ and $\mathfrak{e}_8$ cases do not involve non-invertible defects, and hence we do not discuss them in the main text, though we include the results here for completeness. }
\label{tab:newcasessumm}
\end{table}%
\subsection{Preserving $\cN=6$ SUSY}
We begin by giving the details of the twisted compactification that allows us to obtain 3d $\cN=6$ theories from 4d $\cN=4$ SYM.\footnote{Similar constructions appear in \cite{Ganor:2008hd,Ganor:2010md,Ganor:2012mu}.} Denote the supercharges of $\cN=4$ SYM by $Q_{a \a}$ and $\overline{Q}^a_{\dot a}$ for $a=1, \dots, 4$ and $\a = 1,2$. These transform in the $(\mathbf{2}, {\mathbf{4}})$ and $(\overline{\mathbf{2}}, \overline{\mathbf{4}})$ of $SO(1,3) \times SO(6)_R$, respectively. To the duality group $SL(2, \ZZ)$ we may associate a $U(1)$ bundle, defined such that the transition functions implementing $\tauYM \rightarrow {a \tauYM + b \over c \tauYM + d}$ are replaced with transition functions $e^{i\, \mathrm{arg}(c \tauYM + d)}$ \cite{Kapustin:2006pk}. The supercharges $Q_{a \a}$ and $\overline{Q}^a_{\dot \a}$ have respective charges $\mp \half$ under this $U(1)$ bundle, i.e. under modular transformation by $\gamma$ we have
\bea
\mathsf{\gamma}: \,\,\,Q_{a \a} \rightarrow \left({c \tau + d \over |c \tau + d|} \right)^{-1/2}Q_{a \a} = e^{-{i v \over 2}}Q_{a \a}
\eea
for $v : = \mathrm{arg}(c \tauYM + d)$. Since all of the supercharges are acted on non-trivially, a twisted compactification involving \textit{only} $SL(2, \ZZ)$ would lead to a theory with no residual supersymmetries. 

To obtain a theory preserving some supersymmetry, we include a twist by the R-symmetry $SO(6)_R \cong SU(4)_R$. We pick bases of $\mathbf{4}$ and $\overline{\mathbf{4}}$ such that the diagonal element 
\bea
\cR : = \mathrm{diag} \left(e^{i \phi_1}, \, e^{i \phi_2}, \,e^{i \phi_3}, \,e^{i \phi_4} \right) ~, \hspace{0.8 in} \sum_{a=1}^4 \phi_a = 0
\eea
of $SU(4)_R$ acts on the supercharges via 
\bea
\cR: \,\,\, Q_{a \a}  \rightarrow e^{i \phi_a}Q_{a \a} ~, \hspace{0.5 in} \overline{Q}^a_{\dot a}  \rightarrow e^{-i \phi_a}\overline{Q}^a_{\dot a}~. 
\eea
In total then, we see that the combined action $\gamma\cR$ acts as 
\bea
\gamma \cR: \,\,\, Q_{a \a}  \rightarrow e^{i \phi_a-i {v \over 2}}Q_{a \a}~, \hspace{0.5 in}\overline{Q}^a_{\dot a}  \rightarrow e^{-i \phi_a+i {v \over 2}}\overline{Q}^a_{\dot a}~,
\eea
and allows us to preserve some subset of the supersymmetries by choosing the appropriate values for $\phi_a$. Indeed,  by choosing $\phi_1 = \phi_2 = \phi_3 = v/2$ and $\phi_4 = -3v/2$, we are able to preserve $\cN=6$ supersymmetry.\footnote{Similarly, we may choose $\phi_1 = \phi_2 = v/2$ and $\phi_3 = - v+\phi_4$ with $\phi_4$ unconstrained to preserve 3d $\cN=4$, or $\phi_1 = v/2$, $\phi_2 = \half v- \phi_3 - \phi_4$ with $\phi_{3,4}$ unconstrained to preserve 3d $\cN=2$. In the current work we will explore only the case of $\cN=6$ supersymmetry.} Taking $\gamma$ to be $\mS$, $\mS\mT$, or $\mS\mT\mC$ in $SL(2, \ZZ)$ gives operations $\gamma\cR$ which are respectively $\ZZ_4$, $\ZZ_3$, or $\ZZ_6$ elements. In these cases we will denote the operations by $\gamma_k\cR_k$ for $k=3,4,6$. 

Depending on the theory in question, the element $\gamma_k$ above may not itself be a symmetry, in which case we should replace it by the appropriate non-invertible symmetry. That is, we should dress $\gamma_k$ above with the appropriate combination of $\{ \sigma, \tau\}$, to obtain a non-invertible symmetry $\cN_k$---it is the latter which have been the objects of focus for most of this paper. The operation $\cN_k \cR_k$ acts on local operators in 4d in precisely the same way as $\gamma_k \cR_k$, since the operations $\{ \sigma, \tau\}$ are topological. Thus we conclude that $\cN_k \cR_k$ transforms the supercharges in the same way as before, and can again be used to obtain a 3d $\cN=6$ theory.

\subsection{Compactifications of $\mathfrak{su}(N)$ SYM}

Let us now consider twisted compactifications of $\cN=4$ SYM with gauge algebra $\mathfrak{su}(N)$. The basic strategy is similar to the one developed in \cite{Kaidi:2022lyo}---namely, we aim to study the resulting theories by understanding the structure of their moduli spaces. Recall that $\cN=4$ SYM with gauge algebra $\mathfrak{g}$ has moduli space $\CC^{3N}/\cW(\mathfrak{g})$, where $N$ and $\cW(\mathfrak{g})$ are the rank and Weyl group of $\mathfrak{g}$, respectively.  Upon performing the twisted compactification described above, we obtain a 3d $\cN=6$ theory with moduli space of the form $\CC^{4N}/\Gamma$, with $\Gamma$ a subgroup of $\cW(\mathfrak{g})$. It has been conjectured that $\Gamma$ is always a complex reflection group \cite{Tachikawa:2019dvq}.\footnote{Complex reflection groups are defined by the fact that their actions on $\mathbb{C}^n$ are spanned by complex reflections, i.e. transformations $z\rightarrow e^{\frac{2\pi i}{k}} z$ with $z$ one of the coordinates of $\mathbb{C}^n$. We shall not review aspects of complex reflection groups here, but rather refer the reader to the many expositions on this subject that are now available in the physics literature \cite{Aharony:2016kai,Bonetti:2018fqz,Tachikawa:2019dvq,Kaidi:2022lyo} as well as the resources available in the mathematics literature, e.g. \cite{lehrer2009unitary}.}

To identify $\Gamma$, we must understand the action of $\cN_k \cR_k$ on the moduli space. Let us fix ourselves to gauge group $SU(N)_0$. For $N$ prime, we know from Tables \ref{tab:suNdutable} and \ref{tab:suNtritable} that
\bea
\label{eq:suNtwists}
\cN_4 = \sigma^3 \mS~, \hspace{0.5 in} \cN_3 = \sigma^3 \tau^{-1}\mS \mT~,\hspace{0.5 in} \cN_6 = \mC \cN_3~.
\eea
 For $N$ non-prime, we have not derived the exact form of the defects, but they should again be of the form $\mS$ and $\mS \mT$ dressed with appropriate factors of $\sigma$ and $\tau$. For example, the case of $N=4$ is already implicit in the results in Section \ref{so4k2} since $Spin(6)\cong SU(4)$.

 To understand the effects of $\cN_k \cR_k$ on the moduli space, it suffices to understand the action of $\gamma_k \cR_k$, since $\sigma$ and $\tau$ are topological operations which do not act on local operators, including those parameterizing the moduli space. Furthermore, for simply-laced groups $\mS$ and $\mT$ also leave the moduli space invariant, so the action on the moduli space is entirely due to $\cR_k$ (this ceases to be true in the non-simply-laced case, as we will discuss in Section \ref{sec:nonsimplylacedcomp}). We now focus on the action of $\cR_k$ on the \textit{invariant polynomials} $\{u_n\}$ parameterizing the moduli space. In particular, $\cR_k$ acts on the order-$n$ invariant polynomials as $u_n \rightarrow e^{2 \pi i n /k} u_n$. 

Noting that the invariant polynomials of $\cW(\mathfrak{su}(N))$ are of dimension $2, 3, 4, \dots, N$, we conclude that the following invariant polynomials persist upon twisting, 
\bea
k&=& 3: \hspace{0.5 in} 3, 6, 9, \dots
\no\\
k&=& 4: \hspace{0.5 in} 4,8, 12, \dots
\no\\
k&=& 6: \hspace{0.5 in} 6, 12, 18, \dots
\eea
These are the correct invariant polynomials for the complex reflection groups $G(k, 1, n)$ with $ n = [ N/k]$ \cite{Tachikawa:2019dvq}. This suggests that twisted compactification of $SU(N)_0$ SYM by $\cN_k \cR_k$ gives rise to a 3d $\cN=6$ theory with moduli space of the form $\CC^{4n}/ G(k,1,n)$.  

We can also arrive at this result from a direct analysis of the moduli space, as done in \cite{Kaidi:2022lyo}, which we shall now briefly review. The moduli space of the $\mathcal{N}=4$ theory is spanned by the vevs of the three complex scalar fields in the vector multiplet. Due to the superpotential, these can be simultaneously diagonalized and we can take them to be $\mathrm{diag}(\phi_1, \phi_2, ... , \phi_N)$, where here each $\phi$ collectively stands for three complex scalar fields and takes values in $\mathbb{C}^3$. Different vevs of the fields $\phi$ stand for different points in the moduli space, as long as they are not related to one another by an $\mathfrak{su}(N)$ gauge transformation. The diagonal choice mostly lifts this degeneracy, except for the Weyl transformations, which preserve the diagonal form---indeed, the Weyl transformations are just given by permutations of the $N$ $\phi_i$. As such the moduli space is spanned by the vevs of the fields $\phi_i$ modulo Weyl transformations.

Now we consider the behavior of the $\phi_i$ under twisted compactification. Since the scalar fields are charged under $\cR_k$ they transform as $\phi_i \rightarrow e^{\frac{2\pi i}{k}}\phi_i$ and so naively would be projected out. However, combinations of the $\phi_i$ fields differing by Weyl transformations are actually equivalent. As such, consider the combination $\sum^k_{j=1} e^{\frac{2\pi i j}{k}}\phi_j$, under which the action of $\cR_k$ is equivalent to the Weyl action cyclically permuting the $k$ $\phi_i$ fields. This combination should survive the reduction, and its vev should span part of the $3d$ moduli space. In general, we expect $n$ independent such combinations corresponding to partitioning the $N$ $\phi_i$ fields into groups of size $k$. This will give the space $\mathbb{C}^{4n}$ in $3d$,\footnote{The moduli space in $3d$ is spanned by four complex scalar fields, where the additional complex scalar comes from the vector component along the circle by dualizing the $3d$ vector. These transform in the $\bf{4}$ of the $SO(6)_R$ R-symmetry group of the $\mathcal{N}=6$ supersymmetry group and so the full moduli space should be completely determined by supersymmetry from its $\mathbb{C}^{3n}$ subspace that descends from the $4d$ scalar fields.} spanned by these $n$ combinations, which we shall refer to as $\phi'_{\alpha}$. However, not every distinct value of these combinations should lead to distinct points in the moduli space, as we expect values differing by the Weyl action of $\mathfrak{su}(N)$ to be again equivalent. The latter is given in terms of permutations of the $n$ $\phi'_{\alpha}$, as well as the transformation $\phi'_{\alpha}\rightarrow e^{\frac{2\pi i}{k}}\phi'_{\alpha}$ for each of the individual $\phi'_{\alpha}$. These build the group $G(k,1,n)$ which is a complex reflection group. We then conclude that the resulting $3d$ theory should have the moduli space $\CC^{4n}/ G(k,1,n)$, as expected from the invariant polynomials. This analysis can be done also for the other cases (see \cite{Kaidi:2022lyo} for more explicit examples) and leads to the same results as the ones implied by the invariant polynomials, so from now on we will only present the arguments following from the invariant polynomials. 

In fact, theories with moduli space $\CC^{4n}/ G(k,1,n)$ are already known: they are the ABJ theories $U(n+x)_k \times U(n)_{-k}$, where $N= k n + x$ with $n \in \NN$ and $x < k$ \cite{Aharony:2008gk}. It is then natural to expect that the theories obtained by (non-)invertible twisted compactification of $\mathfrak{su}(N)$ $\cN=4$ SYM are precisely the ABJ theories. 
String theory gives additional evidence for this proposal. Indeed, recall that the ABJ theories can be obtained by considering a stack of $n$ M2-branes probing $\CC^4/\ZZ_k$, together with $x$ M2-branes stuck to the singular locus. On the other hand, the $\cN=4$ SYM setup we just described involves a stack of $N$  D3-branes compactified on a circle with a $\ZZ_k$ twist. T-dualizing the latter configuration is expected to give a $\ZZ_k$ S-fold. If we split $N = k n + x$ with $n \in \NN$ and $x < k$, then we conclude that the T-dual setup should involve $n$ groups of $k$ mirror D-branes, together with $x$ additional D-branes stuck to the S-fold singularity. Lifting to M-theory gives precisely the setup of ABJ. 

Let us emphasize that the above statements hold regardless of whether $\cN_k$ is invertible or non-invertible. For example, in the $SU(N)_0$ theory for $N$ prime we see from (\ref{eq:suNtwists}) that $\cN_3$, $\cN_4$, and $\cN_6$ are all non-invertible, whereas e.g. for $PSU(3)_{1,0}$ we would have $\cN_3$ and $\cN_6$ being invertible. For generic $N$, all three defects are intrinsically non-invertible. We thus see that twisted compactification with non-invertible defects is \textit{required} to reproduce the correct spectrum of ABJ theories upon T-duality.

We close by noting that one could in principle also consider twists by the outer automorphism of $\mathfrak{su}(N)$. This however will not give rise to anything new, since the outer-automorphism operation can be identified with a combination of the center $\mC$ of $SL(2,\ZZ)$ and the appropriate order-2 R-symmetry transformation. This combination was what differentiated the $k=6$ versus $k=3$ twisted compactifications, and was automatically included in the $k=4$ one. To see that the two operations are identical, we first note that they act identically on local operators: in particular, both act on the invariant polynomials of odd degree with a minus sign, while leaving the polynomials of even degree unchanged. Second, the two operations act on line operators in the same way, as can be seen by noting that in the presence of background fields $B$ the operation $\mS^2$ acts as $B\rightarrow - B$. This means that e.g. Wilson lines in the fundamental will be effectively exchanged with those in the anti-fundamental, as would be the case for the outer-automorphism transformation. Finally, from a string theory perspective, a $\ZZ_2$ S-fold is simply an orientifold, which reduces a complex group (e.g. $SU(2N)$) to a real subgroup (e.g. $SO(2N)$ or $USp(2N)$), and hence should be related in some way to the outer-automorphism operation.  Note that similar statements will hold for potential outer-automorphism twists of $\mathfrak{so}(4N+2)$ and $\mathfrak{e}_6$ as well.

\subsection{Compactifications of $\mathfrak{so}(2N)$ SYM}

We next consider non-invertible twisted compactification of $\mathfrak{so}(2N)$ theories. This will for the most part again give rise to ABJ theories, though in the case of $N=4$ we will identify a novel 3d $\cN=6$ theory with moduli space $\CC^8/G_4$, with $G_4$ an exceptional complex reflection group. 

Let us begin by considering the case of $\mathfrak{so}(2N)$ SYM compactified with a twist by $\cN_4 \cR_4$. The precise form of $\cN_4 \cR_4$ depends on whether $N$ is even or odd, as well as the precise global variant, but in general it is an intrinsically non-invertible symmetry.
To understand the 3d theory resulting from twisted compactification, we again look at the effect on the moduli space. The invariant polynomials for $\cW(\mathfrak{so}(2N))$ have degree $2,4,6,8, \dots, 2N-2, N$, and hence the twist by $\cN_4 \cR_4$ preserves 
\bea
k=4: \hspace{0.5 in} 4,8,12, \dots, \left\{\begin{matrix} 2N-4, \, N & & \hspace{0.2 in} N \in 4 \ZZ \\ 2N-4 &&\hspace{0.5 in}  N \in 4 \ZZ + 2 \\  2N-2 & &\hspace{0.5 in}  N \in 2 \ZZ+1 \end{matrix}\right. ~,
\eea
with the answer depending on the value of $N$ modulo 4. For $N = 4 \ell$ with $\ell \in \NN$, the invariants identified above match with the invariants of $G(4,2,2\ell)$. For $N = 4 \ell+2$, the invariants match with those of $G(4, 1, 2\ell)$, while for $N = 2 \ell + 1$ they match with those of $G(4,1,\ell)$. 
We thus expect the results of twisted compactification to be 3d $\cN=6$ theories with moduli spaces of the form $\CC^{8\ell}/G(4,2,2\ell)$, $\CC^{8\ell}/G(4,1,2\ell)$, or $\CC^{4 \ell}/G(4,1,\ell)$, depending again on the value of $N$ modulo 4. In each case, the corresponding moduli space is known to be realized by a theory of ABJ type, where $G(k,p,\ell)$ is associated with the theory $(U(\ell+x)_k\times U(\ell)_{-k})/\ZZ_p$ with $x<k$.\footnote{There is an additional constraint coming from the fact that the $\ZZ_p$ quotient can only be taken if $\frac{x k}{p^2} \in \ZZ$ \cite{Tachikawa:2019dvq}.} This does not fix the theory completely, as it leaves some freedom in the value of $x$. As before, we have used the string theory picture to conjecture the value of $x$. This data is summarized in Table \ref{tab:so2Nk4}.

\begin{table}[tp]
\begin{center}
\begin{tabular}{|c|c|c|c|}
\hline
$N$ & Degrees & $\Gamma$ & Theory \\
\hline
$4 \ell$ & $4,8,12, \dots, 2N-4, N$ & $G(4,2,2\ell)$ & $(U(2\ell)_4\times U(2 \ell)_{-4})/\ZZ_2$\\
$4 \ell+2$ & $4,8,12, \dots, 2N-4$ & $G(4,1,2\ell)$ & $U(2\ell)_4\times U(2 \ell)_{-4}$\\
$2 \ell+1$ & $4,8,12, \dots, 2N-2$ & $G(4,1,\ell)$ & $U(\ell+2)_4\times U(\ell)_{-4}$\\
\hline
\end{tabular}
\end{center}
\caption{Theories obtained by compactifying $\mathfrak{so}(2N)$ SYM via a twist by $\cN_4 \cR_4$. The result depends on the value of $N$ modulo 4. In all cases we obtain theories of ABJ type.}
\label{tab:so2Nk4}
\end{table}%

\begin{table}[tp]
\begin{center}
\begin{tabular}{|c|c|c|c|}
\hline
$N$ & Degrees & $\Gamma$ & Theory \\
\hline
$3 \ell$ & $6,12,18 \dots, 2N-6, N$ & $G(6,2,\ell)$ & $(U(\ell)_6\times U( \ell)_{-6})/\ZZ_2$\\
$3 \ell + 1$ & $6,12,18, \dots, 2N-2$ & $G(6,1,\ell)$ & $U(\ell+2)_6\times U(\ell)_{-6}$\\
$3 \ell+2$ & $6,12,18, \dots, 2N-4$ & $G(6,1,\ell)$ & $U(\ell+2)_6\times U(\ell)_{-6}$\\
\hline
\end{tabular}
\end{center}
\caption{Theories obtained by compactifying $\mathfrak{so}(2N)$ SYM via a twist by $\cN_3 \cR_3$. The result depends on the value of $N$ modulo 3. In all cases we obtain theories of ABJ type.}
\label{tab:so2Nk3}
\end{table}%

\begin{table}[tp]
\begin{center}
\begin{tabular}{|c|c|c|c|}
\hline
$N$ & Degrees & $\Gamma$ & Theory \\
\hline
$6 \ell$ & $6,12,18 \dots, 2N-6, N$ & $G(6,2,2\ell)$ & $(U(2\ell)_6\times U(2\ell)_{-6})/\ZZ_2$\\
$3 \ell + 1$ & $6,12,18, \dots, 2N-2$ & $G(6,1,\ell)$ & $U(\ell+2)_6\times U(\ell)_{-6}$\\
$3 \ell+2$ & $6,12,18, \dots, 2N-4$ & $G(6,1,\ell)$ & $U(\ell+2)_6\times U(\ell)_{-6}$\\
$6 \ell+3$ & $6,12,18, \dots, 2N-6$ & $G(6,1,2\ell)$ & $U(2\ell)_6\times U(2\ell)_{-6}$\\
\hline
\end{tabular}
\end{center}
\caption{Theories obtained by compactifying $\mathfrak{so}(2N)$ SYM via a twist by $\cN_6 \cR_6$. The result depends on the value of $N$ modulo 6. In all cases we obtain theories of ABJ type.}
\label{tab:so2Nk6}
\end{table}%

In an exactly similar manner, one may consider twisted compactification by $\cN_3 \cR_3$ and $\cN_6 \cR_6$. This gives rise to the results shown in Tables \ref{tab:so2Nk3} and \ref{tab:so2Nk6}, respectively. As we can see, the moduli spaces obtained all seem consistent with interpretations as ABJ theories. 

In the special case of $\mathfrak{so}(8)$ SYM, there is an additional twist that we can include, which is namely the  $\mathfrak{so}(8)$ triality transformation $\mU$. For the case of $Spin(8)_{0,0,0}$, the non-invertible symmetry involving $\mathsf{U}$ was found in Section \ref{sec:so4nsec} to be of the form $\cN_3 = \sigma_2 \lambda \tau_3 \sigma_2 \tau_3 \mU \mS \mT$.  For the current discussion, it will be important to understand how $\mU$ acts on the invariant polynomials parameterizing the moduli space.  The invariant polynomials of $\mathfrak{so}(8)$ have dimension 2,4,4, and 6, and will be denoted by $u_2, u_4, \widetilde{u}_4,$ and $u_6$.  With appropriate choice of basis, triality can be chosen to map \cite{Kaidi:2022lyo}
\bea
\mathsf{U}: \qquad u_4 \rightarrow e^{- 2\pi i /3}\, u_4~, \hspace{0.5 in} \widetilde{u}_4 \rightarrow  e^{2\pi i/3}\, \widetilde{u}_4~.
\eea
Recalling that $\cR_3$ acts as 
\bea
u_2 \rightarrow e^{-2 \pi i /3}\,u_2~, \hspace{0.3 in} u_4 \rightarrow e^{2 \pi i /3}\,u_4~, \hspace{0.3 in} \widetilde{u}_4 \rightarrow e^{2 \pi i /3}\, \widetilde{u}_4~,\hspace{0.3 in} u_6 \rightarrow u_6~,
\eea
we then conclude that the combined operation of $\cN_3 \cR_3$ leaves only the invariants of degree $4$ and $6$. There is a candidate complex reflection group realizing these invariants, but it is no longer part of the infinite family of the form $G(m,p,n)$---instead, it is one of the \textit{exceptional} complex reflection groups, known as $G_4$. Our analysis thus predicts that twisted compactification of $\mathfrak{so}(8)$ SYM via the non-invertible symmetry $\cN_3 \cR_3$ gives rise to a 3d $\cN=6$ theory with moduli space $\CC^8/G_4$. No such theory is known, and as far as we are aware it is not possible to access this theory from 4d $\cN=4$ SYM via a twisted compactification by an invertible symmetry.\footnote{It can however potentially be obtained by untwisted compactification of one of the exotic 4d $\cN=3$ theories studied in \cite{Kaidi:2022lyo}.}

\subsection{Compactifications of  $\mathfrak{e}_6$ and $\mathfrak{e}_7$ SYM}
We now comment on the two exceptional cases which admit non-invertible compactifications, namely $\mathfrak{e}_6$ and $\mathfrak{e}_7$ SYM. 
 The invariant polynomials of $\cW(\mathfrak{e}_6)$ are of dimension $2,5,6,8,9,12$, and upon twisting the following invariants persist
\bea
k&=& 3: \hspace{0.5 in} 6, 9, 12~,
\no\\
k&=& 4: \hspace{0.5 in} 8, 12~,
\no\\
k&=& 6: \hspace{0.5 in} 6, 12~.
\eea
These are the correct invariants for, respectively, $G_{25}$, $G_8$, and $G_{5}$. We thus predict that upon twisted compactification by $\cN_k \cR_k$ for $k=3,4,6$, we will get 3d $\cN=6$ theories with moduli spaces labelled by $\Gamma =G_{25}, G_8$, and $G_{5}$. No such 3d $\cN=6$ theories are known. Hence these again provide examples of novel theories obtained by non-invertible twisted compactification. 
Indeed, for the $(E_6)_0$ theory, each of $\cN_4 \cR_4$, $\cN_3 \cR_3$, and $\cN_6 \cR_6$ is non-invertible, and furthermore $\cN_4 \cR_4$ is {intrinsically} non-invertible.

Moving on to the case of $\mathfrak{e}_7$, we first note that the invariant polynomials of $\cW(\mathfrak{e}_7)$ are of dimension 2,6,8,10,12,14,18.  In this case the $\ZZ_6$ twist is the same as a $\ZZ_3$ one, and so our only options are 
\bea
k&=& 3: \hspace{0.5 in} 6,12,18~,
\no\\
k&=& 4: \hspace{0.5 in} 8, 12~.
\eea
These are the correct invariants for $G_{26}$ and $G_8$, respectively. We would thus predict that upon twisted compactification by $\cN_3 \cR_3$ and $\cN_4 \cR_4$, we will get 3d $\cN=6$ theories with moduli spaces labelled by $\Gamma =G_{26}$ and $G_8$, respectively.

\subsection{Compactifications of non-simply-laced SYM}
\label{sec:nonsimplylacedcomp}
We finally consider the non-simply laced cases, in particular $\mathfrak{usp}(2N)$ and $\mathfrak{so}(2N+1)$. As discussed in Section \ref{eq:nonsimplaced}, the only cases with non-invertible defects are the theories with $N$ odd at $\tauYM = {1 \pm i \over 2}$, as well as 
$\mathfrak{usp}(4) \cong \mathfrak{so}(5)$ at $\tauYM = {i \over \sqrt{2}}$. Hence we will focus on these examples. We will denote the non-invertible symmetries in both cases by $\widetilde{\cN}_4 \cR_4$, where in the case of odd $N$ the corresponding $\widetilde{\gamma}_4 = (\widetilde{\mS} \mT)^2$, while in the case of  $\mathfrak{usp}(4)$ the corresponding $\widetilde{\gamma}_4 = \widetilde{\mS} $. 

Let us begin with the case of odd $N$. The dimensions of the invariant polynomials are $2,4,6,8, \dots, 2N$. Under the action of $\widetilde{\cN}_4 \cR_4$, the invariants which persist are 
\bea
k=4: \hspace{0.5 in} 4, 8, 12, \dots~. 
\eea
These are the correct invariants for the group $G(4,1,n)$, and thus we expect that the result theory has moduli space of the form $ \CC^{4n} /G(4,1,n)$. These moduli spaces can be realized by the ABJ theories of type $U(n + 2)_4 \times U(n)_{-4}$ for $N = 2n + 1$. As usual, this conclusion is unchanged for any choice of global variant of the SYM theory, as long as the correct $\widetilde{\cN}_4 \cR_4$ is chosen.

Next we consider the case of $\widetilde{\cN}_4 \cR_4$ in $\mathfrak{usp}(4)$. 
Unlike all of the previous cases, the $\widetilde{\mS}$ transformation has a non-trivial action on the moduli space, on which it acts as the outer automorphism of the Weyl group of $\mathfrak{usp}(4)$ associated with the reflection of the Dynkin diagram. In terms of the Coulomb branch operators of dimensions 2 and 4, it acts as $\Delta_2\rightarrow \Delta_2$, $\Delta_4\rightarrow -\Delta_4$. This means that the combined action  $\widetilde{\cN}_4 \cR_4$  actually projects out \textit{all} of the moduli, thereby leaving a theory with trivial moduli space. We expect that the theory obtained in this way is the theory of 2 M2-branes stuck to a $\ZZ_4$ singularity in M-theory.

\section*{Acknowledgements}

We thank Mario Martone and Kantaro Ohmori for discussions during the early stages of this work, and for collaboration on related topics. We would also like to thank Yichul Choi, Ho Tat Lam, Sahand Seifnashri, Shu-Heng Shao, and Yuji Tachikawa for useful discussions, as well as Yuji Tachikawa for comments on a draft. GZ is supported in part by the Simons Foundation grant 815892. YZ is partially supported by WPI Initiative, MEXT, Japan at IPMU, the University of Tokyo.
\newpage

\appendix

\section{Non-intrinsic $N$-ality defects}
\label{eq:Ndualfusion}

As discussed in the Introduction, the presence or lack of non-invertible symmetries in a given global variant of an $\cN=4$ theory can depend on which patch of the upper half-plane one is working in. When a non-invertible symmetry can be ``removed" by changing to a different patch, the non-invertible symmetry is referred to as \textit{non-intrinsic}. In this appendix we give examples of non-intrinsic symmetries of $\mathfrak{su}(N)$ SYM at $\tauYM=1$. Since the point $\tauYM=1$ is in the same modular orbit as $\tauYM=i \infty$, these non-intrinsic symmetries will all be related to $\mT$ (combined with appropriate factors of $\tau$) in some way or another. 

We begin by noting that $\tauYM=1$ is left fixed by $\mS \mT^{-2}$. We may then identify potential (non-)invertible defects by first applying $\mS \mT^{-2}$ and then asking for the combinations of $\sigma$ and $\tau$ needed to map back to the original theory. Following the methods in Section \ref{sec:suN}, we have for $SU(N)_m$,
\begin{eqnarray}
\begin{split}
    Z_{SU(N)_m}[\tauYM, B]& \stackrel{\mT^{-2}}{\longrightarrow} Z_{SU(N)_{m-2}}[\tauYM-2, B]\\& \stackrel{\mS}{\longrightarrow} Z_{PSU(N)_{0,m-2}}[-1/(\tauYM-2), B]\\& \stackrel{\tau^{-m+2}}{\longrightarrow} Z_{PSU(N)_{0,0}}[-1/(\tauYM-2), B]\\& \stackrel{\sigma^3}{\longrightarrow} Z_{SU(N)_{0}}[-1/(\tauYM-2), B]\\& \stackrel{\tau^{m}}{\longrightarrow} Z_{SU(N)_{m}}[-1/(\tauYM-2), B]~.
\end{split}
\end{eqnarray}
 Hence the $SU(N)_n$ theory at $\tauYM=1$ is invariant under $\tau^m \sigma^3 \tau^{-m+2} \mS\mT^{-2}$. This turns out to be an order $N$ symmetry. To see this, we first note that
\begin{equation}
\begin{split}
    (\tau^m \sigma^3 \tau^{-m+2})^N= \cC^N \tau^{m-1} (\tau \sigma \tau)^N \tau^{-m+1} =\cC^N &\tau^{m-1} (\sigma \tau^{-1} \sigma)^N \tau^{-m+1}= 1
\end{split}
\end{equation}
by the $SL(2,\ZZ_{N})$ algebra. Moreover, we have that
\begin{eqnarray}
(\mS\mT^{-2})^N = \mT (\mT^{-1}\mS \mT^{-1})^N \mT^{-1}=\mT (\mS\mT\mS)^N \mT^{-1} \mC^N = \mT \mS \mT^{N} \mS \mT^{-1} \mC^N = 1~.
\end{eqnarray}
The last equality requires some explanations. Here we have used $\mT^N=1$, which is not strictly true, but holds in this context. Specifically, as $\mT^N$ maps all global variants back to themselves, the only thing that changes is the value of $\tauYM$. However, note that $\mT^N$ here acts on $\mS \mT^{-1} (\tauYM=1)\rightarrow \tauYM=\infty i$, which is invariant under $\mT$. As such $\mT^N$ can be taken to be the identity operation in this context. Another way of phrasing this is that the operation $(\mS\mT^{-2})^N$ maps every global variant to itself while keeping $\tauYM=1$ fixed. We conclude then that $SU(N)_{m}$ SYM at $\tauYM=1$ has a non-invertible $N$-ality symmetry $\tau^m \sigma^3 \tau^{-m+2} \mS\mT^{-2}$.

For $PSU(N)_{2, m}$, we have
\begin{equation}
\begin{split}
    Z_{PSU(N)_{2,m}}[\tauYM, B]& \stackrel{\mT^{-2}}{\longrightarrow} Z_{PSU(N)_{2,m}}[\tauYM-2, B]\\& \stackrel{\mS}{\longrightarrow} Z_{SU(N)_{m}}[-1/(\tauYM-2), -B]\\& \stackrel{\tau^{-m+2}}{\longrightarrow} Z_{SU(N)_{2}}[-1/(\tauYM-2), -B]\\& \stackrel{\sigma^3}{\longrightarrow} Z_{PSU(N)_{2,0}}[-1/(\tauYM-2), B]\\& \stackrel{\tau^{m}}{\longrightarrow} Z_{PSU(N)_{2,m}}[-1/(\tauYM-2), B]~.
\end{split}
\end{equation}
Hence $PSU(N)_{2,m}$ at $\tauYM=1$ is invariant under $\tau^m \sigma^3 \tau^{-m+2} \mS\mT^{-2}$, which is a non-invertible $N$-ality symmetry. 

\begin{table}
	\centering
	\begin{tabular}{| c | c |c |}
	
	\hline
		Theory & Defect  & $n$-ality \\
		\hline\hline
		$SU(N)_m, \, PSU(N)_{2,m}$ & $\tau^m\sigma^3 \tau^{-m+2} \mS\mT^{-2}$ & $N$\\
		
		$PSU(N)_{n,m},~~ (n\neq 1,2\mod N)$ & $\tau^m \sigma \tau^n \sigma \tau^{-n+2} \sigma  \tau^{-m} \mS\mT^{-2}$ &$N$ \\

        $PSU(N)_{1,m}$ & $\tau^{-1} \mS\mT^{-2}$ & 1\\
        \hline
	\end{tabular}
	\caption{(Non-)invertible symmetries of $\mathfrak{su}(N)$ with $N$ being a prime integer at $\tauYM=1$.}
	\label{tab:suNNalitytable}
\end{table}

For $PSU(N)_{n,m}$ with $n\neq 2\mod N$, we have
\begin{equation}
\begin{split}
    Z_{PSU(N)_{n,m}}[\tauYM, B]&\stackrel{\mT^{-2}}{\longrightarrow} Z_{PSU(N)_{n,m}}[\tauYM-2, B]\\&\stackrel{\mS}{\longrightarrow} Z_{PSU(N)_{-x,m(n-2)^2-(n-2)}}[-1/(\tauYM-2), -xB]\\& \stackrel{\tau^{-m}}{\longrightarrow} Z_{PSU(N)_{-x,-n+2}}[-1/(\tauYM-2), -xB]\\& \stackrel{\sigma}{\longrightarrow} Z_{PSU(N)_{0,n-2}}[-1/(\tauYM-2), -B]\\& \stackrel{\tau^{-n+2}}{\longrightarrow} Z_{PSU(N)_{0,0}}[-1/(\tauYM-2), -B]\\& \stackrel{\sigma}{\longrightarrow} Z_{SU(N)_0}[-1/(\tauYM-2), B]\\& \stackrel{\tau^{n}}{\longrightarrow} Z_{SU(N)_n}[-1/(\tauYM-2), B]\\& \stackrel{\sigma}{\longrightarrow} Z_{PSU(N)_{n,0}}[-1/(\tauYM-2), B]\\& \stackrel{\tau^{m}}{\longrightarrow} Z_{PSU(N)_{n,m}}[-1/(\tauYM-2), B]~.
\end{split}
\end{equation}
Hence $PSU(N)_{n,m}$ for $n\neq 2\mod N$ at $\tauYM=1$ is invariant under $\tau^m \sigma \tau^n \sigma \tau^{-n+2} \sigma  \tau^{-m} \mS\mT^{-2}$. It is straightforward to verify that $(\tau^m \sigma \tau^n \sigma \tau^{-n-1} \sigma  \tau^{-m}\mS\mT^{-2})^N=1$. However, when $n=1$ and for any $N$, the symmetry reduces to $\tau^{-1}$ which is an invertible symmetry. Except this special case, $\tau^m \sigma \tau^n \sigma \tau^{-n+2} \sigma  \tau^{-m} \mS\mT^{-2}$ is a non-invertible $N$-ality symmetry. We enumerate the symmetries in Table \ref{tab:suNNalitytable}.  The results are consistent with the statement that the $N$-ality defect is non-intrinsic: for any $N$, there is a global variant, i.e. $PSU(N)_{1,m}$, for which the symmetry becomes invertible.

\bibliographystyle{JHEP}
\bibliography{bib}
\end{document}